\documentclass[twocolumn,aps,prm,superscriptaddress,floatfix]{revtex4-1} %

\usepackage{amsmath}
\usepackage{amssymb}
\usepackage{color}
\usepackage[english]{babel}
\usepackage{enumerate}
\usepackage{float}
\usepackage{placeins}
\usepackage{graphicx}
\usepackage[]{hyperref}
\usepackage{bm}
\usepackage{mhchem}
\usepackage{xcolor}
\hypersetup{
    colorlinks,
    linkcolor=blue,
    citecolor=blue,
    urlcolor=blue,
}

\usepackage{xr}
\newcommand{\supplref}[1]{Fig.~(#1) in Ref.~\cite{kahle_supplementary_2019}}

\newcommand*{\myeqref}[2][Eq.~]{#1(\ref{#2})}

\newcommand*{\myfigref}[2][Fig.~]{#1\ref{#2}}

\newcommand*{\mysecref}[2][Sec.~]{#1\ref{#2}}

\DeclareMathOperator*{\argmax}{arg\,max}

\renewcommand*\vec[1]{\mathbf{#1}}

\begin{document}

\title{Unsupervised landmark analysis for jump detection in molecular dynamics simulations}

\author{Leonid Kahle}
\thanks{These two authors contributed equally to this work.}

\author{Albert Musaelian}
\thanks{These two authors contributed equally to this work.}

\author{Nicola Marzari}
\affiliation{Theory and Simulation of Materials (THEOS), and National Centre for Computational Design and Discovery of Novel Materials (MARVEL), \'{E}cole Polytechnique F\'{e}d\'{e}rale de Lausanne, CH-1015 Lausanne, Switzerland}
\author{Boris Kozinsky}
\affiliation{John A. Paulson School of Engineering and Applied Sciences, Harvard University, Cambridge, Massachusetts 02138, USA}
\date{\today}


\begin{abstract}
Molecular dynamics is a versatile and powerful method to study diffusion in solid-state ionic conductors, requiring minimal prior knowledge of equilibrium or transition states of the system's free energy surface.
However, the analysis of trajectories for relevant but rare events, such as a jump of the diffusing mobile ion, is still rather cumbersome, requiring prior knowledge of the diffusive process in order to get meaningful results.
In this work, we present a novel approach to detect the relevant events in a diffusive system without assuming prior information regarding the underlying process.
We start from a projection of the atomic coordinates into a landmark basis
to identify the dominant features in a mobile ion's environment.
Subsequent clustering in landmark space enables a discretization of 
any trajectory into a sequence of distinct states.
As a final step, the use of the smooth overlap of atomic positions descriptor allows distinguishing between different environments in a straightforward way.
We apply this
algorithm to ten Li-ionic systems and perform
in-depth analyses of cubic \ce{Li7La3Zr2O12}, tetragonal \ce{Li10GeP2S12}, and the $\mathrm{\beta}$-eucryptite \ce{LiAlSiO4}. We compare our results
to existing methods,
underscoring strong points, weaknesses, and insights into the diffusive behavior of the ionic conduction in the materials investigated.
\end{abstract}
\maketitle

\section{Introduction}
\label{sec:intro}
Lithium-ion batteries power an increasingly broad and critical set of technologies~\cite{armand_building_2008}.
Commercially available batteries use organic electrolytes that impose constraints on their safety, power and energy density~\cite{schaefer_electrolytes_2012}
and can introduce chemical instabilities that require the incorporation of fuses and safety vents~\cite{balakrishnan_safety_2006}.
Solid-state electrolytes are widely considered to be a promising alternative for next-generation  batteries: many structural families of candidate solid-state ionic conductors have been identified and are under investigation~\cite{bachman_inorganic_2016}.
While a good solid-state electrolyte must meet several criteria, such as low electronic mobility, easy device integration, and electrochemical and mechanical stability~\cite{manthiram_lithium_2017}, it must first be a fast Li-ion conductor, and consequently optimization of conductivity and analysis of the mechanisms of Li-ion diffusion has been the focus of a large body of literature~\cite{bachman_inorganic_2016}.

Atomistic modeling techniques, and in particular molecular dynamics (MD), have been used to study a wide variety of candidates for solid-state electrolytes and the factors that influence their ionic conductivity.
Classical/empirical force fields were chosen in several studies~\cite{adams_structural_2012, adams_ion_2012, xu_mechanisms_2012, deng_structural_2015, kozinsky_effects_2016, klenk_finite-size_2016, burbano_sparse_2016, dawson_atomic-scale_2018} due to their computational efficiency and access to the time and length scales required to characterize ionic transport.
Accurate, yet expensive, first-principles simulations have also been employed for selected systems~\cite{wood_dynamical_2006, ong_phase_2013, mo_first_2012, xu_one-dimensional_2012, mo_insights_2014, meier_solid-state_2014, wang_design_2015, chu_insights_2016, zhu_li3yps42_2017, marcolongo_ionic_2017,sagotra_influence_2019}.
The necessary compromise between the transferability of first-principles potential energy surfaces and the computational efficiency of force fields has also motivated the development of novel hybrid quantum/classical approaches~\cite{kahle_modeling_2018} to model diffusion.
The estimate of transport coefficients from the Green-Kubo or Einstein relations using molecular dynamics can be done in a straightforward yet expensive way,
though improved methods for obtaining accurate estimates from short trajectories are being developed~\cite{ercole_accurate_2017}.
In addition to computing ionic conductivity, design and characterization of new materials requires detailed understanding of the atomistic mechanisms of ionic transport.
The central challenge is to develop automated methods for accurately analyzing the structure and dynamics of lithium's local atomic environments and for detecting rare transitions and subtle correlations in large amounts of data.

\begin{figure}[t!]
	\centering
    \includegraphics[width=\hsize]{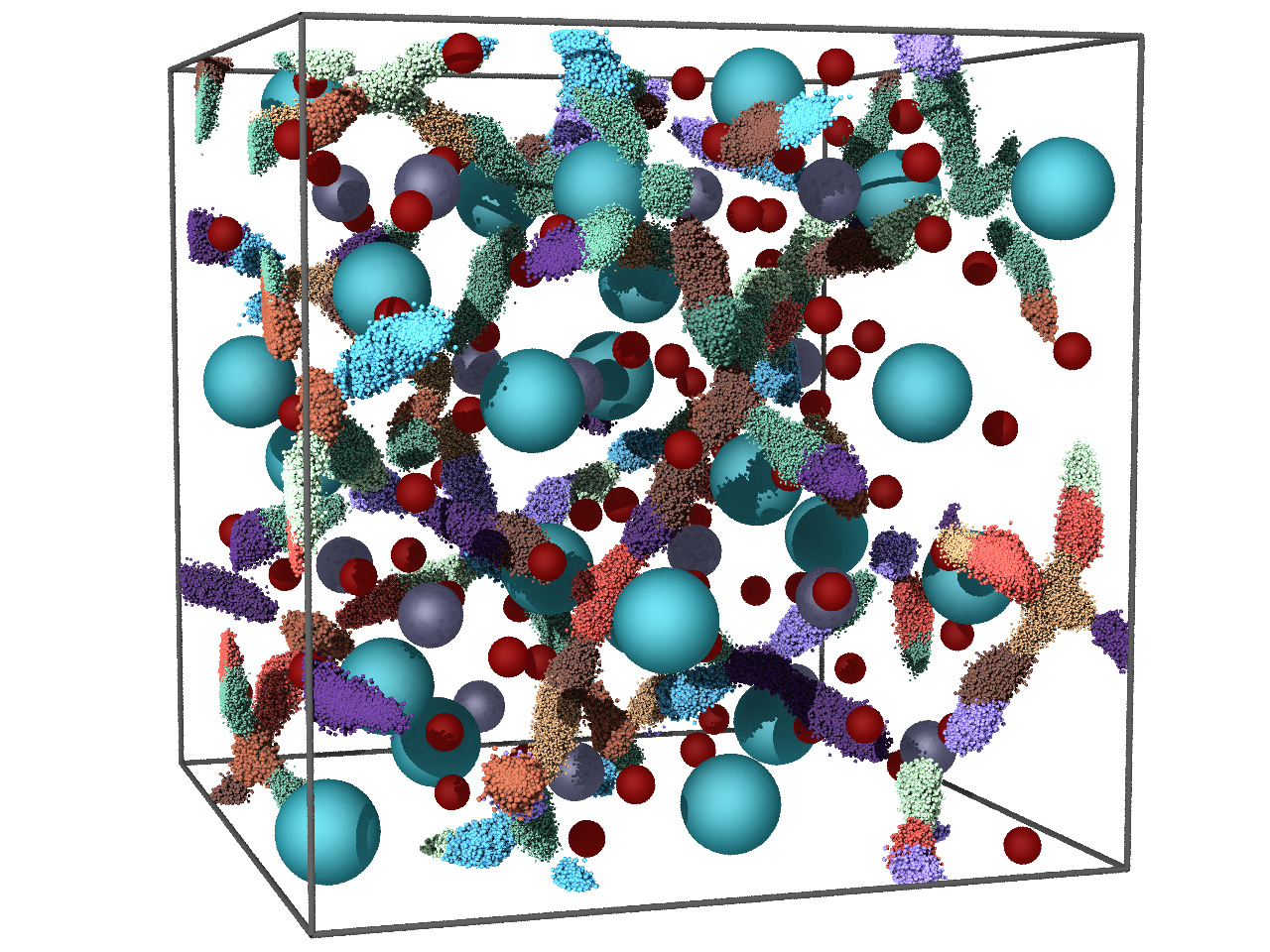}
    \caption{
The site analysis is exemplified above for \ce{Li7La3Zr2O12}, based on the results discussed in Sec.~\ref{sec:llzo}.
The equilibrium positions of lanthanum are shown as large blue spheres, those of zirconium as grey spheres, and those of oxygen as red spheres.
The positions of lithium during the trajectory are collapsed into the same frame and shown as small spheres, with color and reflectivity being chosen according to the site associated with the ion in that frame.}
\label{fig-LLZO-scene}
\end{figure}

In many solid-state Li-ion conductors, Li ions form a mobile, often disordered, sublattice within a non--diffusive sublattice of the other species, which we refer to as the host lattice hereafter.
In the jump-diffusion model, the mobile ions spend the majority of their time in the local minima of the potential energy surface and vibrate within such sites for a sufficiently long time to lose memory of their previous locations while intermittently acquiring sufficient kinetic energy to overcome the barrier separating them from a different potential well.
This formulation of Li-ion diffusion as occupation of and exchange between well-defined crystallographic sites can be used to model diffusion as a Markov-chain model using kinetic Monte Carlo~\cite{van_der_ven_first-principles_2001}.
Also, using this discrete formulation to understand the microscopic origin of diffusion is a common theme in the literature,
and site analysis tools have been used to explore the effects of site volume~\cite{wang_design_2015, kweon_structural_2017} and anion sublattice structure~\cite{wang_design_2015} on ionic conductivity,
to identify conduction pathways and rate-limiting steps~\cite{kozinsky_effects_2016, de_klerk_analysis_2018}, 
and correlated diffusion events~\cite{chen_data_2017}, and to design new descriptors for conductivity~\cite{kweon_structural_2017}.
Ideally, an automated site analysis (see \myfigref{fig-LLZO-scene}) approach should: (1)~automatically identify relevant Li sites, (2)~accurately track migration of mobile ions through those sites, (3)~require no prior knowledge of the material, and (4)~work with the same parameters over a broad range of materials.

Existing approaches fall mostly into three classes: distance-based, topology-based, and density-based methods.
Distance-based methods~\cite{varley_understanding_2017,  kweon_structural_2017, de_klerk_analysis_2018} use preexisting knowledge of the equilibrium positions of all Li-ion sites and consider a Li ion to be resident at a site when it comes within a given cutoff distance from the site's position.
 Cutoffs can be smooth~\cite{kweon_structural_2017} or discrete~\cite{de_klerk_analysis_2018}, but in both cases, they need to be tailored to the structure at hand and are uniform across all sites within it. 
 The positions of sites can also be coupled to the instantaneous positions of nearby host-lattice atoms~\cite{kweon_structural_2017} to decrease sensitivity to thermal noise. Nevertheless, such methods rely on the crystallographic information they are given and also do not account for the varied or non-spherical geometry of sites~\cite{deng_crystal_2018}.
Starting from a prior knowledge of the host structure and possible Li sites, mobile ions can also be automatically assigned to sites based on convex-hull analysis of site polyhedra~\cite{kozinsky_effects_2016, kozinsky_transport_2018}. This topology-based method deals with arbitrary site geometries, eliminates thermal noise and does not require arbitrary distance cutoffs, but does require the site polyhedra to be specified.
Density-based methods~\cite{chen_data_2017} identify regions of high Li-ion density separated by areas of low Li-ion density, as determined by a threshold, and define each high-density region as a site. 
These methods thus do not require prior knowledge about the material and can resolve sites with different geometries. In materials with nearby or rapidly exchanging sites, however, choosing a density threshold that can distinguish such sites from one another can be difficult.
Richards \textit{et al.}~\cite{richards_design_2016} used a k-means clustering of Na-ion positions in \ce{Na10GeP2S12}, initialized with known ionic positions for the similar ionic conductor \ce{Li10GeP2S12}, combining prior information with a density based method.

In order to overcome some of these challenges, this work introduces an algorithm for accurately and automatically analyzing molecular dynamics trajectories and detecting jumps of the mobile ion through the host lattice  with minimal human supervision and no prior knowledge.
This algorithm can be combined with the automatic detection of important structural motifs~\cite{gasparotto_recognizing_2014,gasparotto_recognizing_2018}, leading to a versatile tool for the unsupervised analysis of trajectories and detection of diffusion events. 
The algorithm will be discussed in \mysecref{sec:algorithms};
in \mysecref{sec:results} we apply it to three known ionic conductors and to seven non-diffusing materials and discuss the results;
some details of the implementation are given in \mysecref{sec:implementation}; 
and our final conclusions are presented in \mysecref{sec:conclusion}.

\section{Algorithm}
  \label{sec:algorithms}

Landmarks are persistent local features in an environment and therefore can be used to describe positions in the absence of global information (i.e. real-space coordinates).
Landmark-based navigation explains the homing of social insects~\cite{wehner_visual_1996} and has been applied in the field of autonomous navigation and artificial intelligence~\cite{moller_insect_2000}.
Landmark models employ a vector-based description of the environment via a landmark vector $\bm l$. Such a vector representation is useful for navigation if the distances between the landmark vectors corresponding to two states or positions $A$ and $B$ decrease with reducing distances in real space: $|\bm r_A - \bm r_B| = f(|\bm l_A - \bm l_B|$), where $f$ is a monotonically increasing function of its input.

When analyzing trajectories, the real-space positions are obviously known beforehand.
However, atomic coordinates are inefficient descriptors for most properties since they are not invariant under rigid translation or rotation of the structure.
We will describe the positions of mobile ions through landmarks that encode all the information necessary to detect changes in the ions' environments and are invariant under these transformations.

First, we deduce that the descriptors should only encode local information since the local environment mostly defines the potential energy landscape for the mobile ion, a principle reminiscent of the nearsightedness of electronic matter~\cite{prodan_nearsightedness_2005}.
In addition, we know from Pauling's rules in crystal structures~\cite{pauling_principles_1929} that ionic systems minimize their energy by packing into coordination numbers that are determined, among other factors, by the ratio of the radii of the cations and anions.
Therefore, possible coordination polyhedra in the local environment are meaningful features.
Checking all the possible polyhedra in a crystal is not feasible because of the combinatorial complexity this induces, so we need to restrict the description via a meaningful subset of convex hulls or polyhedra formed by the host lattice. A site description and trajectory discretization via pre-selected convex hulls has been previously developed and applied~\cite{kozinsky_effects_2016} to study Li-ion diffusion in garnets.

Using polyhedra defined by host-lattice atoms as landmarks relies on the assumption that these atoms fluctuate around equilibrium positions, such that well-defined coordination polyhedra persist throughout the simulation. Equivalently, the host lattice is not changing in a way that causes sites to appear or disappear.
Due to this assumption, the present landmark analysis cannot be applied to systems with liquid-like host structure such as polymers, where inter-site hopping happens on a longer time scale than the host motion, and dynamic coordination tracking must be used~\cite{molinari_effect_2018}. It also cannot be applied to systems with a ``paddle-wheel'' diffusion mechanism, where the slow rotation of polyatomic anions creates a constantly changing set of local potential minima, such as shown for proton diffusion in \ce{CsHSO4}~\cite{wood_proton_2007} or lithium- and sodium-ionic diffusion in the closoborate structures~\cite{kweon_structural_2017}.
Similar to the site analysis presented in the literature, our method does not assume that the occupation a given site are Markovian, i.e. whether a mobile ion completely loses memory of its past at any site: We define and find a site based on stable and persistent features in the environment of mobile ions, described by the landmark vectors, without considering information in the time domain. Whether the underlying process is Markovian can be determined by analyzing the resulting statistics~\cite{chen_data_2017,morgan_relationships_2014}.

The basic algorithm has three steps: (1)~definition of suitable landmarks, (2)~expression of the coordinates of the mobile ions during their trajectory in the landmark basis, and (3)~clustering of the landmark vectors to reveal sites and discretize the trajectory of each mobile atom.
We also implemented, as an option, the possibility to: (4)~merge nearby sites that have high exchange rates and that fulfill some distance criteria and (5)~determine site types based on the geometry and chemistry of the local environment.

While the two last steps are optional and independent from each other, we always apply them in the analysis that we show in Sec.~\ref{sec:results}. Step 4 reduces significantly the noise in the data, and step 5 supplies information on the local geometry and chemistry. In the following we explain each of the steps in greater detail and finish with a discussion on why certain design choices were taken.

\subsection{Step 1: Define Landmarks}
\label{subsec:step1}

\begin{figure}[t]
    \includegraphics[width=\hsize]{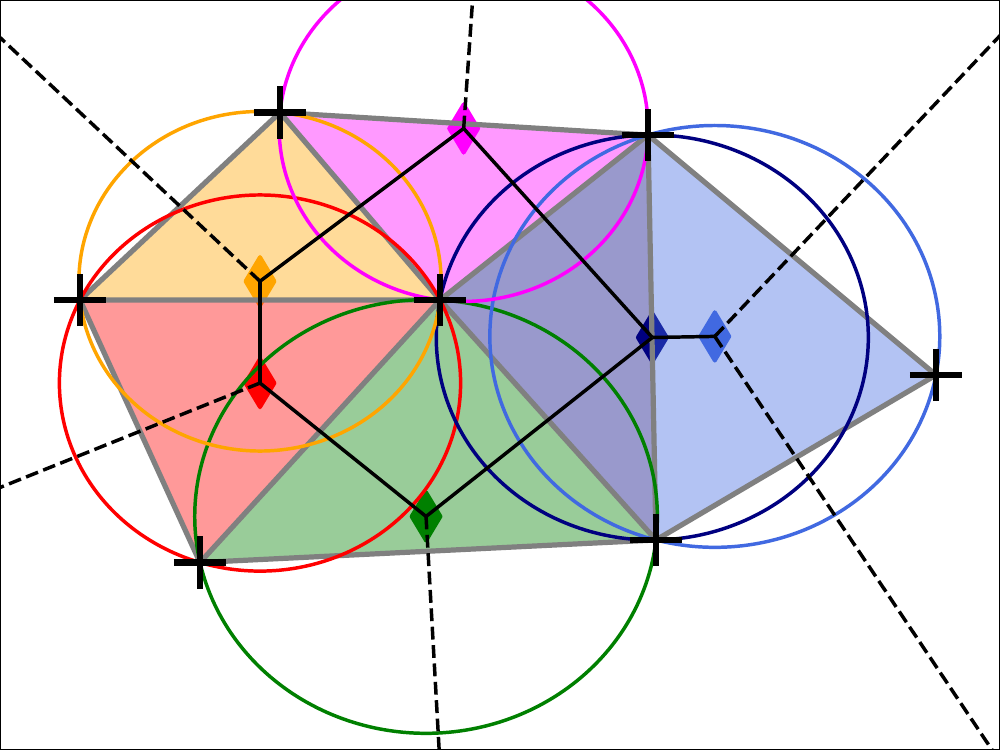}
    \centering
    \caption{Schematic of the Voronoi tessellation in two dimensions of seven seed points (black crosses). The resulting Voronoi facets are shown as black solid lines; dashed black lines are Voronoi facets that are not bounded by a Voronoi node.
    The Voronoi nodes are shown as coloured diamonds, and the associated Delaunay triangles formed by their seeds are filled with the same color. The circumcircles of each Delaunay triangle are shown in the same color, demonstrating that no seed point is inscribed in them and that the associated Voronoi node is at its center.}
    \label{fig:voronoi}
\end{figure}

The landmark analysis we introduce here is based on the Voronoi tessellation of the equilibrium configuration of the host lattice and its geometric dual, the Delaunay triangulation.
Given a set of points in space, termed seeds, a Voronoi tessellation divides space into regions such that all points in each region are closer to the region's seed than to any other seed~\cite{okabe_spatial_2009}.
Formally, the Voronoi region determined by the seed point $s_i \in \mathbb{R}^n$ is given by:
\[
    R_i = \{ x \in \mathbb{R}^n : |\vec{x} - \vec{s}_i| \leq |\vec{x} - \vec{s}_j| \text{ for all } j \neq i \}
\]
Voronoi regions connect at Voronoi facets, as shown in the schematic in \myfigref{fig:voronoi}. Any point on such a facet is equidistant to the seeds of the adjacent Voronoi regions. 
Voronoi nodes are, in a space of $D$ dimensions, points where at least $D$ facets intersect and therefore are equidistant to at least $D+1$ seed points. It follows that each Voronoi node locally maximizes the distance to its adjacent seed points.
The geometric dual of the Voronoi tessellation is the Delaunay triangulation. Such a triangulation or simplicial~\footnote{A simplex in $\mathbb{R}^D$ is the convex hull of $D+1$ points that do not lie on a hyperplane.} decomposition is obtained by connecting seed points that share a Voronoi facet. 
The Delaunay triangulation has the useful property that the circumcircles of all formed triangles have empty interiors, i.e. there are no seed points inside any circumcircle.
It follows from the duality between the two tessellations  that every Voronoi node is associated to exactly one Delaunay simplex. The Voronoi node lies at the center of the circumcircle of the associated Delaunay simplex.
In the remainder of the text we will work in three dimensions unless otherwise specified; in three dimensions a Voronoi node is equidistant to at least four coordinating seeds.

While a Voronoi node is a reasonable guess for a low-energy position since it maximizes the distance to its coordinating seeds, the associated Delaunay simplex corresponds to the coordinating polyhedron of the site or a subset thereof.
A Voronoi node and its coordinating host-lattice atoms are together referred to as a landmark.
The coordinating host-lattice atoms of a landmark are the host-lattice atoms that are vertices of the Delaunay simplex -- dual to the Voronoi node -- in the equilibrium configuration. 

\subsection{Step 2: Landmark Vectors}
\label{subsec:step2}

We start from a molecular dynamics trajectory that gives the real-space positions $\vec{r}_i(t) \in \mathbb{R}^3$
of each atom $i$ at time $t=N\Delta t_s$, an integer multiple of the sampling timestep $\Delta t_s$.
In the remainder, we will use the index $h$ for host-lattice atoms and $m$ for mobile ions.
First, we calculate the time-averaged positions for host-lattice atoms $\bar{\vec{r}}_{h}=\langle\vec{r}_h\rangle_t$ and use
these as seed points for a Voronoi decomposition, resulting in Voronoi nodes $\bar{\vec{r}}_{\text{VN}}^\mathrm{A}$.
The instantaneous position of a mobile particle, $\vec{r}_{m}(t)$, is expressed in terms of a proximity or similarity to each landmark in the system.
That is to say, any real space position  $\vec{r}_{m}$ can be transformed into a vector in the $N$-dimensional landmark space, where $N$ is the number of landmarks in the system, equal to the number of Voronoi nodes and also equal to the number of Delaunay simplices, due to the duality discussed in Sec.~\ref{subsec:step1}.
We index landmarks with capital latin characters.
For a landmark $\mathrm{A}$, we first define the normalized instantaneous distance between a mobile particle $m$ and a host lattice atom $h$, where atom $h$ is one of the coordinating seed atoms of the landmark $\mathrm{A}$:

\begin{equation}
d_{m,h}^{\mathrm{A}}(t) =
        \frac{ %
              \left| \vec{r}_{m}(t) - \vec{r}_{h}(t) \right|%
        }{%
              \left| \bar{\vec{r}}_{\text{VN}}^\mathrm{A} - \bar{\vec{r}}_{h} \right|%
        }, \quad h \in \mathrm{A},
\end{equation}
where
$\vec{r}_{h}(t)$ and $\vec{r}_{m}(t)$ are the instantaneous real-space positions of host-lattice atom $h$ and mobile ion $m$, respectively,
$\bar{\vec{r}}_{h}$ is the time-averaged position of host-lattice atom $h$, and
$\bar{\vec{r}}_{\text{VN}}^\mathrm{A}$ is the position of the landmark's Voronoi node.
The corresponding component of the landmark vector is then computed as:
\begin{equation}
\label{eq:landmark-def}
    l_{\mathrm{A}}^m = \prod_{h}^{N_\text{coord}}\left[f\left(
        d_{m,h}^{\mathrm{A}}\right)\right]^{\frac{1}{N_\text{coord}}},
\end{equation}
where
$h$ ranges over the set of $N_\text{coord}$ coordinating host-lattice atoms and
$f(d)$ is a cutoff function that smoothly goes from 1 to 0.
We base the cutoff function on the logistic function $\sigma(d; d_0,k)$, a sigmoid curve that goes from 0 to 1 around a midpoint $d_0$, with a steepness $k$:
\begin{equation}
    \sigma(d; d_0,k) = \frac{1}{1+e^{-k\left(d-d_0\right)}}.
\end{equation}
To obtain a cutoff function suitable for \myeqref{eq:landmark-def}, we subtract the logistic function from 1:
\begin{align}
    f(d; d_0, k) =& 1 - \sigma(d; d_0, k) = \frac{1}{1+e^{k\left(d-d_0\right)}}.
    \label{eq:cutoff}
\end{align}
The function $f(d; d_0,k)$ varies smoothly from 1 to 0, reaching $\frac{1}{2}$ at the set midpoint $d_0$.
How fast it varies is tuned by the hyperparameter $k$.
As can be seen from \myeqref{eq:landmark-def}, we normalize $f\left(d_{m,h}^{\mathrm{A}}\right)$ for a varying $N_\text{coord}$. 
In three dimensions and in the present framework, $N_\text{coord}$ is always $4$, since we use a simplicial decomposition to determine the landmarks.
However, the framework could be changed to include a varying number of coordinating host-lattice atoms, motivating this normalization. 
The cutoff function in \myeqref{eq:cutoff} was preferred due to its continuity and simplicity.
Because distances are normalized to the equilibrium distance between the Voronoi node and the host atoms,
the magnitude of each landmark vector component depends on neither the volume nor shape of the corresponding landmark's polyhedron.
This allows landmark analysis to distinguish between sites whose coordination polyhedra have very different volumes,
as well as accurately tracking mobile particles through highly distorted sites.

\subsection{Step 3: Landmark Clustering}
\label{subsec:step3}

\begin{figure*}[t!]
	\centering
    \includegraphics[width=\textwidth]{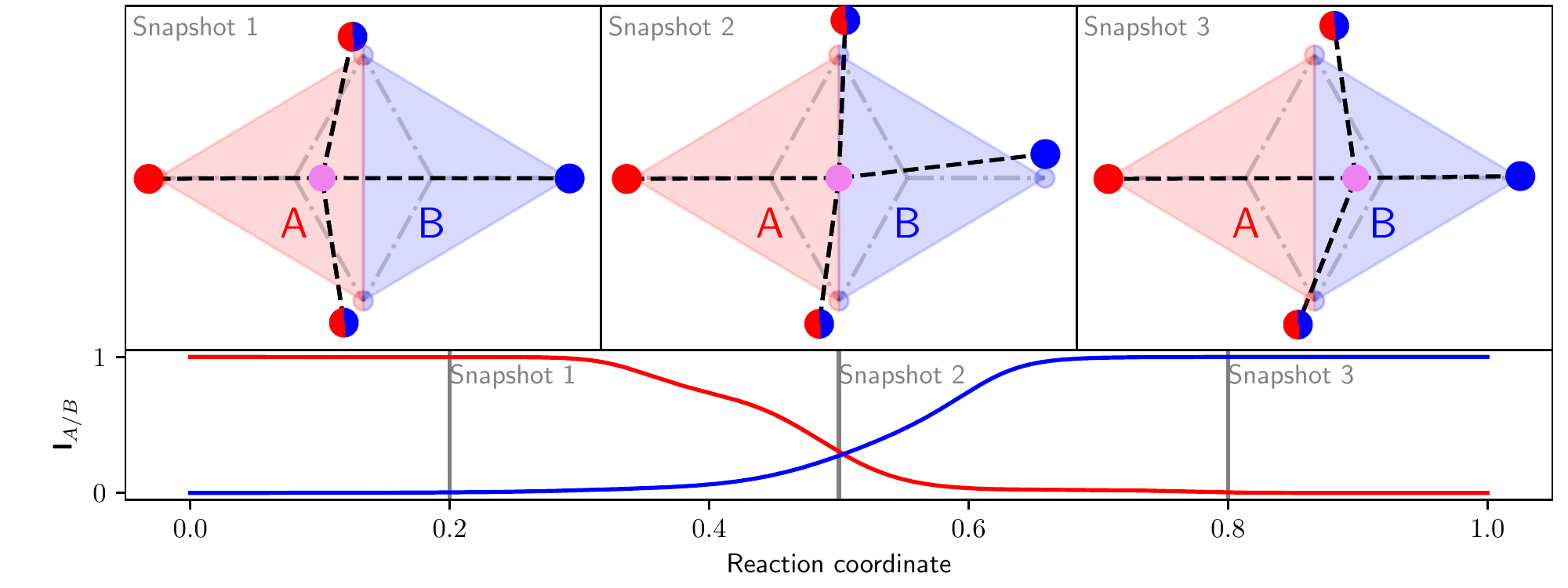}
    \caption{
    Simplified schematic to illustrate our algorithm:
    A mobile ion (in violet) is jumping from site A to site B along a straight line. The reaction coordinate of the jump takes a value of 0 when the ion is at the Voronoi node of A and a value of 1 when it is at the Voronoi node of site B.
    The distance to the neighboring host atoms is marked with a dashed black line.
    The host-lattice atoms are shown at an instantaneous position (equation of motion of an harmonic oscillator, initialized randomly) and are colored red if they are part of landmark A, blue if part of landmark B, and half red, half blue if they belong to both.
    We show the Delaunay triangulation based on the equilibrium positions of the host lattice as semi-translucent red and blue triangles. 
    The lower panel shows the landmark vector components of the mobile ion corresponding to A and B in red and blue respectively against the reaction coordinate.
    The vertical grey lines indicate the three snapshots shown in the top panel.
    We see that during the transition component A is decreasing while component B is increasing smoothly. At the transition point, the landmark components are approximately equal.
    }
    \label{fig:lmk-schema}
\end{figure*}

\begin{figure*}
	\centering
    \includegraphics[width=0.8\textwidth]{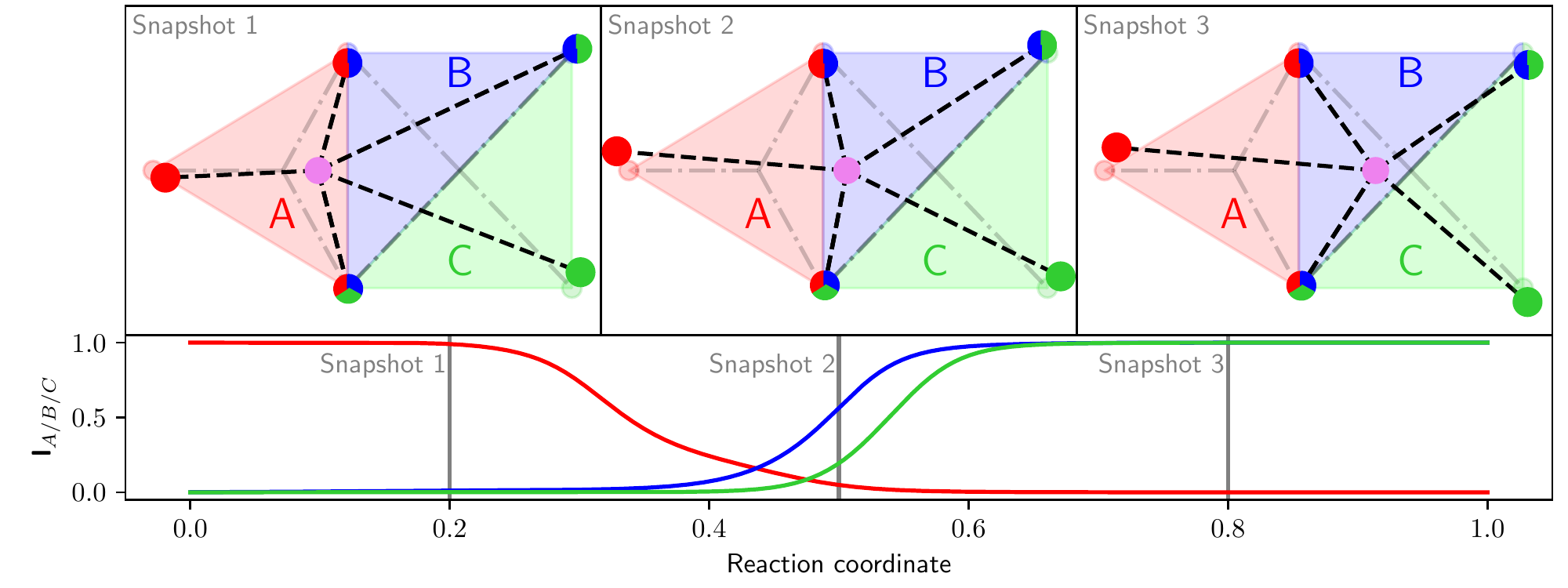}
    \caption{
    Simplified schematic to illustrate our algorithm for a non-simplicial site, similar to \myfigref{fig:lmk-schema}. The example contains two sites, one comprising landmark A and the other landmarks B and C.
    A mobile ion (in violet) is jumping from site A to site B/C along a straight line. The reaction coordinate of the jump takes a value of 0 when the ion is at the Voronoi node of A and a value of 1 when it is at the Voronoi nodes of site B/C.
    The distance to the neighboring host-lattice atoms is marked with a dashed black line.
    The host-lattice atoms are shown at an instantaneous position (equation of motion of an harmonic oscillator, initialized randomly), and are colored red if they coordinate landmark A, blue if they coordinate landmark B, and green if they coordinate landmark C.
    We show the Delaunay triangulation based on the equilibrium positions of the host lattice as semi-translucent red, blue, and green triangles.
    The lower panel shows the landmark vector components of the mobile ion corresponding to A, B and C in red, blue, and green, respectively, against the reaction coordinate.
    The vertical grey lines indicate the three snapshots shown in the top panel.
    We see that during the transition, component A is decreasing, while B and C are increasing similarly. The presence of the mobile ion at site on the right is therefore indicated by high values for both the B and C landmark vector components.
    }
    \label{fig:lmk-schema-octa}
\end{figure*}

The magnitude of each component of the landmark vector indicates the extent to which a mobile atom's position is dominated by that landmark.
If, for example, a mobile atom occupies a tetrahedral site, its landmark vector would have one large value at the corresponding landmark's component and some low-magnitude noise for neighboring landmarks. 
During a transition between sites, there are no dominant contributions, as shown schematically in two dimensions in \myfigref{fig:lmk-schema}.

If an atom occupies an octahedral site, however, the landmark vector will have four major contributions, corresponding to the four tetrahedrons resulting from the Delaunay triangulation of the octahedron.
We show this schematically for two dimensions in \myfigref{fig:lmk-schema-octa}.

Because we have chosen a smooth function of position for the landmark vector components, the landmark vectors are a continuous function of trajectory time.
By definition, landmark vectors are invariant under rigid translations or rotations of the system and as such are ideally suited as descriptors for dominant recurring features.
A clustering of the landmark vectors can be used to group similar landmark vectors and therefore discretize our trajectory in landmark space.
We use density-based clusters of landmark vectors, where each cluster is described by a high-density region in landmark space, corresponding to a frequent feature in the local environment of the mobile ion.
Therefore, we define sites as clusters in landmark space.

We use a custom hierarchical clustering algorithm (described in more detail in Appendix~\ref{sec:lvec-clustering}) with a simple cosine similarity metric:
\begin{equation}
    S(\vec{l}_A, \vec{l}_B) = \frac{\vec{l}_A \cdot \vec{l}_B}{|\vec{l}_A||\vec{l}_B|},
\end{equation}
where $\vec{l}_{A/B}$ are landmark vectors. The clustering algorithm scales linearly with the number of landmark vectors.

The clustering algorithm is run on the landmark vectors computed from the real-space positions of all mobile atoms every $n$ frames, where $n$ is sufficiently small and corresponds to a time span that is below the jump rate.
A mobile atom is said to be occupying site $i$ at time $t$ if its corresponding landmark vector at that time is a member of the $i$-th landmark cluster.
If the mobile atom's landmark vector is not a member of any cluster, the atom is said to be unassigned at that time.
The time sequence of such site assignments for a given mobile atom is its discretized trajectory; every change of site in that discretized trajectory is defined as a jump event.

The center of each site is defined as the spatial average of all real-space positions of mobile ions assigned to it.

\subsection{Step 4 (optional): Merge Sites}
\label{subsec:step4}

While one of the main strengths of the landmark analysis is its ability to distinguish between very close sites, that level of resolution often identifies multiple sites where only one should exist.
This is mainly due to a lack of data for the clustering.
This issue is particularly prominent in host lattices containing sites with greater than four-fold coordination whose coordination polyhedra are highly distorted from the corresponding regular polyhedra.
To merge such split sites, a post-processing clustering of the sites themselves can be applied, taking into account information from the time domain.
We define $M$ as the stochastic matrix observed from the exchanges of ions between sites:

\begin{equation*}
    [m_{AB}] = \begin{cases}
        0                      & \text{if } |\vec{r}_A - \vec{r}_B| > \text{cutoff} \\
        p_{A \rightarrow B} & \text{otherwise}
    \end{cases}
\end{equation*}%
where $\vec{r}_A$ is the center of site $A$ and $p_{A \rightarrow B}$
is the probability that an ion occupies site $B$, conditional on the ion's having occupied site $A$ in the previous frame (for $A\neq B$).
For $A=B$ it is the probability that an ion remains at site $A$ until the next frame.
We apply Markov Clustering~\cite{van_dongen_MCL_2008} to the weighted graph defined by the stochastic matrix $M$,
resulting in clusters of highly-connected subgraphs. Sites belonging to the same subgraph are merged (additional details are given in Appendix~\ref{sec:markov-clustering}).

\subsection{Step 5 (optional): Site Type Analysis}

Sites are commonly defined by their Wyckoff points, and symmetry-equivalent sites can be interpreted as one site type.
Such analysis depends on preexisting crystallographic data and also neglects that the energetics of a site are defined by the local geometry and chemistry.
In line with our goal of making unsupervised site analysis possible, we developed a method for determining the type of the sites identified by the steps described from \mysecref{subsec:step1} to \mysecref{subsec:step4}. 
Different sites whose environments cannot be distinguished are said to be of the same site type.

We describe local atomic environments using the smooth overlap of atomic positions (SOAP)~\cite{bartok_representing_2013} as implemented in the QUIP molecular dynamics framework~\cite{noauthor_libatoms/quip_2018}.
Briefly, a SOAP descriptor is a vector that describes the local geometry around a point in a rotation-, translation-, and permutation-invariant way.
The descriptor changes smoothly with the Cartesian coordinates of the structure.
For these reasons, SOAP descriptors have become a powerful tool to express local geometry for machine-learning applications~\cite{de_comparing_2016} and the detection of structural motifs~\cite{gasparotto_recognizing_2014, gasparotto_recognizing_2018}.

Multiple SOAP vectors must be computed for each site to provide sufficient data density for subsequent clustering. Computing these vectors for a site requires some procedure for sampling the real-space positions of both the site and its surrounding host-lattice atoms. We implemented two sampling schemes.
%
(1)~Real-space averaging: the real-space positions of all mobile atoms when they occupy the site are collected, and $n$ average real-space positions are computed for the site, where $n$ is a parameter chosen by the user. SOAP is computed on the averaged sites.
(2)~SOAP-space averaging: SOAP vectors are computed for all real-space positions with the host-lattice atoms at their corresponding instantaneous positions.
Then, $n$ average descriptor vectors are computed in SOAP~space.

After reducing the dimensionality of the SOAP vectors with Principal Component Analysis, we cluster them using density-peak clustering~\cite{rodriguez_clustering_2014} with a Euclidean distance metric. A simple parameter estimation scheme is used to determine the number of clusters (see Appendix~\ref{sec:den-peak-params}). Each cluster of descriptor vectors corresponds to a site type. Each site is assigned to the type corresponding to the descriptor cluster to which the majority of its descriptors were assigned. Small majorities (less than 70-80\% agreement) typically indicate insufficient data, poorly chosen SOAP parameters, or very similar environments.

\subsection{Discussion of design choices}

The main motivation for a landmark based approach is its ability to significantly reduce noise resulting from thermal vibrations in the system while reducing the dimensionality and discretizing the trajectory of the mobile ions. 
The design described in this section is driven by physical intuition and trial-and-error. While developing the present approach, we attempted and discarded a number of approaches due to poor performance in trial systems.
(1)~Directly clustering the Cartesian coordinates of the mobile ions (the density-based approach discussed in \mysecref{sec:intro}) was found to work poorly in some systems. We show this in more detail in \mysecref{sec:llzo}.
(2)~An analysis based on the $N$ nearest neighbors of the mobile atom was tried but discarded, since we could not determine $N$ without relying on the knowledge of the structure under investigation, in particular the expected size of the coordination shell of Li.
(3)~We tried various landmark representations, the most simple being the distance to each host-lattice atom, therefore taking the instantaneous positions of host-lattice atoms as landmarks. The results for different systems were not satisfactory.

We conclude this section by pointing out that passing from Cartesian coordinates to landmark vectors can significantly reduce the noise that comes mostly from thermal vibrations in the system.
Different formulations of landmarks can be envisioned, and while the present framework performs well for Li-ionic diffusion, a different landmark framework might be needed to describe, for example, Grotthus-like proton diffusion in superprotonic \ce{CsHSO4}~\cite{wood_proton_2007}.

\section{Results and discussion}
  \label{sec:results}
We apply the algorithms above to ten representative materials,
\ce{Li7La3Z2O12}, \ce{LiAlSiO4}, \ce{Li10GeP2S12}, 
\ce{Li32Al16B16O64},
\ce{Li24Sc8B16O48},
\ce{Li24Ba16Ta8N32},
\ce{Li20Re4N16},
\ce{Li12Rb8B4P16O56},
\ce{Li6Zn6As6O24} and
\ce{Li24Zn4O16}.
For the subsequent analysis, we also calculate radial distribution functions, 
mean-square displacements, and ionic densities.
The mobile-ion densities $n_\text{M}({\bm r})$ are calculated from molecular dynamics trajectories as:
\begin{equation}
 n_M(\bm r) =\left\langle \sum_m^M \delta(\bm r-\bm r_m(t)) \right\rangle_t,
 \label{eq:density}
\end{equation}
where the index $m$ runs over all mobile ions $M$ and the angular brackets $\langle\cdot\rangle_t$ indicate a time average over the trajectory,
which is equal to an ensemble average under the assumption of ergodicity.
When applying \myeqref{eq:density} we replace the delta function by a Gaussian with a standard deviation of 0.3\r{A}, and the summation is performed on a grid of ten points per \r{A} in every direction.
The tracer diffusion coefficient of the mobile species $D_{\mathrm{tr}}^M$ is computed from the mean-square displacement of the mobile ions as a function of time:
\begin{equation}
D_\mathrm{tr}^M =\lim_{\tau \rightarrow \infty} \frac{1}{6\tau}  \frac{1}{N_M} \sum_m^M   \left\langle |\bm r_m(\tau+t)-\bm r_m(t)|^2 \right\rangle_t,
\label{eq:einstein}
\end{equation}
where $\bm r_m(t)$ is the position of the mobile ion at time $t$.
In practice, we fit a line to the mean-square displacement in the diffusive regime.
The error of the tracer diffusion coefficients is estimated with a block analysis~\cite{allen_computer_1987}.
The radial distribution function
$g(r)_{M-S}$ of the mobile ions $M$ with species $S$ is calculated as:
\begin{align}
\notag g_{M-S}(r) =& \frac{\rho(r)}{f(r)} \\
 =& \frac{1}{f(r)} \frac{1}{N_M} \sum_m^M 
\sum_s^S \left\langle \delta \left(r - \left| \bm r_m(t) - \bm r_s(t) \right| \right) \right\rangle_t,
\end{align}
where $f(r)$ is the ideal-gas average number density at the same overall density.
In addition, we integrate the average number density $\rho(r)$ to give the average coordination number as a function of distance~\cite{frenkel_understanding_1996}.

\FloatBarrier

\subsection{Analysis of Li$_7$La$_3$Zr$_2$O$_{12}$}
  \label{sec:llzo}

\begin{figure}[t]
\includegraphics[width=\hsize]{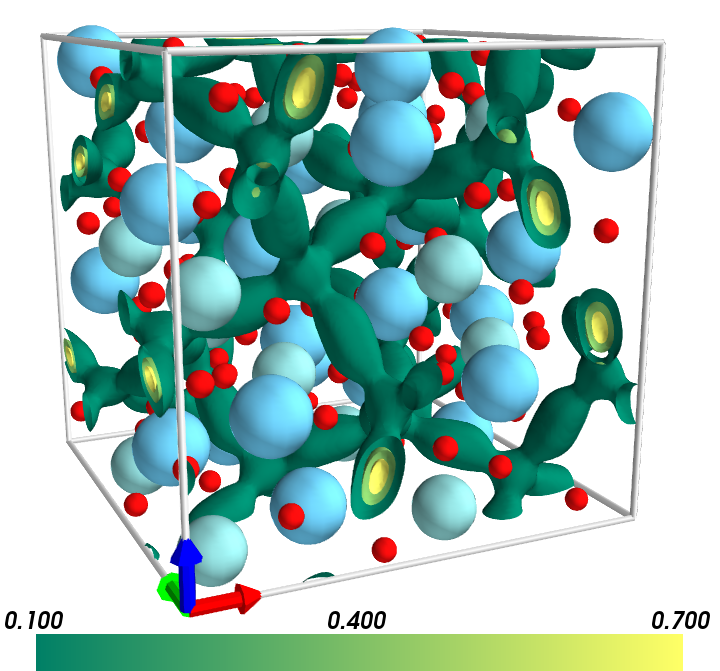}
\caption{The Li-ion density in LLZO is shown above as three isosurfaces going from green (low density) to yellow (high density).
The equilibrium positions of lanthanum are shown in blue, those of zirconium in turquoise, and those of oxygen in red.
The Li-ion densities reveal the three-dimensional percolation pathways in this material.
}
\label{fig:dens-llzo}
\end{figure}

\begin{figure}[t]
\includegraphics[width=\hsize]{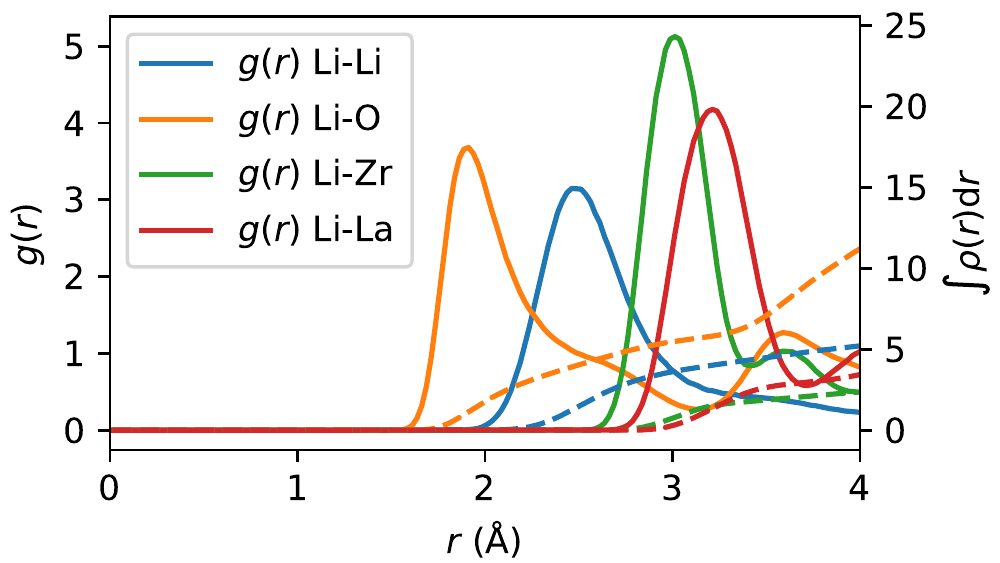}\\
\includegraphics[width=\hsize]{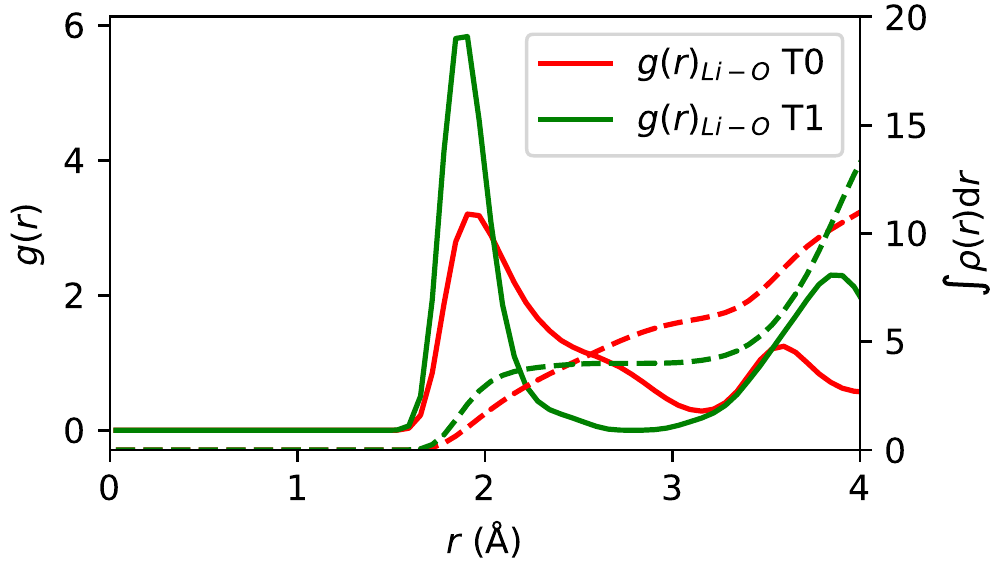}
\caption{
(Top) The Li-Li (blue), Li-O (orange), Li-Zr (green), and
Li-La (red) radial distribution functions $g(r)$ are shown as solid lines.
The integral, representing the coordination as a function of distance, is plotted against the right axis using dashed lines and the same color encoding.
(Bottom) Radial distribution function for lithium-oxygen pairs for the two distinct site types we found. 
The red lines correspond to a site of type 0, the green lines to a site type of 1, and the integrals are shown with dashed lines and the same color encoding.}
\label{fig:rdf-llzo}
\end{figure}

\begin{figure}[t]
\includegraphics[width=\hsize]{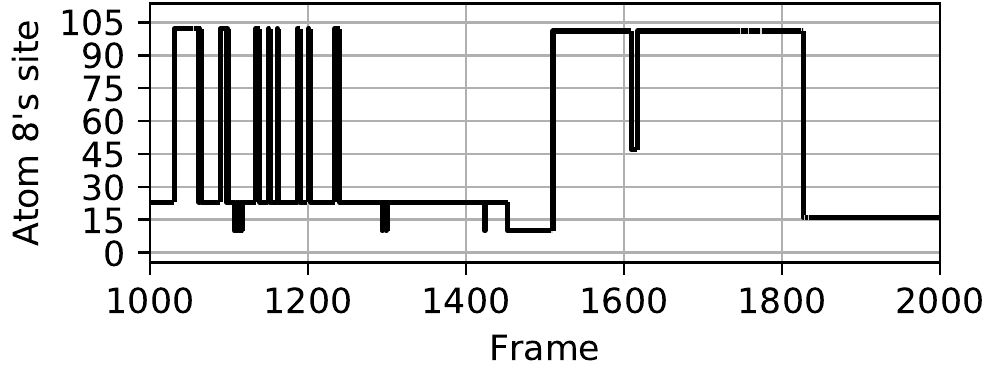}
\caption{Site trajectory in LLZO at 500~K for a representative mobile lithium ion over 1000~frames. For every frame, we determine the most likely site the ion is occupying. We plot the ion's occupation over time, where discontinuities are interpreted as jumps. 
The site index on the $y$~axis is arbitrary, and the distances in index space, i.e., the vertical distance in above plot, are not reflective of the actual jump distances.}
\label{fig:site-trajectory-llzo}
\end{figure}

\begin{figure}[b]
\includegraphics[width=\hsize]{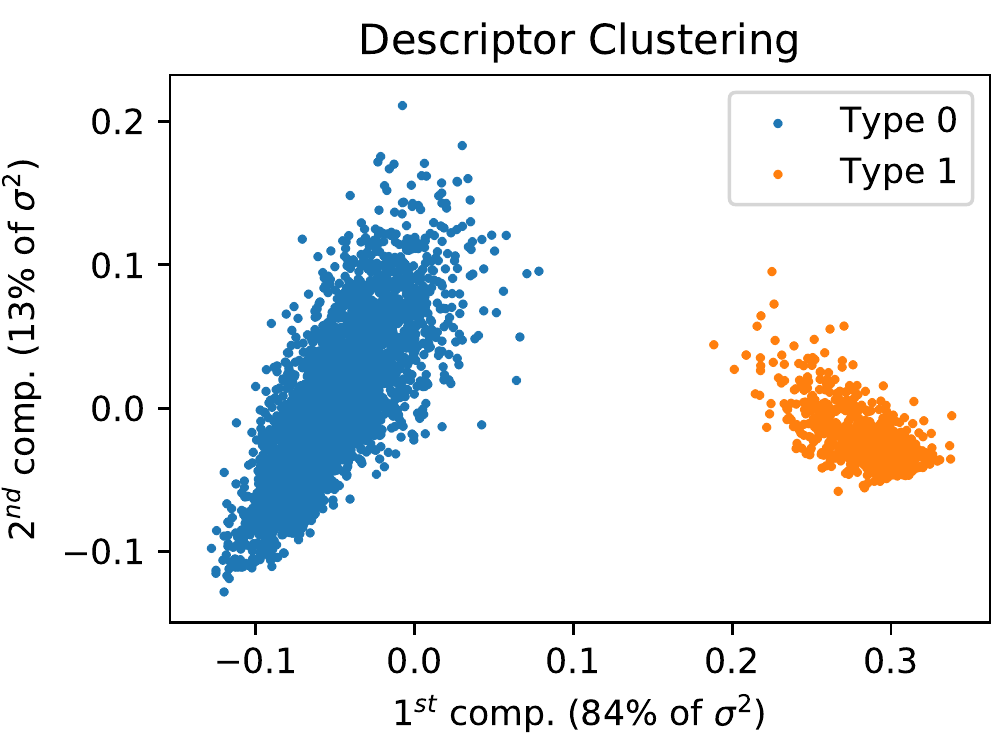}
\caption{SOAP descriptor clustering for LLZO. Each point is an average SOAP vector and is colored according to its assigned cluster (site type).
The first and second principal components are plotted along the $X$ and $Y$ axes respectively.
}
\label{fig:soaps-llzo}
\end{figure}

\begin{figure}[t]
\includegraphics[width=\hsize]{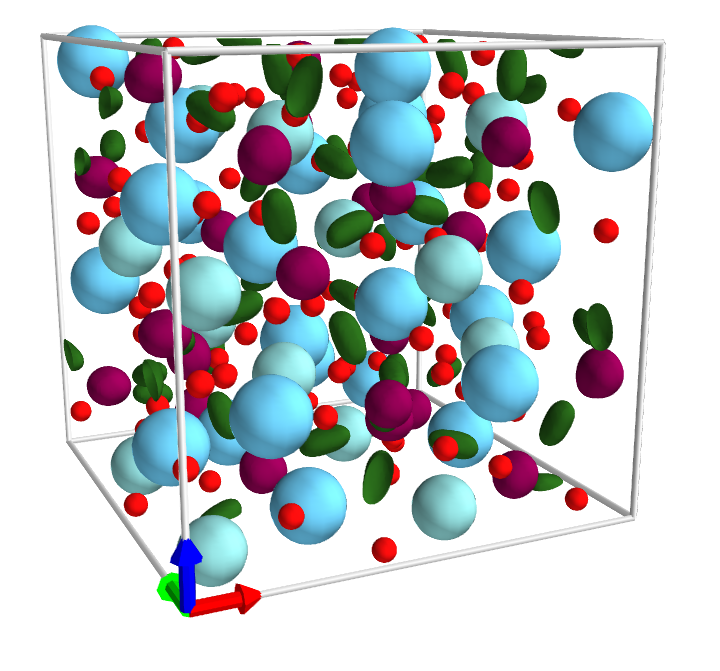}
\caption{Li-ion density in LLZO is shown above for the same isovalue (0.1) for Type 0 (octahedral environment) in green and Type 1 (tetrahedral environment) in bordeaux.}
\label{fig:dens-type-llzo}
\end{figure}

\begin{figure}[b]
\includegraphics[width=\hsize]{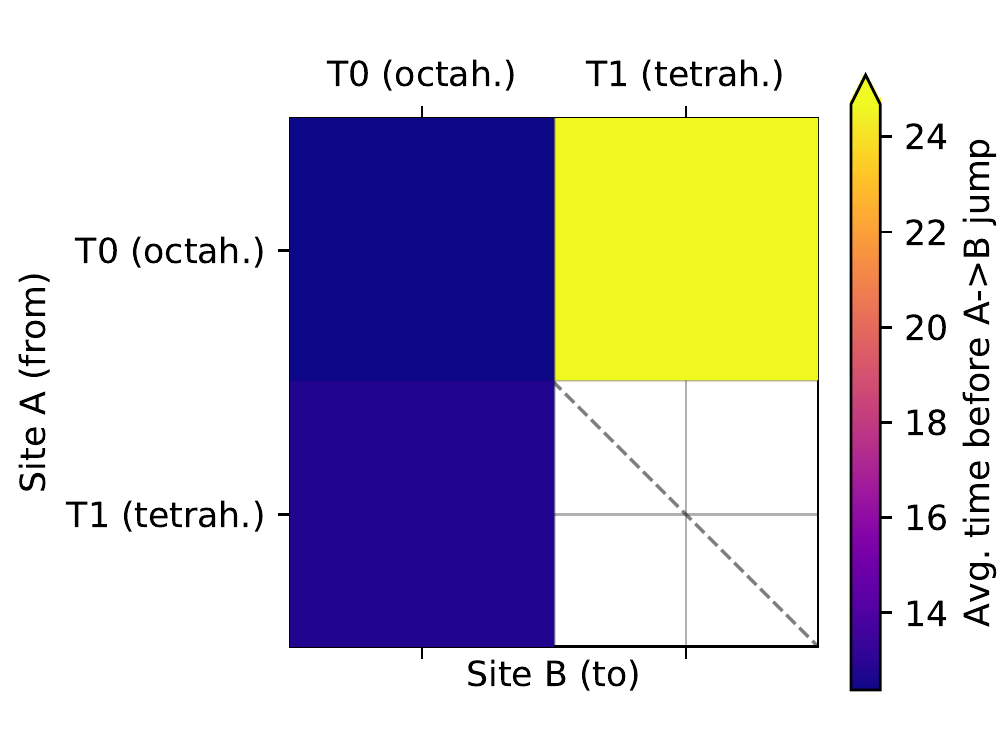}
\caption{We show the jump lag or residence time in LLZO, which is the time
an ion spends in site A before jumping to site B, averaged over sites of the same type. The color encodes the residence time, with no color (white) meaning that no jump has been observed. The time is given in multiples of the interval between frames.}
\label{fig:jump-lag-llzo}
\end{figure}

\begin{figure}[t]
\includegraphics[width=\hsize]{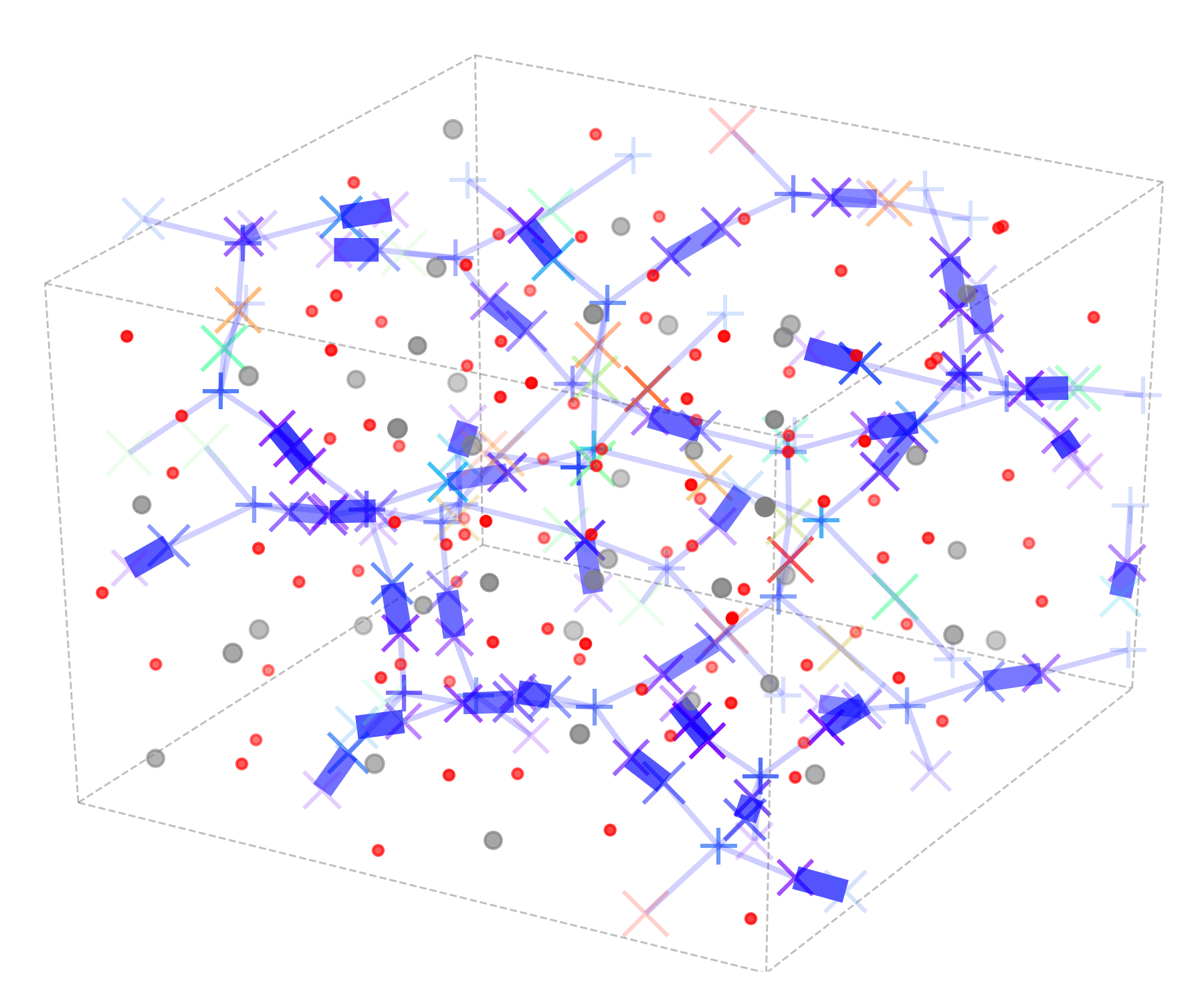}
\caption{The diffusive pathways in LLZO at 500 K.
The centers of sites of type 0 (octahedral environment) and type 1 (tetrahedral environment) are shown as crosses and plusses, respectively.
The color of the sites encodes the average residence time.
Edges connect sites that have exchanged mobile ions, with the edge width related linearly to the observed flux of particles.
The equilibrium host lattice positions of oxygen (red), lanthanum (grey), and zirconium (light grey) are shown as small spheres.
The entire network of diffusion has one connected component.} 
\label{fig:final-analysis-llzo}
\end{figure}

Garnet-type structures were proposed as lithium-ionic conductors by Thangadurai \textit{et al.}~\cite{thangadurai_novel_2003}.
The general formula of garnets is \ce{Li5La3M2O12} $\left( \mathrm{M= Ta, Nb}\right)$~\cite{knauth_inorganic_2009},
but aliovalent substitutions of $M$ can change the lithium content.
Xie \textit{et al.}~\cite{xie_lithium_2011} studied in more detail the distribution of Li$^+$ in garnets.
Their results indicate that increasing the lithium concentration in garnets leads to an increase in occupation of octahedral sites, which is
confirmed by simulations~\cite{kozinsky_effects_2016} and also in experiments~\cite{ocallaghan_switching_2008}.
It has been established~\cite{ocallaghan_lithium_2007,adams_ion_2012, thangadurai_garnet-type_2014} for the garnet structure that Li ions can occupy tetragonal 24d sites, octahedral 48g sites and 96h distorted octahedral sites. The latter stem from a site splitting of the 48g sites to increase the Li-Li distances and occur at higher lithium concentrations.
In this work, we study the Li-ion distribution of Zr-based cubic garnets with the stochiometric formula \ce{Li7La3Zr2O12}, referred to as LLZO in the remainder.

We sample the dynamics in the cell of 192 atoms in the canonical ensemble via a GLE thermostat~\cite{ceriotti_langevin_2009} at a temperature of 500~K, 
using a lattice constant of 12.9872~\r{A}, and a polarizable force field.
We use LAAMPS~\cite{plimpton_fast_1995} to perform the simulation for 10~ns, with the parameters of the force field taken from the work by Mottet \textit{et al.}~\cite{mottet_doping_2019},
which accurately reproduces the kinetics of the diffusing process in LLZO.

The estimate of the diffusion coefficient via \myeqref{eq:einstein} reveal that Li-ions indeed are diffusive in LLZO,
with a tracer diffusion coefficient of $D_{tr}^{Li}=2.4 \times \mathrm{10^{-6}\, cm^2 \, s^{-1}}$.
Application of the Nernst-Einstein equation gives the ionic conductivity $\sigma$:
\begin{equation}
\sigma = \frac{Z^2 e^2 N}{k_BT}  \frac{D_{tr}}{H},
\end{equation}
where $\left(Ze\right)$ the carrier's charge, $N$ the carrier density, $k_B$ the Boltzmann constant, $T$ the absolute temperature and $H$ is the ratio between the tracer and charge diffusion coefficient,
commonly referred to as Haven ratio: $H=\frac{D_{tr}}{D_\sigma}$.
To account for the strong evidence for correlated motion in this material~\cite{jalem_concerted_2013,meier_solid-state_2014}, we set the Haven ratio to $H=0.4$,
reported in a study~\cite{morgan_benjamin_j._lattice-geometry_2017} for this Li-ion concentration in LLZO.
We find $\mathrm{\sigma=0.58\,S cm^{-1}}$, which is one order of magnitude larger than the values reported by Murugan \textit{et al.}~\cite{murugan_fast_2007}
This is within the acceptable range, especially for a classical force-field, and not of concern since the focus of this work is the analysis method.
The diffusive pathways can be illustrated by the Li-ion density, shown for three isosurfaces in~\myfigref{fig:dens-llzo}.
By visual inspection, the densities look similar to those presented by Adams and Rao in their computational study~\cite{adams_ion_2012}.
The splitting of 48g sites into 96h sites~\cite{thangadurai_garnet-type_2014} cannot be seen from the isosurfaces at any isovalues,
which is consistent with the conclusions drawn by Chen \textit{et al.}~\cite{chen_data_2017} that density-based clustering of real-space positions cannot resolve the two distinct 96h sites in LLZO from the 48g site.
The radial distribution function $g(r)$, shown in the upper panel of~\myfigref{fig:rdf-llzo}, shows how the Li ions in our simulation are,
as expected, coordinated closest by oxygen and then by other Li ions.

We use the site analysis presented to discretize the trajectory of lithium ions into meaningful states, as illustrated for one lithium ion in \myfigref{fig:site-trajectory-llzo}.
The subsequent SOAP analysis produces two clearly resolvable clusters, which are detected by the clustering algorithm.
We show the first two principal components in \myfigref{fig:soaps-llzo}, with a color encoding representing the cluster detected.
It is evident that the SOAP descriptor produces data that clusters well in this projection and that the clustering algorithm correctly assigns the clusters.
The algorithm detects 24 sites of one kind (type 1) and 83 of another (type 0). 
We attribute the tetrahedral environment to the former, and the octahedral environment to the latter, since the expected values are 24 sites for the tetrahedral environment and 96 for the octahedral one.
The last number is due to the site splitting inside each of the 48 octahedral cavities, leading to two sites inside each octahedral cavity.
The under-prediction of the number of octahedral sites is due to the merging, in some cases, of octahedral sites into a single site.
We stress that the numbers of sites presented as final results are after the site-merging step, presented in \mysecref{subsec:step4}.
The proximity and fast ion exchange between the octahedral sites in the same cavity explains why our algorithm does not give the correct answer,
but it is remarkably close to the correct result, without any encoding of prior information about possible site splitting.
Comparing to the study by Chen \textit{et al.}~\cite{chen_data_2017}, we can conclude that the landmark analysis is able to better distinguish minima in close proximity. We speculate that the main reason is a higher tolerance for thermal vibrations of the host lattice, that can lead to energetic minima being spread in real space.

To ensure that the analysis of LLZO provides reasonable and expected results, we calculate the Li-O radial distribution function $g(r)$ separately for each site type;
these are shown in the bottom panel of \myfigref{fig:rdf-llzo}.
Li ions attributed to sites of type 1 have an environment characterized by a distinct nearest-neighbor peak stemming from four-fold coordination of lithium with oxygen, as evidenced by the integral plateauing at a value of four. The first peak for type 0, shown in red, has a shape compatible with a distorted octahedron, due to the appearance of a shoulder, and the weak, but distinguishable, plateau of the integral at a value of approximately 6. This is further evidence that the site types have been correctly attributed to the tetrahedral and octahedral site environments of LLZO, and that the SOAP descriptor can be used to cluster site types correctly.
Additionally, we resolve the Li-ion density by site type in \myfigref{fig:dens-type-llzo}.
The isosurfaces are compatible, by visual inspection, with reported work~\cite{adams_ion_2012}.
We also calculate the jump lag, which is the average residence time at a site A before jumping to a site B.
If we average over all sites belonging to the same type, as shown in~\myfigref{fig:jump-lag-llzo}, we see that jumps between the octahedral sites are fastest.
From the site splitting of the 48g into 96h sites follows that two sites are present inside each octahedron, and that there is a free energy barrier between the split sites. 
Our results are therefore in agreement with \textit{ab~initio} calculations~\cite{xu_mechanisms_2012, jalem_concerted_2013, miara_effect_2013, meier_solid-state_2014},
that show that the minima in the Li-ion potential energy surface are displaced from the original central site in the octahedral site. 
The presence of two distinct but very close sites manifests in very high exchange rates between these two.
LLZO also displays fast jumps from the tetrahedral into the octahedral environment, whereas the reverse jump takes three to four times longer.
No jumps between tetrahedral environments are observed, as expected, since an ion needs to traverse octahedral sites to reach a different tetrahedral site.
While the jump probabilities, or lag times, are non-symmetric, the fluxes are symmetric, which is necessary to observe local detailed balance.

The diffusive pathways estimated from the algorithms are shown in \myfigref{fig:final-analysis-llzo}.
The connectivity analysis reveals the existence of a single dominant pathway that allows mobile ions to diffuse through the entire simulation cell.
The edge widths in the figure are proportional to the observed flux of particles, and we see that, where the octahedral site splitting is correctly determined, there is a large flux of ions between split octahedral sites compared to the smaller flux between the tetrahedral and octahedral environments.

Summarizing our results for this material, our site analysis finds the splitting of the 48g to 96h sites, which sets it apart from any density-based analysis.
An analysis based on distance criteria to crystallographic sites would have worked as well or better, but obviously requires prior knowledge.

\begin{figure}[b]
\includegraphics[width=\hsize]{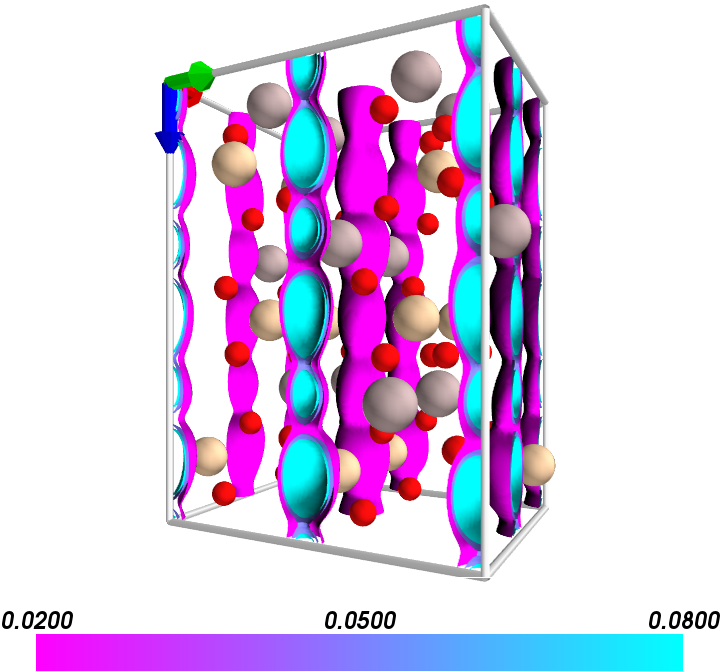}
\caption{
Li-ion density in LASO, shown above as 3 isosurfaces going from violet (low density) to sky blue (high density).
The equilibrium positions of oxygen are shown in red, of silicon in grey, and of aluminum in beige.
Silicon and aluminum appear in alternating planes perpendicular to the c-axis.
}
\label{fig:dens-laso}
\end{figure}

\begin{figure}[t]
\includegraphics[width=\hsize]{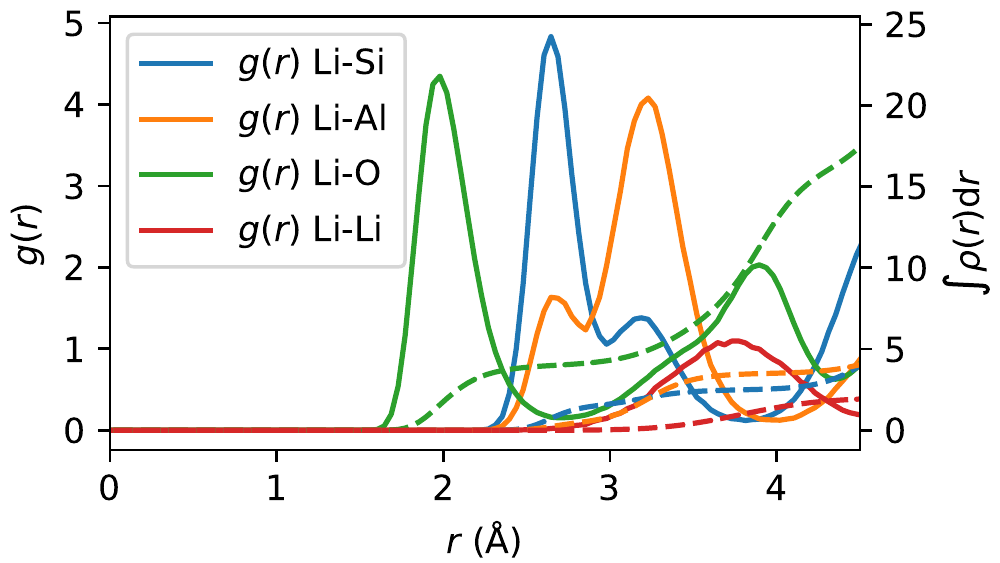} \\
\includegraphics[width=\hsize]{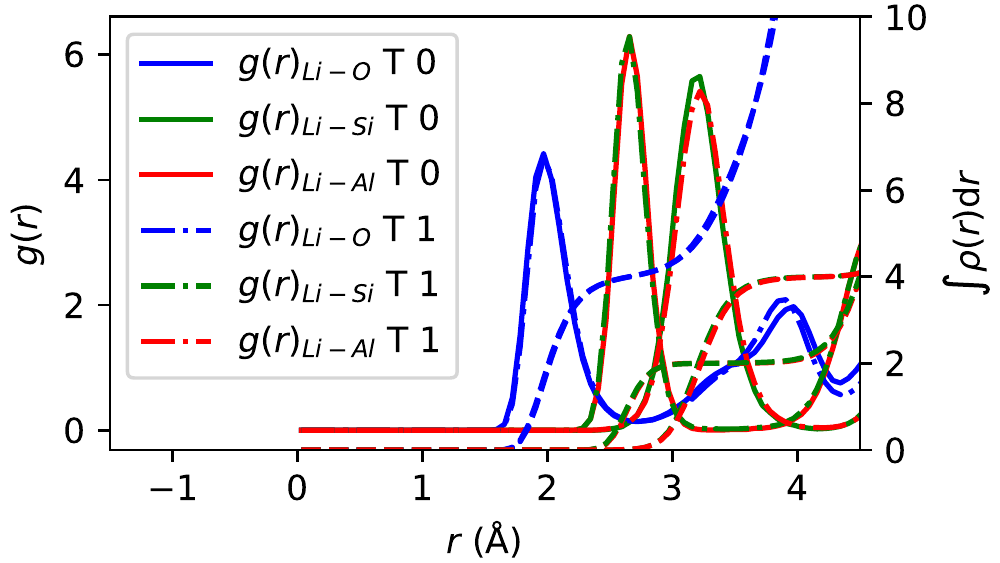}
\caption{Upper panel: 
The Li-Si (blue), Li-Al (orange), Li-O (green), and
Li-Li (red) radial distributions $g(r)$ are shown as solid lines. 
The integral of the average number density is plotted against the right axis as dashed lines in the same color.
Lower panel: Li-O (blue), Li-Si (green), and Li-Al (red) radial distribution functions for the two distinct site types we found.
The solid lines correspond to a site of type 0, the dash-dotted lines to a site of type 1.}
\label{fig:rdf-laso}
\end{figure}

\begin{figure}[t]
\includegraphics[width=\hsize]{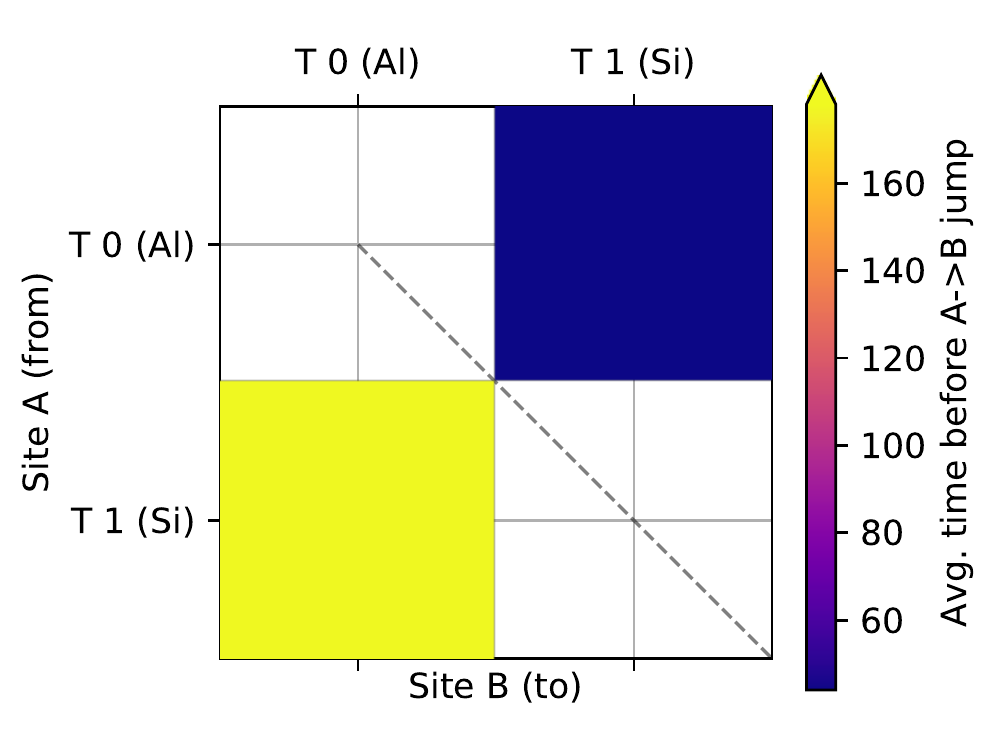}
\caption{Similarly to \myfigref{fig:jump-lag-llzo} we show the average residence time before a jump for the two distinct site types in LASO. White corresponds to no jumps occurring.}
\label{fig:jump-lag-laso}
\end{figure}

\begin{figure}[t]
\includegraphics[width=0.9\hsize]{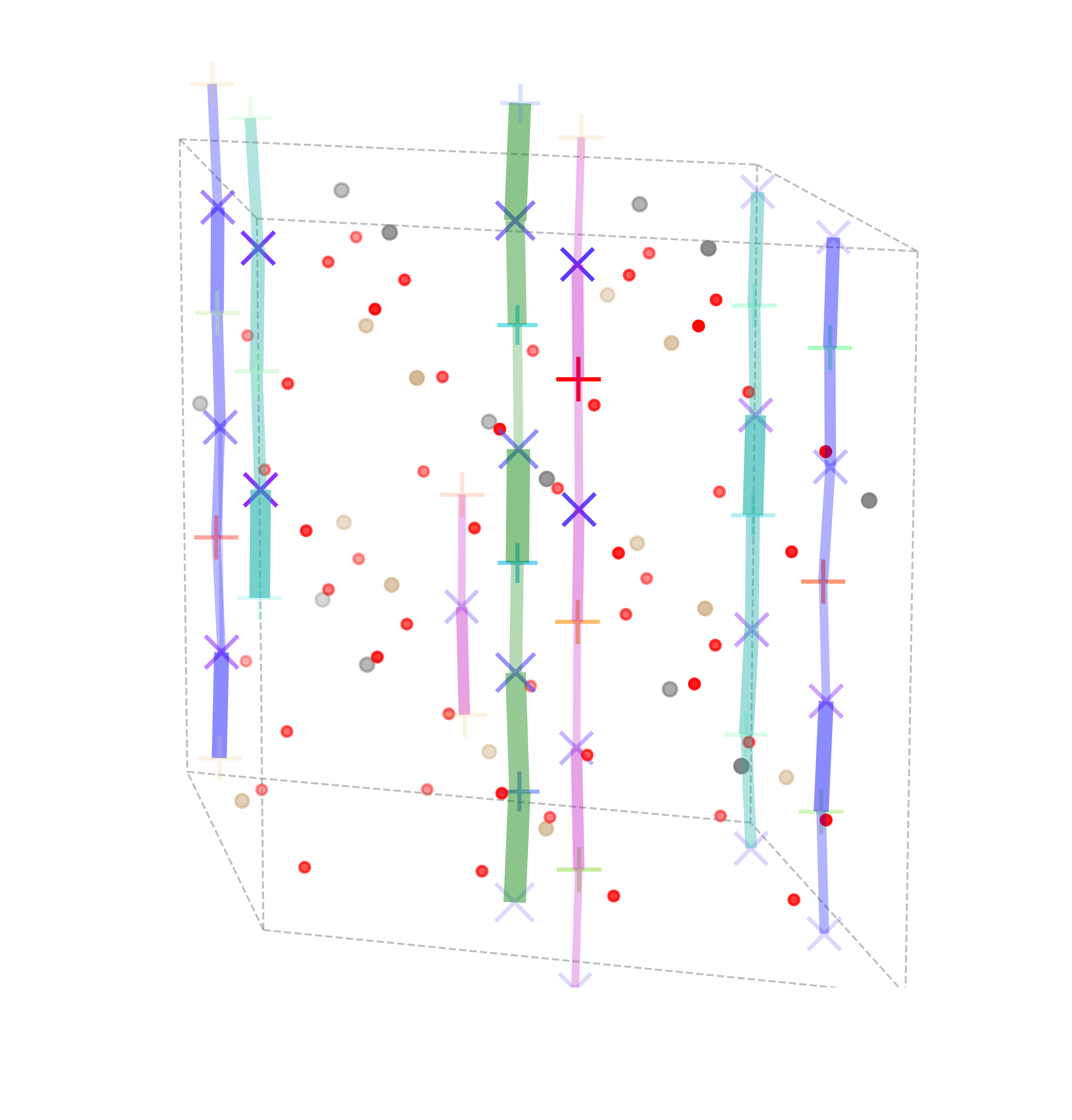}
\caption{The diffusive pathways in LASO at 750 K:
sites of type 0 are shown as crosses and sites of type 1 as pluses.
Edges are drawn between sites that exchange ions, similar to \myfigref{fig:final-analysis-llzo}.
Unlike LLZO, the diffusion network has four disconnected components, indicated by differently colored edges;
The channels in LASO do not exchange ions in our simulations.
}
\label{fig:final-analysis-laso}
\end{figure}

\subsection{Analysis of LiAlSiO$_4$}
\label{sec:laso}

The structure of the $\beta$-eucryptite \ce{LiAlSiO4}~\cite{william_w._pillars_crystal_1973}, referred to as LASO hereafter, is taken from COD~\cite{grazulis_crystallography_2012} entry 9000368.
It has been studied for its anisotropic expansion coefficient~\cite{lichtenstein_anisotropic_1998,lichtenstein_anisotropic_2000}
and its ionic conductivity~\cite{schink_vacancy-induced_1983, donduft_li+_1988, sartbaeva_ionic_2005, chen_thermal_2018}.
The structure can be described as an ordered $\beta$-quartz solid solution, with alternating aluminum and silicon planes.
Location and occupation of the sites for lithium have been contested.
In the original reference~\cite{william_w._pillars_crystal_1973}, the difficulties in determining the lithium sites in previous and in the same work are explained very well.
For example, earlier work~\cite{winkler_synthese_1948} concluded that the Li sites are coplanar with the Al sites, while Pillars and Peacor~\cite{william_w._pillars_crystal_1973} show that the lithium sites are also present in the Si plane.
Later work~\cite{donduft_li+_1988} shows that both sites are available to lithium and establishes the unidimensional chain of these sites as the mechanism for ionic diffusion in this material.
There is now a better understanding of this structure and the sites available to lithium, but the original CIF-file in the COD originating from the experiments by Pillars and Peacor~\cite{william_w._pillars_crystal_1973} does not list all sites.
Any analysis that relies on this knowledge would therefore have failed.
Our molecular dynamics simulations and subsequent site analysis yield results that are compatible with the latest literature regarding the ionic transport in this material.

We simulate with first principles \ce{Li12Al12Si12O48}, starting from the reported CIF-file~\cite{grazulis_crystallography_2012}.
A full atomic and cell optimization results in a volume increase of 3.6\% without changing the cell angles in a significant way.
We perform the subsequent dynamical simulations using Born-Oppenheimer molecular dynamics in the canonical ensemble at a temperature of 750~K for 291~ps, 
with further details given in Appendix~\ref{sec:calc-params}.

We show in \myfigref{fig:dens-laso} the Li-ion densities sampled during the dynamics.
The unidimensional channels of ionic diffusion are compatible with published results~\cite{donduft_li+_1988, chen_thermal_2018}.
The diffusion coefficient is hard to converge for the short dynamics we obtained for this system, and so quantifying the diffusion coefficient and its error cannot be done rigorously.
We plot the coordinates of Li-ions as a function of time in \supplref{2} to show that motion along the z-coordinate is observed during the simulation, compatible with long-range diffusion.
The RDFs of Li with all present species, shown in the upper panel of \myfigref{fig:rdf-laso}, display a first coordination shell composed by four oxygen ions, compatible with literature findings.
A second and third shell are composed of silicon and aluminum, with the amplitude of Si being stronger in the second shell, and Al in the third shell.
This hints that Li ions prefer sites in the Si plane to those in the Al plane.

When running the  site analysis, we find 24 sites of two different types, twelve of type 0 and twelve of type 1, compatible with the latest literature results~\cite{donduft_li+_1988}.
The parameters are as given in Appendix~\ref{sec:sa-params}, except for a cutoff midpoint of 1.3 instead of 1.5, which is more robust with respect to the total number of sites obtained.
The clustering analysis in \supplref{5} shows that types can be distinguished easily.
An analysis of the RDF for the individual site types, shown in the bottom panel of~\myfigref{fig:rdf-laso}, reveals that the discriminant is the different coordination of Al and Si, which is detected by the SOAP descriptor.
For type 0, the second shell is composed of two aluminum atoms; four silicon atoms are in the third shell.
For type 1, the numbers are the same, but silicon is replaced by aluminum and vice versa.
The RDF in the upper image of \myfigref{fig:rdf-laso} hints at the fact that the Li ions prefer to occupy type 1 sites where the Si ions are closer than the Al ions.
From the site analysis of our simulation, we calculate the mean occupation ratio to be 77\% for site type 1 and 23\% for site type 0.
Literature reports give occupancies of 68\% and 22\%~\cite{press_neutron_1980}, respectively, or a 3:1 ratio~\cite{lichtenstein_anisotropic_2000}.
In~\myfigref{fig:jump-lag-laso}, we show the jump lag between the type sites.
Jumps from type 1 to type 0 are about 3.5 times faster, which is necessary to preserve detailed balance.
We observe no jumps within the site types, which is expected since the sites' types are alternating along the diffusion channels (see~\myfigref{fig:final-analysis-laso}).

We can thus show with first-principles molecular dynamics and an unsupervised analysis that the lithium ions occupy two different site types in LASO.
This is done without any knowledge of the possible sites, since in the original CIF file only twelve sites (for twelve lithium ions) are given.
This example highlights the challenges for algorithms that rely on prior knowledge of crystallographic sites.
Such information might be missing or wrong, for example, because of the difficulties of resolving low occupancy sites for light elements when using XRD or neutron diffraction, or because of simulation conditions (e.g., temperature) differing from the experimental setup.
Relying on all the sites being known can obviously be problematic in some cases.
An unsupervised approach requiring minimal knowledge of the structures should be preferred in such cases.
Unlike the case of LLZO, a density-based clustering on the lithium-ion positions would very likely also have given the same results,
as can be conjectured from the lithium-ion densities in~\myfigref{fig:dens-laso}, where the highest isovalue clearly shows disconnected regions of high ionic density.

\begin{figure}[t]
\includegraphics[width=\hsize]{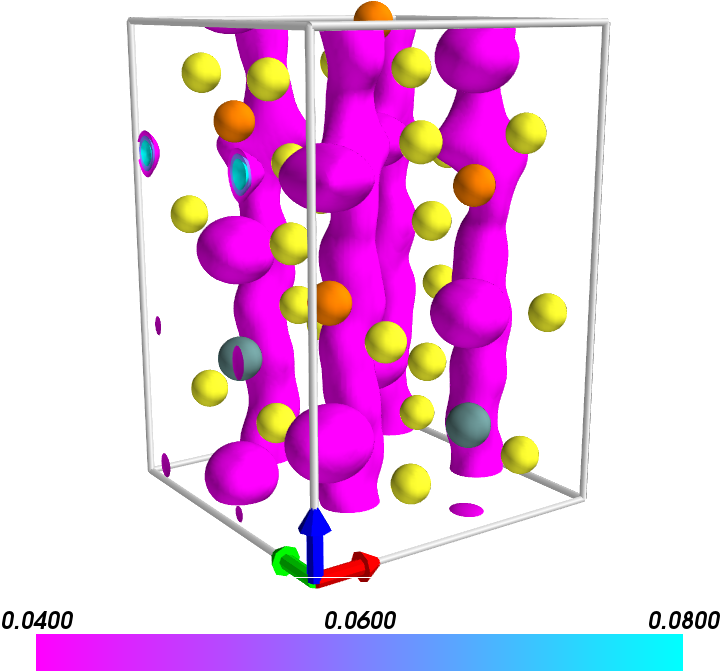}
\caption{Li-ion density in LGPS is shown above as three isosurfaces from violet (low density) to sky blue (high density).
The equilibrium positions of sulfur are shown in yellow, of phosphorus in orange, and of germanium in green.
The Li-ion densities reveal the unidimensional ion-conducting channels in this material.
}
\label{fig:dens-lgps}
\end{figure}

\begin{figure}[t!]
\includegraphics[width=\hsize]{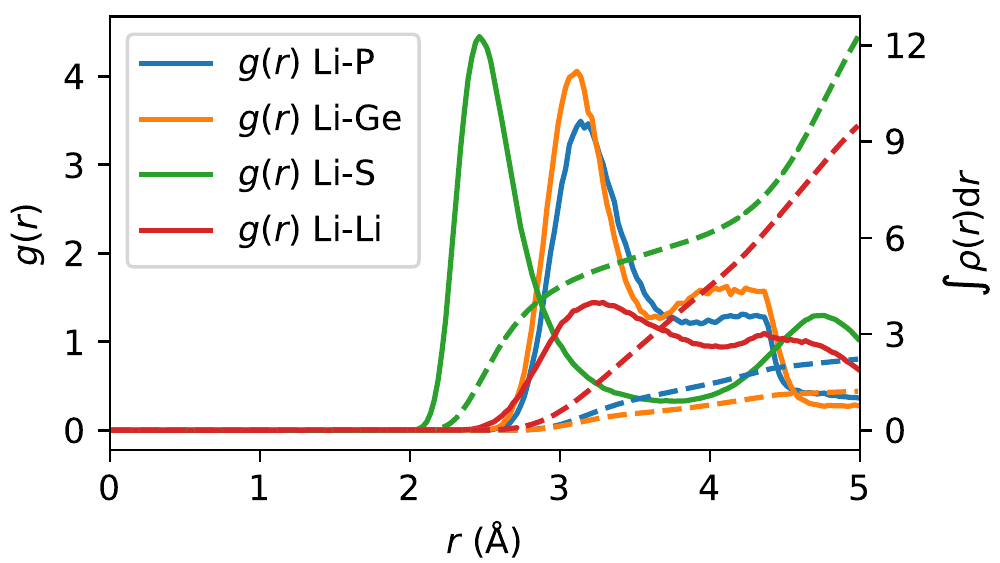}\\
\includegraphics[width=\hsize]{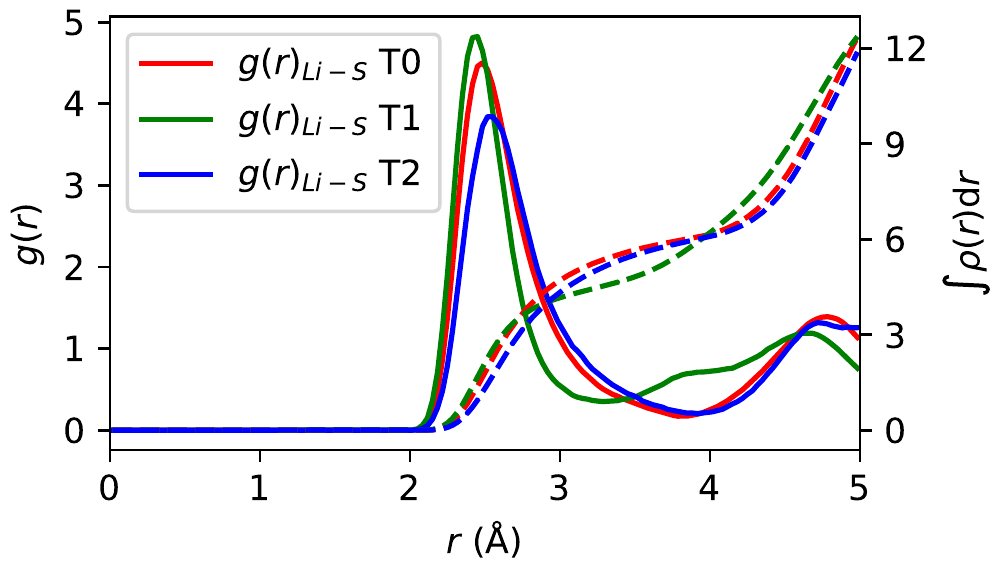}\\
\includegraphics[width=\hsize]{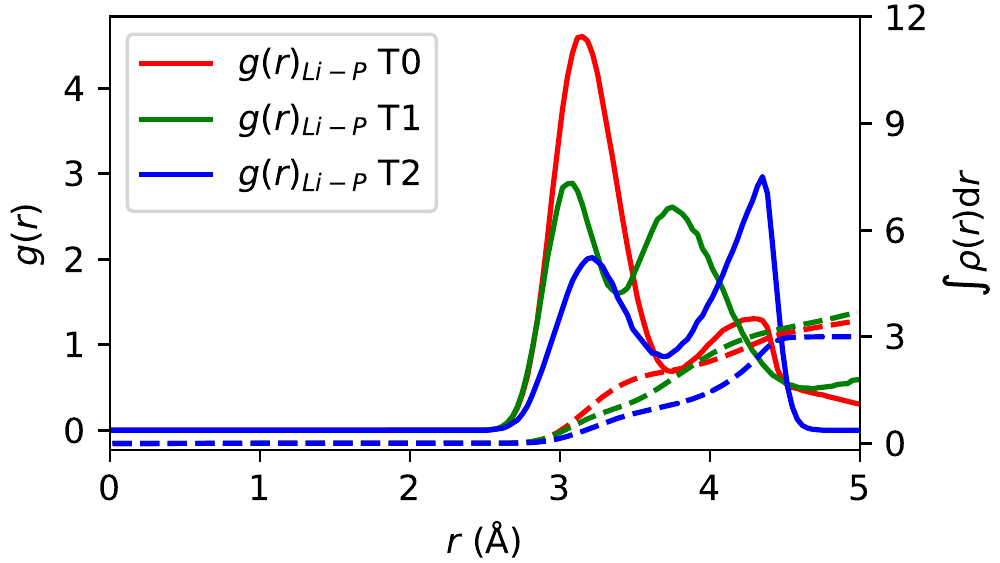}
\caption{
(Top) The Li-P (blue), Li-Ge (orange), Li-S (green) and Li-Li (red) radial distribution functions are shown as solid lines.
The Li-Li distribution displays the expected liquid-like lack of structure.
Center and bottom panels show the RDF of Li-S and Li-P, respectively.
In both panels, the RDF is shown for Li occupying sites of type 0 in red, sites of type 1 in green, and and sites of type 2 in blue.
}
\label{fig:rdf-lgps}
\end{figure}

\begin{figure}[t]
\includegraphics[width=0.9\hsize]{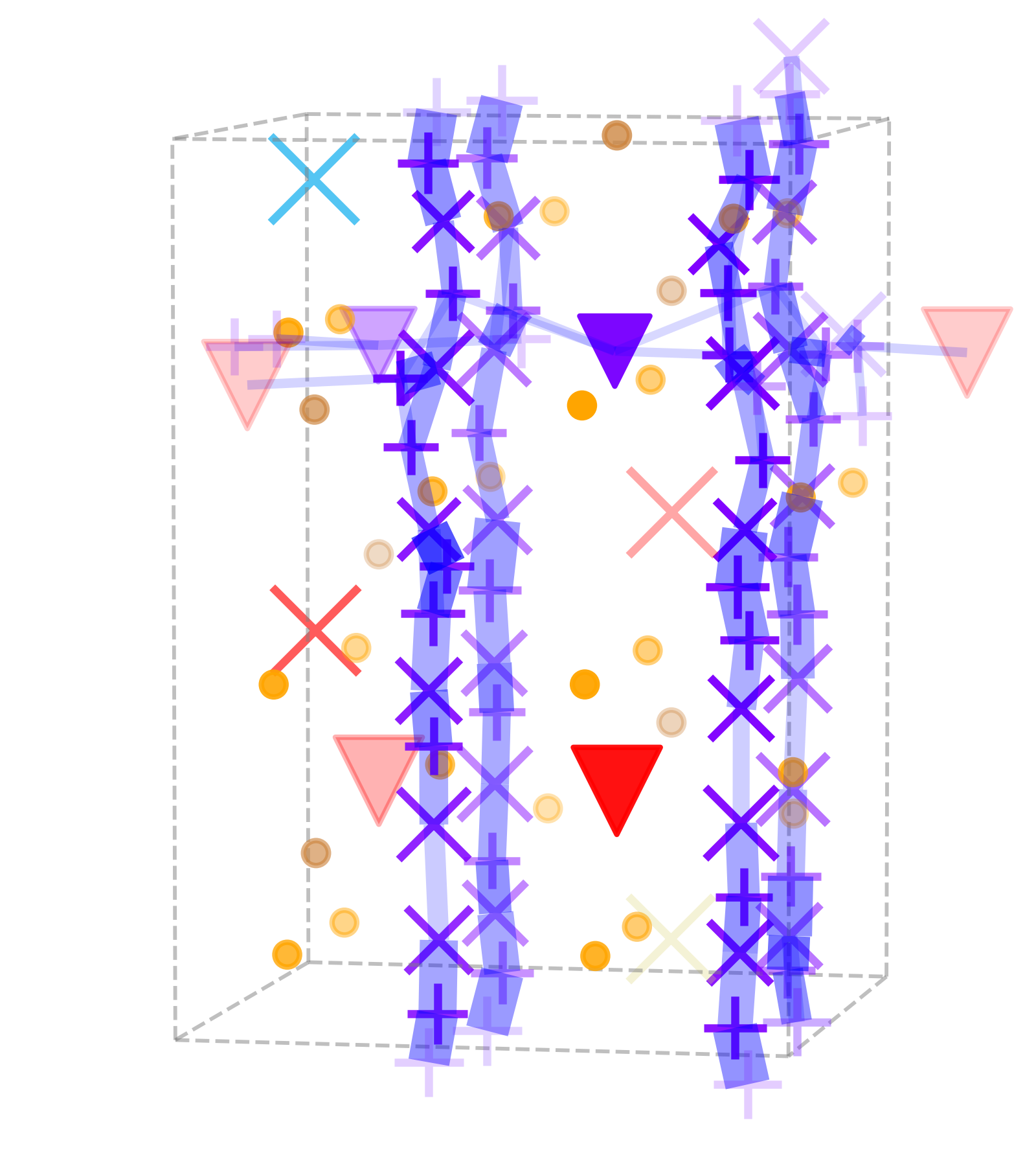}
\caption{The diffusive pathways in LGPS at 500~K.
Sites of type 0 are shown as crosses,  sites of type 1 as pluses, and sites of type 2 as triangles.
Unlike LASO, the ion-conducting channels do exchange ions, leading to a single connected component, illustrated by one color for the entire network.}
\label{fig:final-analysis-lgps}
\end{figure}

\subsection{Analysis of tetragonal Li$_{10}$GeP$_2$S$_{12}$}

The  superionic conductor \ce{Li10GeP2S12} (LGPS) in its tetragonal phase was first reported by Kamaya \textit{et al.}~\cite{kamaya_lithium_2011}
Its unprecedented ionic conductivity at room temperature motivated studies of its diffusion mechanisms using atomistic simulation techniques~\cite{adams_structural_2012,mo_first_2012,xu_one-dimensional_2012,marcolongo_ionic_2017}.
The original paper~\cite{kamaya_lithium_2011} reports three site types in the unit cell: tetrahedrally coordinated 16h, tetrahedrally coordinated 8f, and octahedrally coordinated 4d sites, all coordinated with sulfur, with only the latter possessing full occupancy.
The 16h and 8f sites denote edge-sharing \ce{LiS4}  tetrahedra that form one-dimensional channels along the c-axis, the main diffusive pathways~\cite{kamaya_lithium_2011, mo_first_2012}.
Adams and Rao~\cite{adams_structural_2012} found evidence for an additional four-fold coordinated site -- termed 4c -- using classical simulations, which was validated in subsequent experiments by Kuhn~\textit{et al.}~\cite{kuhn_single-crystal_2013}

We analyze the first-principles molecular dynamics trajectories for LGPS that were produced for a recent work discussing the failure of the Nernst-Einstein relation in this structure~\cite{marcolongo_ionic_2017}.
We refer to the reference for computational details, and state only that the trajectories were run with the cp.x code of the \texttt{Quantum ESPRESSO} distribution~\cite{giannozzi_quantum_2009}
 with a PBE-exchange correlation functional~\cite{perdew_generalized_1996}.
Using a unit cell of 50 atoms, 428 picoseconds of dynamics were obtained in the microcanonical ensemble, after an equilibration run at a target temperature of 500~K.
We find a Li-ion density, shown in \myfigref{fig:dens-lgps}, that is compatible with literature results on the unidimensional channels~\cite{adams_structural_2012} that dominate the diffusion in this material.
The diffusion in this material, calculated from the mean-square displacement, shown in \supplref{3}, is $D_{tr}^{Li}=\mathrm{3.25\times 10^{-6}\,cm^2\,s^{-1}}$,
compatible with literature results.
For example, Kuhn \textit{et al.}~\cite{kuhn_tetragonal_2013} report a value of the Li-ion tracer diffusion coefficient of $D_{tr}^{Li}\approx 10^{-6}\,\mathrm{cm^2\,s^{-1}}$ at 500~K,
which is close to our estimate and certainly within the likely error bounds of FPMD that stem from, among other factors, short simulations in small unit cells.

The landmark analysis is applied to the equilibrated trajectory to determine statistics.
We treat germanium and phosphorus atoms as one species since the 4d tetrahedral site is occupied by either species to avoid identifying extraneous site types due to the arbitrary choice of occupation of these sites.
We will refer to both phosphorus and germanium as phosphorus hereafter.
After site detection and SOAP clustering, shown in \supplref{4}, we find 30 sites of type 0, 24 sites of type 1, and four sites of type 2.
We see that type 2 corresponds to the octahedral environment of the 4d site.
To understand the difference between the different site types, we calculate the RDF for every site type, shown in the middle and bottom panels of \myfigref{fig:rdf-lgps} for sulfur and phosphorus. A visual depiction of where the sites are located is shown in \myfigref{fig:final-analysis-lgps}.

In the RDFs between lithium and phosphorus, key differences appear between the different site types.
While site type 1 is compatible with four-fold coordination with sulfur, site types 0 and 2 tend to plateau towards a coordination with six sulfur atoms, which is expected only for the latter site type.
There is no evidence for a six-fold coordinated site type inside the ion-conducting channel of LGPS.
We should note that --- to our knowledge --- no analysis has yet been done on dynamically short-lived features of the coordination of lithium with sulfur in LGPS,
so it is possible that the features we perceive in our analysis are not detected when studying averages.
However, we also observe that the algorithm is less robust than for the other studied examples.
The number of sites as well as the clustering to types depend in this case more strongly on the parameters chosen for the site analysis.
The very similar atomic environment of different site types leads to large overlap of clusters of the SOAP vectors, shown in \supplref{4}.
The superionic behavior of Li ions in LGPS impedes the precise definition of a site for any single mobile ion in the dynamic potential energy landscape.
LGPS, representative of superionic systems, can be seen as a worst-case scenario for the present site analysis.

When analyzing LGPS it also becomes evident that the classification via SOAP vectors can yield different results than the Wyckoff symbols resulting from symmetry analysis.
Different Wyckoff positions can be classified as the same site type if their chemical and geometric environments are too similar to differentiate.
Further, as a result of symmetry breaking during molecular dynamics, two sites with the same Wyckoff position can be classified as different types, especially in non-ergodic simulations.
This is not necessarily a weakness of the analysis, but something to be aware of.
We note that despite the obvious difficulties in detecting sites and site types reported in the literature,
our analysis found four off-channel four-fold coordinated sites, which were termed 4c sites by Adams and Rao~\cite{adams_structural_2012}.
These sites had been missed in earlier FPMD simulations~\cite{mo_first_2012}, since they were not reported by preceding experiments.
Thus an unsupervised and unbiased analysis can help when experimental data is lacking or incomplete.

\setlength{\tabcolsep}{0.5em}

\begin{table}[t]
	\centering
    
    \begin{tabular}{@{\extracolsep{\fill}}lll}
           Structure & Sites & Sites \\ 
           & Wyckoff (CIF)  & (Landmark + SOAP) \\ 
        \hline
        \ce{Li32Al16B16O64} & 16e + 16e  & 16 + 16 \checkmark \\ 
        \ce{Li24Sc8B16O48}  & N/A        & 8 + 16  \\
        \ce{Li24Ba16Ta8N32}   & 8e + 16f & 8 + 16  \checkmark \\ 
        \ce{Li20Re4N16}      & 4a + 16g  & 8 + 16  \\
        \ce{Li12Rb8B4P16O56} & 4d + 8g   & 4 + 8   \checkmark \\
        \ce{Li6Zn6As6O24}   & 3b + 3b    & 3 + 3   \checkmark \\ 
        \ce{Li24Zn4O16}     & 16f + 8d   & 20 + 8  \\
    \end{tabular}
    \caption{
        Comparison between the presented landmark analysis and the sites listed in CIF files taken from structural databases.
        A checkmark indicates structures where the sites in the CIF file and the results of the unsupervised analysis agree, both in number of site types and number of sites of each type. This is the case in all but two structures, \ce{Li20Re4N16} and \ce{Li24Zn4O16}, which are discussed in the text.
    }
    \label{tab:lmk-vs-cif}
\end{table}

\begin{figure}[t]
    \includegraphics[width=\hsize]{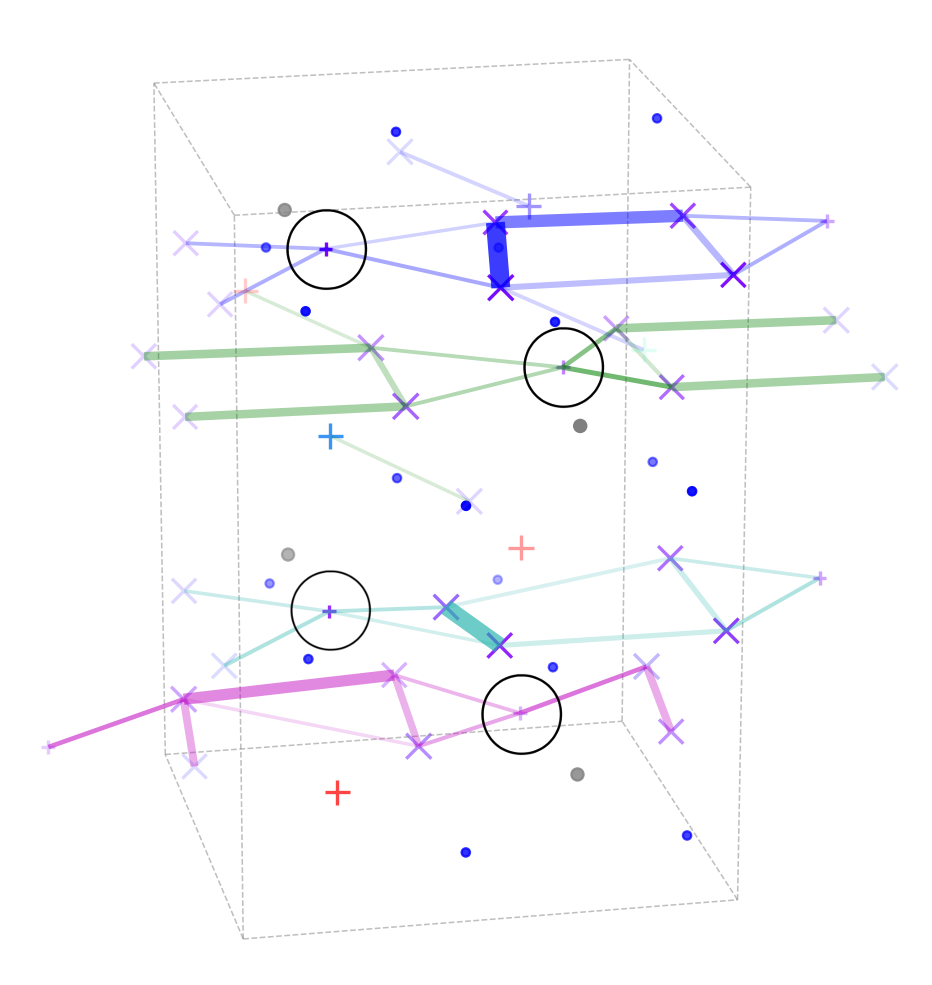}
    \includegraphics[width=\hsize]{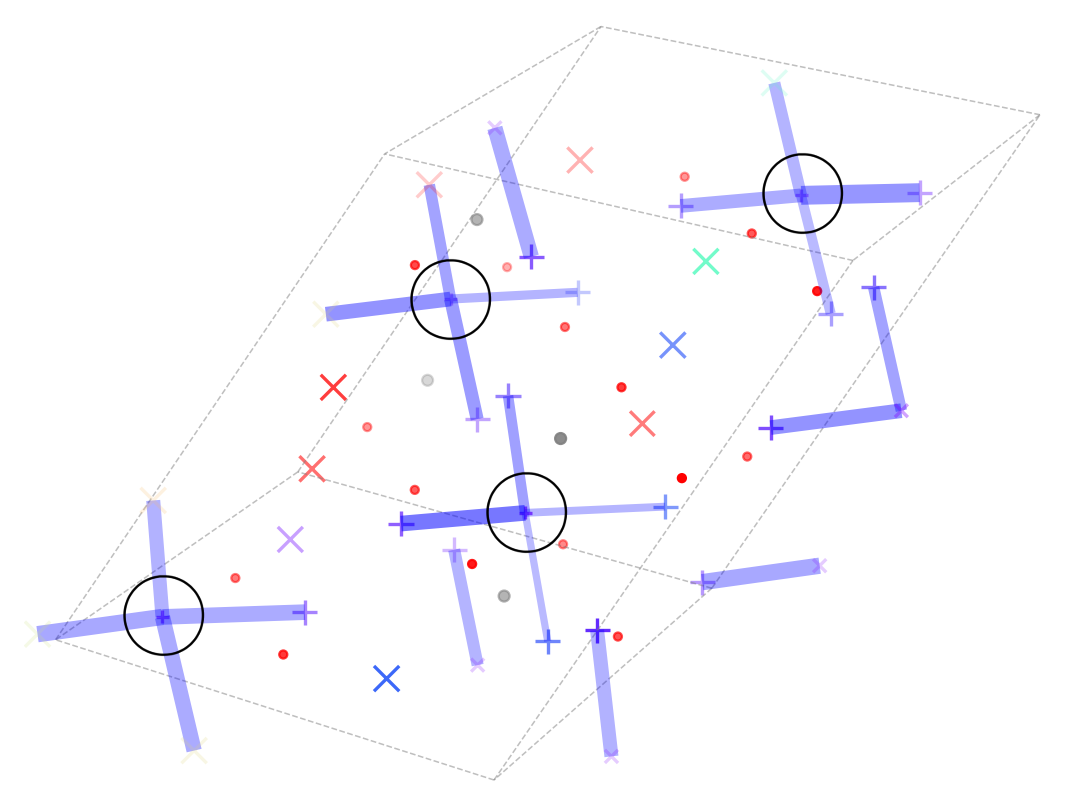}
    \caption{Landmark analysis of \ce{Li20Re4N16} (top) and \ce{Li24Zn4O16} (bottom). 
    The sites circled in black are absent in the CIF files.}
    \label{fig:non-cond-extra-sites}
\end{figure}

\subsection{Non-diffusive structures}
\label{subsec:non-diff}
To further validate the method, we additionally study seven non-conductive structures.
The structures were selected from an ongoing screening effort intended to find new solid-state electrolytes, and chosen from the least conductive systems that had two site types in their CIF file.
For every structure, a molecular dynamics simulation is run at a temperature of 1000~K, with simulation lengths long enough to estimate the diffusivity of the material.
All other simulation parameters are the same as those presented for LASO in \mysecref{sec:laso}; details can be found in Appendix~\ref{sec:calc-params}.
The landmark analysis is run on every second frame of the trajectory (about every 60~fs).
We use the same default landmark analysis parameters for all of the materials, with further details given in Appendix~\ref{sec:sa-params}.

The results can be seen for each of the seven materials in Table~\ref{tab:lmk-vs-cif}.
For all but two materials, the landmark analysis produces the same number of sites and the same division of those sites into types as given in the corresponding CIF files.
For these materials, unlike LGPS, the Wyckoff analysis and the SOAP analysis coincide.
In \ce{Li20Re4N16} and \ce{Li24Zn4O16}, however -- like in \ce{LiAlSiO4} -- unsupervised landmark analysis identifies sites that are not present in the CIF files from ICSD (see \myfigref{fig:non-cond-extra-sites}).
In \ce{Li20Re4N16}, these four sites complete the planar connected components in the material; they are transitional sites with low occupancy and residence time.
Their existence is confirmed by an analysis of the Li-ion densities observed in the trajectories.
In \ce{Li24Zn4O16}, Li-ions from neighboring sites occasionally and briefly jump to the additional sites and then back.
The additional sites again have low occupancy and residence time and are confirmed by a density analysis of the real-space coordinates.

\section{Implementation}
  \label{sec:implementation}

The landmark analysis presented here is implemented as a component of \texttt{SITATOR}~\cite{sitator},
a modular, extensible, open-source Python framework for analyzing networks of sites in molecular dynamics simulations of solid-state materials.
\texttt{SITATOR} provides two fundamental data structures: \texttt{SITENETWORK}, which represents possible sites for some mobile atoms in a host lattice and \texttt{SITETRAJECTORY},
which stores discretized trajectories for those mobile atoms.
A \texttt{SITENETWORK} can also store arbitrary site and edge attributes. 
\texttt{SITATOR} includes an optimized implementation of landmark analysis as well as pre-processing utilities for trajectories and tools for analyzing and visualizing the results of site analyses.

\section{Conclusions}
  \label{sec:conclusion}

We presented a novel method to perform a site analysis of molecular dynamics trajectories to analyze ionic diffusion in solid-state structures.
The method is robust and can run over a large range of materials with a minimal set of parameters and little human intervention.
As we have shown, our landmark analysis performs well where other methods fail, whether because of very high exchange rates and/or close proximity between sites (as in LLZO), or because needed prior information is missing (as in LASO, \ce{Li20Re4N16}, and \ce{Li24Zn4O16} where several sites that were occupied during our simulations are not given in the experimental CIF file).
As became evident for LGPS, superionic conductors with a liquid-like, highly disordered lithium sublattice are hard to analyze with the tool, and the signals from the analysis need to be studied in further detail in subsequent work.
A suggestion for subsequent work is automatically computing the configurational entropy descriptor $\tilde{S}$ described by Kweon \textit{et al.}~\cite{kweon_structural_2017} 
While the method presented here will not necessarily outperform carefully chosen analysis tools with parameters specific to the system under investigation, it has advantages when comparing different systems and in high-throughput applications, such as the search for microscopic descriptors for ionic diffusion in the solid state.
Another suggestion is to study the collective motion in common superionic conductors from occupation statistics given by the landmark analysis.
Concerted motion is important for ionic diffusion in a wide class of systems~\cite{he_origin_2017}, and an analysis that can be used with the same set of parameters on a wide range of materials can be used to quantify collective effects rigorously.

\section*{Acknowledgments}
We thank Aris Marcolongo for supplying the LGPS trajectories.
We also express our gratitude to Matthieu Mottet who provided force-field parameters for LLZO and advice on running the simulations.
We would like to thank Felix Musil, Piero Gasparotto, and Michele Ceriotti for their support with the SOAP descriptor and the \texttt{QUIPPY} interface as well as fruitful discussions.
We gratefully acknowledge financial support from the Swiss National Science Foundation (SNSF)  Project No. 200021-159198.
This work was supported by a grant from the Swiss National Supercomputing Centre (CSCS) under project ID mr0.


\appendix

\section{Landmark vector clustering algorithm}
    \label{sec:lvec-clustering}
    
When determining the landmarks in a system, we use the efficient implementation of the Voronoi decomposition from \texttt{ZEO++}~\cite{willems_algorithms_2012, martin_addressing_2012},
 which accounts for periodic boundary conditions.
A custom hierarchical agglomerative algorithm is used to cluster the landmark vectors. The algorithm is designed for streaming:
no pairwise distance matrix is ever computed or stored, and the landmark vectors can be streamed from disk in the order they were written, avoiding random access.
Clusters are represented by their average landmark vectors, called centers.
At each iteration clusters whose centers are sufficiently similar are merged.
After a small number of iterations, a steady state is reached when no clusters can be merged; this is taken as the final clustering.
The original landmark vectors are then each assigned to the cluster whose center they are most similar to.

Two parameters control the characteristics of the landmark clustering:
the clustering threshold, which determines how aggressively new clusters (sites) should be added, and the minimum cluster size,
which filters out sites whose occupancy is extremely low (such clusters likely represent thermal noise or transitional states).
These parameters allow the user to tune spatial and temporal resolution.
Specifically:

\begin{enumerate}
    \item Set the initial cluster centers $\vec{c}_i$ to the landmark vectors. The order of the landmark vectors does affect the clustering, but in practice we have found the effect to be minimal. We process the landmark vectors in the order they were generated: chronologically and in whatever order the mobile ions were numbered.
    \item Take the first existing cluster center $\vec{c}_0$ as the first new cluster center $\vec{c}'_0$. Then, for each remaining cluster center $\vec{c}_i$, $i \in [1, N)$:

        \begin{enumerate}
            \item Find the new cluster center $\vec{c}'_j$ to which the old cluster center $\vec{c}_i$ is most similar:
                \begin{equation*}
                    j = \argmax_{j \in [0, N')}{S(\vec{c}_i, \vec{c}'_j)}
                \end{equation*}
            
            where $N'$ is the current number of new cluster centers and $S(\cdot, \cdot)$ is the normalized cosine metric:
                \begin{equation*}
                    S(\vec{c}_i, \vec{c}'_j) = \frac{\vec{c}_i \cdot \vec{c}'_j}{|\vec{c}_i||\vec{c}'_j|}
                \end{equation*}
                
            \item If
                \begin{equation*}
                    S(\vec{c}_i, \vec{c}'_{j}) > \text{clustering threshold}
                \end{equation*}
                then merge the old cluster $\vec{c}_i$ into the new cluster $\vec{c}'_{j}$:
                \begin{equation*}
                    \vec{c}'_{j} = \frac{n \vec{c}'_{j} + \vec{c}_i}{n + 1}
                \end{equation*}
                where $n$ is the total number of old clusters that have been merged to form $\vec{c}'_{j}$ so far.
                
                \noindent Otherwise, keep $\vec{c}_i$ as the center of its own cluster:
                \begin{align*}
                    \vec{c}'_{N'} &= \vec{c}_i \\
                    N' &= N' + 1
                \end{align*}
        \end{enumerate}
        
    \item Repeat the previous step until no further clusters can be merged; the $\vec{c}_i$, $i \in [0, N)$ are the final clusters.
    
    \item Assign the landmark vectors to clusters. The assignment threshold controls how dissimilar a landmark vector can be to its cluster's center before it is marked as unassigned. This parameter controls the trade-off between spatial accuracy and the proportion of unassigned mobile atom positions: high values will give greater spatial precision, while lower values will ensure that almost all mobile atoms are assigned to sites at all times.

    For each landmark vector $\vec{l}$:
    
        \begin{enumerate}
            \item Find the most similar cluster center:
                \begin{equation*}
                    s = \max_{j \in [0, N)}{S(\vec{l}, \vec{c}_{j})}
                \end{equation*}
            \item If $s > \text{assignment threshold}$, then mark $\vec{l}$ as a member of the corresponding cluster with confidence $s$.
        
                  Otherwise, mark $\vec{l}$ as unassigned.
        \end{enumerate}
    
    \item Remove clusters smaller than the minimum cluster size.
    \item Repeat step 4 with the remaining clusters, yielding the final cluster assignments.
    
\end{enumerate}

\section{Markov clustering}
    \label{sec:markov-clustering}

We apply Markov Clustering~\cite{van_dongen_MCL_2008} to the matrix $\bm M$  to simulate biased random walks through a graph, giving preference to high-probability routes.
Once the process converges, a set of internally highly connected subgraphs remains.
The sites in each resulting subgraph, if there are more than one, are merged into a single site. Their real-space positions are averaged, and the mobile ions that occupied any of the merged sites now occupy the new site.

We use typical Markov Clustering parameters of $2.0$ for both expansion and inflation.
We do not add artificial self loops to the graph since $\bm M$ already contains appropriate nonzero values on the diagonal.

\section{Parameter estimation for Density-Peak clustering}
    \label{sec:den-peak-params}

Density-peak clustering~\cite{rodriguez_clustering_2014} defines the number of clusters as the number of data points with extreme outlier values of $\rho$ (density) and $\delta$ (distance to nearest neighbor with larger $\rho$), as determined by a user-specified threshold.
Rodriguez and Laio~\cite{rodriguez_clustering_2014} suggest a simple heuristic for determining this threshold that we adopt and automate. First, we compute the values $\gamma_i = \rho_i\delta_i$ and sort them into decreasing order. In a well behaved clustering problem, a plot of $\gamma$ then has a recognizable ``elbow,'' and the points before the elbow -- before the curve rapidly flattens out -- are the outliers. Thus the problem of determining the thresholds is equivalent to finding the elbow of this curve.

We use a simplified version of the knee-finding algorithm presented by Satopaa \textit{et al.}~\cite{satopaa_finding_2011} A straight line is taken between $(0, \gamma_0)$ and $(n, \gamma_n)$, and the point $(i, \gamma_i)$ with the maximum distance to that line is taken as the elbow. The $\rho$ and $\delta$ values corresponding to that point are then used as the thresholds for the density-peak clustering.

\section{Molecular dynamics parameters}
    \label{sec:calc-params}
The simulations for LASO and the seven non-diffusive structures are performed with the pw.x module in the \texttt{Quantum ESPRESSO} distribution~\cite{giannozzi_quantum_2009},
 using pseudopotentials and cutoffs from the SSSP Efficiency library 1.0~\cite{prandini_precision_2018}.
The exchange-correlation used in the DFT is PBE~\cite{perdew_generalized_1996}.
The materials informatics platform AiiDA~\cite{pizzi_aiida:_2016} is used to ensure full reproducibility of the results and achieve a high degree of automation.

We always perform a variable-cell relaxation prior to the molecular dynamics, with a uniform k-point grid of 0.2~\r{A}$^{-1}$ and no electronic smearing since we consider only electronic insulators.
The energy and force convergence thresholds are  $0.5\times 10^{-4}$ and $0.25\times 10^{-5}$ in atomic units, respectively.
We set the pressure threshold to 0.5~kbar.
A meta-convergence threshold on the volume, which specifies the relative volume change between subsequent relaxations, is set to 0.01.

We create supercells with the criterion that the minimal distance between opposite faces is always larger than 6.5~\r{A}.
We run the molecular dynamics simulations with a stochastic velocity rescaling thermostat~\cite{bussi_canonical_2007} which we implemented into \texttt{Quantum ESPRESSO}, with a characteristic time of the thermostat set to 0.2~ps at constant volume and number of particles (NVT ensemble).
The timestep is set to 1.45~fs, and snapshots of the trajectory are taken every 20 time steps.

The origin of the non-diffusive structures is given in Table~\ref{tab:nondiffusive-data}, together with the simulation time.
The structures are taken from the Inorganic Crystallography Open Database (ICSD)~\cite{belsky_new_2002} and the Open Crystallography database (COD)~\cite{grazulis_crystallography_2012}.
\setlength{\tabcolsep}{0.5em}
\begin{table}[t]
    \centering
    \begin{tabular}{ l  c  c  c }
         Structure  & $T_{sim}$ (ps)     & DB & DB-ID \\
        \hline
        \ce{Li32Al16B16O64}   & 72            & ICSD           & 50612  \\ 
        \ce{Li24Sc8B16O48}    & 218           & COD            & 2218562 \\  
        \ce{Li24Ba16Ta8N32}   & 58            & ICSD           & 75031 \\ 
        \ce{Li20Re4N16}       & 159           & ICSD           & 92468  \\ 
        \ce{Li12Rb8B4P16O56}  & 116           & ICSD           & 424352  \\ 
        \ce{Li6Zn6As6O24}     & 226           & ICSD           & 86184   \\ 
        \ce{Li24Zn4O16}       & 407           & ICSD           & 62137 \\ 
    \end{tabular}
    \caption{For every structure analyzed in \mysecref{subsec:non-diff}, we list the simulation length $T_{sim}$ in picoseconds, the database (DB) from which the structure was retrieved, and the structure's database ID.}
    \label{tab:nondiffusive-data}
\end{table}

\section{Site analysis parameters}
    \label{sec:sa-params}

Unless otherwise indicated, the landmark analysis uses a cutoff midpoint of $d_0=1.5$ and steepness of $k=30$, a minimum site occupancy of 1\%, and landmark clustering and assignment thresholds of $0.9$.
For computing SOAP descriptors, unless otherwise specified, we use a Gaussian width of 0.5~\r{A} on the atomic positions, a cutoff transition width of 0.5~\r{A}, and spherical harmonics up to $n_{max}=l_{max}=6$.
The radial cutoff is set to always include the nearest neighbor shell of all other species (excluding the mobile species).
We calculate SOAP vectors for mobile ions every tenth frame, and average every 10 SOAP vectors to reduce noise.
The principal components of the averaged SOAP vectors are extracted using Principal Component analysis (PCA) to retain at least 95\% of the variance, and the clustering is performed in this reduced space.

\bibliography{bibliography}

\begin{thebibliography}{90}%
\makeatletter
\providecommand \@ifxundefined [1]{%
 \@ifx{#1\undefined}
}%
\providecommand \@ifnum [1]{%
 \ifnum #1\expandafter \@firstoftwo
 \else \expandafter \@secondoftwo
 \fi
}%
\providecommand \@ifx [1]{%
 \ifx #1\expandafter \@firstoftwo
 \else \expandafter \@secondoftwo
 \fi
}%
\providecommand \natexlab [1]{#1}%
\providecommand \enquote  [1]{``#1''}%
\providecommand \bibnamefont  [1]{#1}%
\providecommand \bibfnamefont [1]{#1}%
\providecommand \citenamefont [1]{#1}%
\providecommand \href@noop [0]{\@secondoftwo}%
\providecommand \href [0]{\begingroup \@sanitize@url \@href}%
\providecommand \@href[1]{\@@startlink{#1}\@@href}%
\providecommand \@@href[1]{\endgroup#1\@@endlink}%
\providecommand \@sanitize@url [0]{\catcode `\\12\catcode `\$12\catcode
  `\&12\catcode `\#12\catcode `\^12\catcode `\_12\catcode `\%12\relax}%
\providecommand \@@startlink[1]{}%
\providecommand \@@endlink[0]{}%
\providecommand \url  [0]{\begingroup\@sanitize@url \@url }%
\providecommand \@url [1]{\endgroup\@href {#1}{\urlprefix }}%
\providecommand \urlprefix  [0]{URL }%
\providecommand \Eprint [0]{\href }%
\providecommand \doibase [0]{http://dx.doi.org/}%
\providecommand \selectlanguage [0]{\@gobble}%
\providecommand \bibinfo  [0]{\@secondoftwo}%
\providecommand \bibfield  [0]{\@secondoftwo}%
\providecommand \translation [1]{[#1]}%
\providecommand \BibitemOpen [0]{}%
\providecommand \bibitemStop [0]{}%
\providecommand \bibitemNoStop [0]{.\EOS\space}%
\providecommand \EOS [0]{\spacefactor3000\relax}%
\providecommand \BibitemShut  [1]{\csname bibitem#1\endcsname}%
\let\auto@bib@innerbib\@empty
\bibitem [{\citenamefont {Armand}\ and\ \citenamefont
  {Tarascon}(2008)}]{armand_building_2008}%
  \BibitemOpen
  \bibfield  {author} {\bibinfo {author} {\bibfnamefont {M.}~\bibnamefont
  {Armand}}\ and\ \bibinfo {author} {\bibfnamefont {J.-M.}\ \bibnamefont
  {Tarascon}},\ }\href {\doibase 10.1038/451652a} {\bibfield  {journal}
  {\bibinfo  {journal} {Nature}\ }\textbf {\bibinfo {volume} {451}},\ \bibinfo
  {pages} {652} (\bibinfo {year} {2008})}\BibitemShut {NoStop}%
\bibitem [{\citenamefont {Schaefer}\ \emph {et~al.}(2012)\citenamefont
  {Schaefer}, \citenamefont {Lu}, \citenamefont {Moganty}, \citenamefont
  {Agarwal}, \citenamefont {Jayaprakash},\ and\ \citenamefont
  {Archer}}]{schaefer_electrolytes_2012}%
  \BibitemOpen
  \bibfield  {author} {\bibinfo {author} {\bibfnamefont {J.~L.}\ \bibnamefont
  {Schaefer}}, \bibinfo {author} {\bibfnamefont {Y.}~\bibnamefont {Lu}},
  \bibinfo {author} {\bibfnamefont {S.~S.}\ \bibnamefont {Moganty}}, \bibinfo
  {author} {\bibfnamefont {P.}~\bibnamefont {Agarwal}}, \bibinfo {author}
  {\bibfnamefont {N.}~\bibnamefont {Jayaprakash}}, \ and\ \bibinfo {author}
  {\bibfnamefont {L.~A.}\ \bibnamefont {Archer}},\ }\href {\doibase
  10.1007/s13204-011-0044-x} {\bibfield  {journal} {\bibinfo  {journal}
  {Applied Nanoscience}\ }\textbf {\bibinfo {volume} {2}},\ \bibinfo {pages}
  {91} (\bibinfo {year} {2012})}\BibitemShut {NoStop}%
\bibitem [{\citenamefont {Balakrishnan}\ \emph {et~al.}(2006)\citenamefont
  {Balakrishnan}, \citenamefont {Ramesh},\ and\ \citenamefont
  {Prem~Kumar}}]{balakrishnan_safety_2006}%
  \BibitemOpen
  \bibfield  {author} {\bibinfo {author} {\bibfnamefont {P.~G.}\ \bibnamefont
  {Balakrishnan}}, \bibinfo {author} {\bibfnamefont {R.}~\bibnamefont
  {Ramesh}}, \ and\ \bibinfo {author} {\bibfnamefont {T.}~\bibnamefont
  {Prem~Kumar}},\ }\href {\doibase 10.1016/j.jpowsour.2005.12.002} {\bibfield
  {journal} {\bibinfo  {journal} {Journal of Power Sources}\ }\textbf {\bibinfo
  {volume} {155}},\ \bibinfo {pages} {401} (\bibinfo {year} {2006})},\ \bibinfo
  {note} {00340}\BibitemShut {NoStop}%
\bibitem [{\citenamefont {Bachman}\ \emph {et~al.}(2016)\citenamefont
  {Bachman}, \citenamefont {Muy}, \citenamefont {Grimaud}, \citenamefont
  {Chang}, \citenamefont {Pour}, \citenamefont {Lux}, \citenamefont {Paschos},
  \citenamefont {Maglia}, \citenamefont {Lupart}, \citenamefont {Lamp},
  \citenamefont {Giordano},\ and\ \citenamefont
  {Shao-Horn}}]{bachman_inorganic_2016}%
  \BibitemOpen
  \bibfield  {author} {\bibinfo {author} {\bibfnamefont {J.~C.}\ \bibnamefont
  {Bachman}}, \bibinfo {author} {\bibfnamefont {S.}~\bibnamefont {Muy}},
  \bibinfo {author} {\bibfnamefont {A.}~\bibnamefont {Grimaud}}, \bibinfo
  {author} {\bibfnamefont {H.-H.}\ \bibnamefont {Chang}}, \bibinfo {author}
  {\bibfnamefont {N.}~\bibnamefont {Pour}}, \bibinfo {author} {\bibfnamefont
  {S.~F.}\ \bibnamefont {Lux}}, \bibinfo {author} {\bibfnamefont
  {O.}~\bibnamefont {Paschos}}, \bibinfo {author} {\bibfnamefont
  {F.}~\bibnamefont {Maglia}}, \bibinfo {author} {\bibfnamefont
  {S.}~\bibnamefont {Lupart}}, \bibinfo {author} {\bibfnamefont
  {P.}~\bibnamefont {Lamp}}, \bibinfo {author} {\bibfnamefont {L.}~\bibnamefont
  {Giordano}}, \ and\ \bibinfo {author} {\bibfnamefont {Y.}~\bibnamefont
  {Shao-Horn}},\ }\href {\doibase 10.1021/acs.chemrev.5b00563} {\bibfield
  {journal} {\bibinfo  {journal} {Chemical Reviews}\ }\textbf {\bibinfo
  {volume} {116}},\ \bibinfo {pages} {140} (\bibinfo {year}
  {2016})}\BibitemShut {NoStop}%
\bibitem [{\citenamefont {Manthiram}\ \emph {et~al.}(2017)\citenamefont
  {Manthiram}, \citenamefont {Yu},\ and\ \citenamefont
  {Wang}}]{manthiram_lithium_2017}%
  \BibitemOpen
  \bibfield  {author} {\bibinfo {author} {\bibfnamefont {A.}~\bibnamefont
  {Manthiram}}, \bibinfo {author} {\bibfnamefont {X.}~\bibnamefont {Yu}}, \
  and\ \bibinfo {author} {\bibfnamefont {S.}~\bibnamefont {Wang}},\ }\href
  {\doibase 10.1038/natrevmats.2016.103} {\bibfield  {journal} {\bibinfo
  {journal} {Nature Reviews Materials}\ }\textbf {\bibinfo {volume} {2}},\
  \bibinfo {pages} {16103} (\bibinfo {year} {2017})}\BibitemShut {NoStop}%
\bibitem [{\citenamefont {Adams}\ and\ \citenamefont
  {Rao}(2012{\natexlab{a}})}]{adams_structural_2012}%
  \BibitemOpen
  \bibfield  {author} {\bibinfo {author} {\bibfnamefont {S.}~\bibnamefont
  {Adams}}\ and\ \bibinfo {author} {\bibfnamefont {R.~P.}\ \bibnamefont
  {Rao}},\ }\href {\doibase 10.1039/C2JM16688G} {\bibfield  {journal} {\bibinfo
   {journal} {Journal of Materials Chemistry}\ }\textbf {\bibinfo {volume}
  {22}},\ \bibinfo {pages} {7687} (\bibinfo {year}
  {2012}{\natexlab{a}})}\BibitemShut {NoStop}%
\bibitem [{\citenamefont {Adams}\ and\ \citenamefont
  {Rao}(2012{\natexlab{b}})}]{adams_ion_2012}%
  \BibitemOpen
  \bibfield  {author} {\bibinfo {author} {\bibfnamefont {S.}~\bibnamefont
  {Adams}}\ and\ \bibinfo {author} {\bibfnamefont {R.~P.}\ \bibnamefont
  {Rao}},\ }\href {\doibase 10.1039/C1JM14588F} {\bibfield  {journal} {\bibinfo
   {journal} {J. Mater. Chem.}\ }\textbf {\bibinfo {volume} {22}},\ \bibinfo
  {pages} {1426} (\bibinfo {year} {2012}{\natexlab{b}})}\BibitemShut {NoStop}%
\bibitem [{\citenamefont {Xu}\ \emph {et~al.}(2012{\natexlab{a}})\citenamefont
  {Xu}, \citenamefont {Park}, \citenamefont {Lee}, \citenamefont {Kim},
  \citenamefont {Park},\ and\ \citenamefont {Ma}}]{xu_mechanisms_2012}%
  \BibitemOpen
  \bibfield  {author} {\bibinfo {author} {\bibfnamefont {M.}~\bibnamefont
  {Xu}}, \bibinfo {author} {\bibfnamefont {M.~S.}\ \bibnamefont {Park}},
  \bibinfo {author} {\bibfnamefont {J.~M.}\ \bibnamefont {Lee}}, \bibinfo
  {author} {\bibfnamefont {T.~Y.}\ \bibnamefont {Kim}}, \bibinfo {author}
  {\bibfnamefont {Y.~S.}\ \bibnamefont {Park}}, \ and\ \bibinfo {author}
  {\bibfnamefont {E.}~\bibnamefont {Ma}},\ }\href {\doibase
  10.1103/PhysRevB.85.052301} {\bibfield  {journal} {\bibinfo  {journal}
  {Physical Review B}\ }\textbf {\bibinfo {volume} {85}},\ \bibinfo {pages}
  {052301} (\bibinfo {year} {2012}{\natexlab{a}})}\BibitemShut {NoStop}%
\bibitem [{\citenamefont {Deng}\ \emph {et~al.}(2015)\citenamefont {Deng},
  \citenamefont {Eames}, \citenamefont {Chotard}, \citenamefont {Lalère},
  \citenamefont {Seznec}, \citenamefont {Emge}, \citenamefont {Pecher},
  \citenamefont {Grey}, \citenamefont {Masquelier},\ and\ \citenamefont
  {Islam}}]{deng_structural_2015}%
  \BibitemOpen
  \bibfield  {author} {\bibinfo {author} {\bibfnamefont {Y.}~\bibnamefont
  {Deng}}, \bibinfo {author} {\bibfnamefont {C.}~\bibnamefont {Eames}},
  \bibinfo {author} {\bibfnamefont {J.-N.}\ \bibnamefont {Chotard}}, \bibinfo
  {author} {\bibfnamefont {F.}~\bibnamefont {Lalère}}, \bibinfo {author}
  {\bibfnamefont {V.}~\bibnamefont {Seznec}}, \bibinfo {author} {\bibfnamefont
  {S.}~\bibnamefont {Emge}}, \bibinfo {author} {\bibfnamefont {O.}~\bibnamefont
  {Pecher}}, \bibinfo {author} {\bibfnamefont {C.~P.}\ \bibnamefont {Grey}},
  \bibinfo {author} {\bibfnamefont {C.}~\bibnamefont {Masquelier}}, \ and\
  \bibinfo {author} {\bibfnamefont {M.~S.}\ \bibnamefont {Islam}},\ }\href
  {\doibase 10.1021/jacs.5b04444} {\bibfield  {journal} {\bibinfo  {journal}
  {Journal of the American Chemical Society}\ }\textbf {\bibinfo {volume}
  {137}},\ \bibinfo {pages} {9136} (\bibinfo {year} {2015})}\BibitemShut
  {NoStop}%
\bibitem [{\citenamefont {Kozinsky}\ \emph {et~al.}(2016)\citenamefont
  {Kozinsky}, \citenamefont {Akhade}, \citenamefont {Hirel}, \citenamefont
  {Hashibon}, \citenamefont {Els\"asser}, \citenamefont {Mehta}, \citenamefont
  {Logeat},\ and\ \citenamefont {Eisele}}]{kozinsky_effects_2016}%
  \BibitemOpen
  \bibfield  {author} {\bibinfo {author} {\bibfnamefont {B.}~\bibnamefont
  {Kozinsky}}, \bibinfo {author} {\bibfnamefont {S.~A.}\ \bibnamefont
  {Akhade}}, \bibinfo {author} {\bibfnamefont {P.}~\bibnamefont {Hirel}},
  \bibinfo {author} {\bibfnamefont {A.}~\bibnamefont {Hashibon}}, \bibinfo
  {author} {\bibfnamefont {C.}~\bibnamefont {Els\"asser}}, \bibinfo {author}
  {\bibfnamefont {P.}~\bibnamefont {Mehta}}, \bibinfo {author} {\bibfnamefont
  {A.}~\bibnamefont {Logeat}}, \ and\ \bibinfo {author} {\bibfnamefont
  {U.}~\bibnamefont {Eisele}},\ }\href {\doibase
  10.1103/PhysRevLett.116.055901} {\bibfield  {journal} {\bibinfo  {journal}
  {Physical Review Letters}\ }\textbf {\bibinfo {volume} {116}},\ \bibinfo
  {pages} {055901} (\bibinfo {year} {2016})}\BibitemShut {NoStop}%
\bibitem [{\citenamefont {Klenk}\ and\ \citenamefont
  {Lai}(2016)}]{klenk_finite-size_2016}%
  \BibitemOpen
  \bibfield  {author} {\bibinfo {author} {\bibfnamefont {M.~J.}\ \bibnamefont
  {Klenk}}\ and\ \bibinfo {author} {\bibfnamefont {W.}~\bibnamefont {Lai}},\
  }\href {\doibase 10.1016/j.ssi.2016.03.002} {\bibfield  {journal} {\bibinfo
  {journal} {Solid State Ionics}\ }\textbf {\bibinfo {volume} {289}},\ \bibinfo
  {pages} {143} (\bibinfo {year} {2016})}\BibitemShut {NoStop}%
\bibitem [{\citenamefont {Burbano}\ \emph {et~al.}(2016)\citenamefont
  {Burbano}, \citenamefont {Carlier}, \citenamefont {Boucher}, \citenamefont
  {Morgan},\ and\ \citenamefont {Salanne}}]{burbano_sparse_2016}%
  \BibitemOpen
  \bibfield  {author} {\bibinfo {author} {\bibfnamefont {M.}~\bibnamefont
  {Burbano}}, \bibinfo {author} {\bibfnamefont {D.}~\bibnamefont {Carlier}},
  \bibinfo {author} {\bibfnamefont {F.}~\bibnamefont {Boucher}}, \bibinfo
  {author} {\bibfnamefont {B.~J.}\ \bibnamefont {Morgan}}, \ and\ \bibinfo
  {author} {\bibfnamefont {M.}~\bibnamefont {Salanne}},\ }\href {\doibase
  10.1103/PhysRevLett.116.135901} {\bibfield  {journal} {\bibinfo  {journal}
  {Physical Review Letters}\ }\textbf {\bibinfo {volume} {116}},\ \bibinfo
  {pages} {135901} (\bibinfo {year} {2016})}\BibitemShut {NoStop}%
\bibitem [{\citenamefont {Dawson}\ \emph {et~al.}(2018)\citenamefont {Dawson},
  \citenamefont {Canepa}, \citenamefont {Famprikis}, \citenamefont
  {Masquelier},\ and\ \citenamefont {Islam}}]{dawson_atomic-scale_2018}%
  \BibitemOpen
  \bibfield  {author} {\bibinfo {author} {\bibfnamefont {J.~A.}\ \bibnamefont
  {Dawson}}, \bibinfo {author} {\bibfnamefont {P.}~\bibnamefont {Canepa}},
  \bibinfo {author} {\bibfnamefont {T.}~\bibnamefont {Famprikis}}, \bibinfo
  {author} {\bibfnamefont {C.}~\bibnamefont {Masquelier}}, \ and\ \bibinfo
  {author} {\bibfnamefont {M.~S.}\ \bibnamefont {Islam}},\ }\href {\doibase
  10.1021/jacs.7b10593} {\bibfield  {journal} {\bibinfo  {journal} {Journal of
  the American Chemical Society}\ }\textbf {\bibinfo {volume} {140}},\ \bibinfo
  {pages} {362} (\bibinfo {year} {2018})}\BibitemShut {NoStop}%
\bibitem [{\citenamefont {Wood}\ and\ \citenamefont
  {Marzari}(2006)}]{wood_dynamical_2006}%
  \BibitemOpen
  \bibfield  {author} {\bibinfo {author} {\bibfnamefont {B.~C.}\ \bibnamefont
  {Wood}}\ and\ \bibinfo {author} {\bibfnamefont {N.}~\bibnamefont {Marzari}},\
  }\href {\doibase 10.1103/PhysRevLett.97.166401} {\bibfield  {journal}
  {\bibinfo  {journal} {Physical Review Letters}\ }\textbf {\bibinfo {volume}
  {97}},\ \bibinfo {pages} {166401} (\bibinfo {year} {2006})}\BibitemShut
  {NoStop}%
\bibitem [{\citenamefont {Ong}\ \emph {et~al.}(2013)\citenamefont {Ong},
  \citenamefont {Mo}, \citenamefont {Richards}, \citenamefont {Miara},
  \citenamefont {Lee},\ and\ \citenamefont {Ceder}}]{ong_phase_2013}%
  \BibitemOpen
  \bibfield  {author} {\bibinfo {author} {\bibfnamefont {S.~P.}\ \bibnamefont
  {Ong}}, \bibinfo {author} {\bibfnamefont {Y.}~\bibnamefont {Mo}}, \bibinfo
  {author} {\bibfnamefont {W.~D.}\ \bibnamefont {Richards}}, \bibinfo {author}
  {\bibfnamefont {L.}~\bibnamefont {Miara}}, \bibinfo {author} {\bibfnamefont
  {H.~S.}\ \bibnamefont {Lee}}, \ and\ \bibinfo {author} {\bibfnamefont
  {G.}~\bibnamefont {Ceder}},\ }\href {\doibase 10.1039/C2EE23355J} {\bibfield
  {journal} {\bibinfo  {journal} {Energy Environ. Sci.}\ }\textbf {\bibinfo
  {volume} {6}},\ \bibinfo {pages} {148} (\bibinfo {year} {2013})}\BibitemShut
  {NoStop}%
\bibitem [{\citenamefont {Mo}\ \emph {et~al.}(2012)\citenamefont {Mo},
  \citenamefont {Ong},\ and\ \citenamefont {Ceder}}]{mo_first_2012}%
  \BibitemOpen
  \bibfield  {author} {\bibinfo {author} {\bibfnamefont {Y.}~\bibnamefont
  {Mo}}, \bibinfo {author} {\bibfnamefont {S.~P.}\ \bibnamefont {Ong}}, \ and\
  \bibinfo {author} {\bibfnamefont {G.}~\bibnamefont {Ceder}},\ }\href
  {\doibase 10.1021/cm203303y} {\bibfield  {journal} {\bibinfo  {journal}
  {Chemistry of Materials}\ }\textbf {\bibinfo {volume} {24}},\ \bibinfo
  {pages} {15} (\bibinfo {year} {2012})}\BibitemShut {NoStop}%
\bibitem [{\citenamefont {Xu}\ \emph {et~al.}(2012{\natexlab{b}})\citenamefont
  {Xu}, \citenamefont {Ding},\ and\ \citenamefont
  {Ma}}]{xu_one-dimensional_2012}%
  \BibitemOpen
  \bibfield  {author} {\bibinfo {author} {\bibfnamefont {M.}~\bibnamefont
  {Xu}}, \bibinfo {author} {\bibfnamefont {J.}~\bibnamefont {Ding}}, \ and\
  \bibinfo {author} {\bibfnamefont {E.}~\bibnamefont {Ma}},\ }\href {\doibase
  10.1063/1.4737397} {\bibfield  {journal} {\bibinfo  {journal} {Applied
  Physics Letters}\ }\textbf {\bibinfo {volume} {101}},\ \bibinfo {pages}
  {031901} (\bibinfo {year} {2012}{\natexlab{b}})}\BibitemShut {NoStop}%
\bibitem [{\citenamefont {Mo}\ \emph {et~al.}(2014)\citenamefont {Mo},
  \citenamefont {Ong},\ and\ \citenamefont {Ceder}}]{mo_insights_2014}%
  \BibitemOpen
  \bibfield  {author} {\bibinfo {author} {\bibfnamefont {Y.}~\bibnamefont
  {Mo}}, \bibinfo {author} {\bibfnamefont {S.~P.}\ \bibnamefont {Ong}}, \ and\
  \bibinfo {author} {\bibfnamefont {G.}~\bibnamefont {Ceder}},\ }\href
  {\doibase 10.1021/cm501563f} {\bibfield  {journal} {\bibinfo  {journal}
  {Chemistry of Materials}\ }\textbf {\bibinfo {volume} {26}},\ \bibinfo
  {pages} {5208} (\bibinfo {year} {2014})}\BibitemShut {NoStop}%
\bibitem [{\citenamefont {Meier}\ \emph {et~al.}(2014)\citenamefont {Meier},
  \citenamefont {Laino},\ and\ \citenamefont
  {Curioni}}]{meier_solid-state_2014}%
  \BibitemOpen
  \bibfield  {author} {\bibinfo {author} {\bibfnamefont {K.}~\bibnamefont
  {Meier}}, \bibinfo {author} {\bibfnamefont {T.}~\bibnamefont {Laino}}, \ and\
  \bibinfo {author} {\bibfnamefont {A.}~\bibnamefont {Curioni}},\ }\href
  {\doibase 10.1021/jp5002463} {\bibfield  {journal} {\bibinfo  {journal} {The
  Journal of Physical Chemistry C}\ }\textbf {\bibinfo {volume} {118}},\
  \bibinfo {pages} {6668} (\bibinfo {year} {2014})}\BibitemShut {NoStop}%
\bibitem [{\citenamefont {Wang}\ \emph {et~al.}(2015)\citenamefont {Wang},
  \citenamefont {Richards}, \citenamefont {Ong}, \citenamefont {Miara},
  \citenamefont {Kim}, \citenamefont {Mo},\ and\ \citenamefont
  {Ceder}}]{wang_design_2015}%
  \BibitemOpen
  \bibfield  {author} {\bibinfo {author} {\bibfnamefont {Y.}~\bibnamefont
  {Wang}}, \bibinfo {author} {\bibfnamefont {W.~D.}\ \bibnamefont {Richards}},
  \bibinfo {author} {\bibfnamefont {S.~P.}\ \bibnamefont {Ong}}, \bibinfo
  {author} {\bibfnamefont {L.~J.}\ \bibnamefont {Miara}}, \bibinfo {author}
  {\bibfnamefont {J.~C.}\ \bibnamefont {Kim}}, \bibinfo {author} {\bibfnamefont
  {Y.}~\bibnamefont {Mo}}, \ and\ \bibinfo {author} {\bibfnamefont
  {G.}~\bibnamefont {Ceder}},\ }\href {\doibase 10.1038/nmat4369} {\bibfield
  {journal} {\bibinfo  {journal} {Nature Materials}\ }\textbf {\bibinfo
  {volume} {14}},\ \bibinfo {pages} {1026} (\bibinfo {year}
  {2015})}\BibitemShut {NoStop}%
\bibitem [{\citenamefont {Chu}\ \emph {et~al.}(2016)\citenamefont {Chu},
  \citenamefont {Nguyen}, \citenamefont {Hy}, \citenamefont {Lin},
  \citenamefont {Wang}, \citenamefont {Xu}, \citenamefont {Deng}, \citenamefont
  {Meng},\ and\ \citenamefont {Ong}}]{chu_insights_2016}%
  \BibitemOpen
  \bibfield  {author} {\bibinfo {author} {\bibfnamefont {I.-H.}\ \bibnamefont
  {Chu}}, \bibinfo {author} {\bibfnamefont {H.}~\bibnamefont {Nguyen}},
  \bibinfo {author} {\bibfnamefont {S.}~\bibnamefont {Hy}}, \bibinfo {author}
  {\bibfnamefont {Y.-C.}\ \bibnamefont {Lin}}, \bibinfo {author} {\bibfnamefont
  {Z.}~\bibnamefont {Wang}}, \bibinfo {author} {\bibfnamefont {Z.}~\bibnamefont
  {Xu}}, \bibinfo {author} {\bibfnamefont {Z.}~\bibnamefont {Deng}}, \bibinfo
  {author} {\bibfnamefont {Y.~S.}\ \bibnamefont {Meng}}, \ and\ \bibinfo
  {author} {\bibfnamefont {S.~P.}\ \bibnamefont {Ong}},\ }\href {\doibase
  10.1021/acsami.6b00833} {\bibfield  {journal} {\bibinfo  {journal} {ACS
  Applied Materials \& Interfaces}\ }\textbf {\bibinfo {volume} {8}},\ \bibinfo
  {pages} {7843} (\bibinfo {year} {2016})}\BibitemShut {NoStop}%
\bibitem [{\citenamefont {Zhu}\ \emph {et~al.}(2017)\citenamefont {Zhu},
  \citenamefont {Chu},\ and\ \citenamefont {Ong}}]{zhu_li3yps42_2017}%
  \BibitemOpen
  \bibfield  {author} {\bibinfo {author} {\bibfnamefont {Z.}~\bibnamefont
  {Zhu}}, \bibinfo {author} {\bibfnamefont {I.-H.}\ \bibnamefont {Chu}}, \ and\
  \bibinfo {author} {\bibfnamefont {S.~P.}\ \bibnamefont {Ong}},\ }\href
  {\doibase 10.1021/acs.chemmater.6b04049} {\bibfield  {journal} {\bibinfo
  {journal} {Chemistry of Materials}\ }\textbf {\bibinfo {volume} {29}},\
  \bibinfo {pages} {2474} (\bibinfo {year} {2017})}\BibitemShut {NoStop}%
\bibitem [{\citenamefont {Marcolongo}\ and\ \citenamefont
  {Marzari}(2017)}]{marcolongo_ionic_2017}%
  \BibitemOpen
  \bibfield  {author} {\bibinfo {author} {\bibfnamefont {A.}~\bibnamefont
  {Marcolongo}}\ and\ \bibinfo {author} {\bibfnamefont {N.}~\bibnamefont
  {Marzari}},\ }\href {\doibase 10.1103/PhysRevMaterials.1.025402} {\bibfield
  {journal} {\bibinfo  {journal} {Physical Review Materials}\ }\textbf
  {\bibinfo {volume} {1}},\ \bibinfo {pages} {025402} (\bibinfo {year}
  {2017})}\BibitemShut {NoStop}%
\bibitem [{\citenamefont {Sagotra}\ \emph {et~al.}(2019)\citenamefont
  {Sagotra}, \citenamefont {Chu},\ and\ \citenamefont
  {Cazorla}}]{sagotra_influence_2019}%
  \BibitemOpen
  \bibfield  {author} {\bibinfo {author} {\bibfnamefont {A.~K.}\ \bibnamefont
  {Sagotra}}, \bibinfo {author} {\bibfnamefont {D.}~\bibnamefont {Chu}}, \ and\
  \bibinfo {author} {\bibfnamefont {C.}~\bibnamefont {Cazorla}},\ }\href
  {\doibase 10.1103/PhysRevMaterials.3.035405} {\bibfield  {journal} {\bibinfo
  {journal} {Physical Review Materials}\ }\textbf {\bibinfo {volume} {3}},\
  \bibinfo {pages} {035405} (\bibinfo {year} {2019})}\BibitemShut {NoStop}%
\bibitem [{\citenamefont {Kahle}\ \emph {et~al.}(2018)\citenamefont {Kahle},
  \citenamefont {Marcolongo},\ and\ \citenamefont
  {Marzari}}]{kahle_modeling_2018}%
  \BibitemOpen
  \bibfield  {author} {\bibinfo {author} {\bibfnamefont {L.}~\bibnamefont
  {Kahle}}, \bibinfo {author} {\bibfnamefont {A.}~\bibnamefont {Marcolongo}}, \
  and\ \bibinfo {author} {\bibfnamefont {N.}~\bibnamefont {Marzari}},\ }\href
  {\doibase 10.1103/PhysRevMaterials.2.065405} {\bibfield  {journal} {\bibinfo
  {journal} {Physical Review Materials}\ }\textbf {\bibinfo {volume} {2}},\
  \bibinfo {pages} {065405} (\bibinfo {year} {2018})}\BibitemShut {NoStop}%
\bibitem [{\citenamefont {Ercole}\ \emph {et~al.}(2017)\citenamefont {Ercole},
  \citenamefont {Marcolongo},\ and\ \citenamefont
  {Baroni}}]{ercole_accurate_2017}%
  \BibitemOpen
  \bibfield  {author} {\bibinfo {author} {\bibfnamefont {L.}~\bibnamefont
  {Ercole}}, \bibinfo {author} {\bibfnamefont {A.}~\bibnamefont {Marcolongo}},
  \ and\ \bibinfo {author} {\bibfnamefont {S.}~\bibnamefont {Baroni}},\ }\href
  {\doibase 10.1038/s41598-017-15843-2} {\bibfield  {journal} {\bibinfo
  {journal} {Scientific Reports}\ }\textbf {\bibinfo {volume} {7}},\ \bibinfo
  {pages} {15835} (\bibinfo {year} {2017})}\BibitemShut {NoStop}%
\bibitem [{\citenamefont {Van~der Ven}\ \emph {et~al.}(2001)\citenamefont
  {Van~der Ven}, \citenamefont {Ceder}, \citenamefont {Asta},\ and\
  \citenamefont {Tepesch}}]{van_der_ven_first-principles_2001}%
  \BibitemOpen
  \bibfield  {author} {\bibinfo {author} {\bibfnamefont {A.}~\bibnamefont
  {Van~der Ven}}, \bibinfo {author} {\bibfnamefont {G.}~\bibnamefont {Ceder}},
  \bibinfo {author} {\bibfnamefont {M.}~\bibnamefont {Asta}}, \ and\ \bibinfo
  {author} {\bibfnamefont {P.~D.}\ \bibnamefont {Tepesch}},\ }\href {\doibase
  10.1103/PhysRevB.64.184307} {\bibfield  {journal} {\bibinfo  {journal}
  {Physical Review B}\ }\textbf {\bibinfo {volume} {64}},\ \bibinfo {pages}
  {184307} (\bibinfo {year} {2001})}\BibitemShut {NoStop}%
\bibitem [{\citenamefont {Kweon}\ \emph {et~al.}(2017)\citenamefont {Kweon},
  \citenamefont {Varley}, \citenamefont {Shea}, \citenamefont {Adelstein},
  \citenamefont {Mehta}, \citenamefont {Heo}, \citenamefont {Udovic},
  \citenamefont {Stavila},\ and\ \citenamefont {Wood}}]{kweon_structural_2017}%
  \BibitemOpen
  \bibfield  {author} {\bibinfo {author} {\bibfnamefont {K.~E.}\ \bibnamefont
  {Kweon}}, \bibinfo {author} {\bibfnamefont {J.~B.}\ \bibnamefont {Varley}},
  \bibinfo {author} {\bibfnamefont {P.}~\bibnamefont {Shea}}, \bibinfo {author}
  {\bibfnamefont {N.}~\bibnamefont {Adelstein}}, \bibinfo {author}
  {\bibfnamefont {P.}~\bibnamefont {Mehta}}, \bibinfo {author} {\bibfnamefont
  {T.~W.}\ \bibnamefont {Heo}}, \bibinfo {author} {\bibfnamefont {T.~J.}\
  \bibnamefont {Udovic}}, \bibinfo {author} {\bibfnamefont {V.}~\bibnamefont
  {Stavila}}, \ and\ \bibinfo {author} {\bibfnamefont {B.~C.}\ \bibnamefont
  {Wood}},\ }\href {\doibase 10.1021/acs.chemmater.7b02902} {\bibfield
  {journal} {\bibinfo  {journal} {Chemistry of Materials}\ }\textbf {\bibinfo
  {volume} {29}},\ \bibinfo {pages} {9142} (\bibinfo {year}
  {2017})}\BibitemShut {NoStop}%
\bibitem [{\citenamefont {de~Klerk}\ \emph {et~al.}(2018)\citenamefont
  {de~Klerk}, \citenamefont {van~der Maas},\ and\ \citenamefont
  {Wagemaker}}]{de_klerk_analysis_2018}%
  \BibitemOpen
  \bibfield  {author} {\bibinfo {author} {\bibfnamefont {N.~J.}\ \bibnamefont
  {de~Klerk}}, \bibinfo {author} {\bibfnamefont {E.}~\bibnamefont {van~der
  Maas}}, \ and\ \bibinfo {author} {\bibfnamefont {M.}~\bibnamefont
  {Wagemaker}},\ }\href {\doibase 10.1021/acsaem.8b00457} {\bibfield  {journal}
  {\bibinfo  {journal} {{ACS} Applied Energy Materials}\ }\textbf {\bibinfo
  {volume} {1}},\ \bibinfo {pages} {3230} (\bibinfo {year} {2018})}\BibitemShut
  {NoStop}%
\bibitem [{\citenamefont {Chen}\ \emph {et~al.}(2017)\citenamefont {Chen},
  \citenamefont {Lu},\ and\ \citenamefont {Ciucci}}]{chen_data_2017}%
  \BibitemOpen
  \bibfield  {author} {\bibinfo {author} {\bibfnamefont {C.}~\bibnamefont
  {Chen}}, \bibinfo {author} {\bibfnamefont {Z.}~\bibnamefont {Lu}}, \ and\
  \bibinfo {author} {\bibfnamefont {F.}~\bibnamefont {Ciucci}},\ }\href
  {\doibase 10.1038/srep40769} {\bibfield  {journal} {\bibinfo  {journal}
  {Scientific Reports}\ }\textbf {\bibinfo {volume} {7}},\ \bibinfo {pages}
  {40769} (\bibinfo {year} {2017})}\BibitemShut {NoStop}%
\bibitem [{\citenamefont {Varley}\ \emph {et~al.}(2017)\citenamefont {Varley},
  \citenamefont {Kweon}, \citenamefont {Mehta}, \citenamefont {Shea},
  \citenamefont {Heo}, \citenamefont {Udovic}, \citenamefont {Stavila},\ and\
  \citenamefont {Wood}}]{varley_understanding_2017}%
  \BibitemOpen
  \bibfield  {author} {\bibinfo {author} {\bibfnamefont {J.~B.}\ \bibnamefont
  {Varley}}, \bibinfo {author} {\bibfnamefont {K.}~\bibnamefont {Kweon}},
  \bibinfo {author} {\bibfnamefont {P.}~\bibnamefont {Mehta}}, \bibinfo
  {author} {\bibfnamefont {P.}~\bibnamefont {Shea}}, \bibinfo {author}
  {\bibfnamefont {T.~W.}\ \bibnamefont {Heo}}, \bibinfo {author} {\bibfnamefont
  {T.~J.}\ \bibnamefont {Udovic}}, \bibinfo {author} {\bibfnamefont
  {V.}~\bibnamefont {Stavila}}, \ and\ \bibinfo {author} {\bibfnamefont
  {B.~C.}\ \bibnamefont {Wood}},\ }\href {\doibase
  10.1021/acsenergylett.6b00620} {\bibfield  {journal} {\bibinfo  {journal}
  {ACS Energy Letters}\ }\textbf {\bibinfo {volume} {2}},\ \bibinfo {pages}
  {250} (\bibinfo {year} {2017})}\BibitemShut {NoStop}%
\bibitem [{\citenamefont {Deng}\ \emph {et~al.}(2018)\citenamefont {Deng},
  \citenamefont {Eames}, \citenamefont {Nguyen}, \citenamefont {Pecher},
  \citenamefont {Griffith}, \citenamefont {Courty}, \citenamefont {Fleutot},
  \citenamefont {Chotard}, \citenamefont {Grey}, \citenamefont {Islam},\ and\
  \citenamefont {Masquelier}}]{deng_crystal_2018}%
  \BibitemOpen
  \bibfield  {author} {\bibinfo {author} {\bibfnamefont {Y.}~\bibnamefont
  {Deng}}, \bibinfo {author} {\bibfnamefont {C.}~\bibnamefont {Eames}},
  \bibinfo {author} {\bibfnamefont {L.~H.~B.}\ \bibnamefont {Nguyen}}, \bibinfo
  {author} {\bibfnamefont {O.}~\bibnamefont {Pecher}}, \bibinfo {author}
  {\bibfnamefont {K.~J.}\ \bibnamefont {Griffith}}, \bibinfo {author}
  {\bibfnamefont {M.}~\bibnamefont {Courty}}, \bibinfo {author} {\bibfnamefont
  {B.}~\bibnamefont {Fleutot}}, \bibinfo {author} {\bibfnamefont {J.-N.}\
  \bibnamefont {Chotard}}, \bibinfo {author} {\bibfnamefont {C.~P.}\
  \bibnamefont {Grey}}, \bibinfo {author} {\bibfnamefont {M.~S.}\ \bibnamefont
  {Islam}}, \ and\ \bibinfo {author} {\bibfnamefont {C.}~\bibnamefont
  {Masquelier}},\ }\href {\doibase 10.1021/acs.chemmater.7b05237} {\bibfield
  {journal} {\bibinfo  {journal} {Chemistry of Materials}\ }\textbf {\bibinfo
  {volume} {30}},\ \bibinfo {pages} {2618} (\bibinfo {year}
  {2018})}\BibitemShut {NoStop}%
\bibitem [{\citenamefont {Kozinsky}(2018)}]{kozinsky_transport_2018}%
  \BibitemOpen
  \bibfield  {author} {\bibinfo {author} {\bibfnamefont {B.}~\bibnamefont
  {Kozinsky}},\ }in\ \href {\doibase 10.1007/978-3-319-50257-1_54-1} {\emph
  {\bibinfo {booktitle} {Handbook of {Materials} {Modeling}: {Applications}:
  {Current} and {Emerging} {Materials}}}},\ \bibinfo {editor} {edited by\
  \bibinfo {editor} {\bibfnamefont {W.}~\bibnamefont {Andreoni}}\ and\ \bibinfo
  {editor} {\bibfnamefont {S.}~\bibnamefont {Yip}}}\ (\bibinfo  {publisher}
  {Springer International Publishing},\ \bibinfo {address} {Cham},\ \bibinfo
  {year} {2018})\ pp.\ \bibinfo {pages} {1--20}\BibitemShut {NoStop}%
\bibitem [{\citenamefont {Richards}\ \emph {et~al.}(2016)\citenamefont
  {Richards}, \citenamefont {Tsujimura}, \citenamefont {Miara}, \citenamefont
  {Wang}, \citenamefont {Kim}, \citenamefont {Ong}, \citenamefont {Uechi},
  \citenamefont {Suzuki},\ and\ \citenamefont {Ceder}}]{richards_design_2016}%
  \BibitemOpen
  \bibfield  {author} {\bibinfo {author} {\bibfnamefont {W.~D.}\ \bibnamefont
  {Richards}}, \bibinfo {author} {\bibfnamefont {T.}~\bibnamefont {Tsujimura}},
  \bibinfo {author} {\bibfnamefont {L.~J.}\ \bibnamefont {Miara}}, \bibinfo
  {author} {\bibfnamefont {Y.}~\bibnamefont {Wang}}, \bibinfo {author}
  {\bibfnamefont {J.~C.}\ \bibnamefont {Kim}}, \bibinfo {author} {\bibfnamefont
  {S.~P.}\ \bibnamefont {Ong}}, \bibinfo {author} {\bibfnamefont
  {I.}~\bibnamefont {Uechi}}, \bibinfo {author} {\bibfnamefont
  {N.}~\bibnamefont {Suzuki}}, \ and\ \bibinfo {author} {\bibfnamefont
  {G.}~\bibnamefont {Ceder}},\ }\href {\doibase 10.1038/ncomms11009} {\bibfield
   {journal} {\bibinfo  {journal} {Nature Communications}\ }\textbf {\bibinfo
  {volume} {7}},\ \bibinfo {pages} {11009} (\bibinfo {year}
  {2016})}\BibitemShut {NoStop}%
\bibitem [{\citenamefont {Gasparotto}\ and\ \citenamefont
  {Ceriotti}(2014)}]{gasparotto_recognizing_2014}%
  \BibitemOpen
  \bibfield  {author} {\bibinfo {author} {\bibfnamefont {P.}~\bibnamefont
  {Gasparotto}}\ and\ \bibinfo {author} {\bibfnamefont {M.}~\bibnamefont
  {Ceriotti}},\ }\href {\doibase 10.1063/1.4900655} {\bibfield  {journal}
  {\bibinfo  {journal} {The Journal of Chemical Physics}\ }\textbf {\bibinfo
  {volume} {141}},\ \bibinfo {pages} {174110} (\bibinfo {year}
  {2014})}\BibitemShut {NoStop}%
\bibitem [{\citenamefont {Gasparotto}\ \emph {et~al.}(2018)\citenamefont
  {Gasparotto}, \citenamefont {Meissner},\ and\ \citenamefont
  {Ceriotti}}]{gasparotto_recognizing_2018}%
  \BibitemOpen
  \bibfield  {author} {\bibinfo {author} {\bibfnamefont {P.}~\bibnamefont
  {Gasparotto}}, \bibinfo {author} {\bibfnamefont {R.~H.}\ \bibnamefont
  {Meissner}}, \ and\ \bibinfo {author} {\bibfnamefont {M.}~\bibnamefont
  {Ceriotti}},\ }\href {\doibase 10.1021/acs.jctc.7b00993} {\bibfield
  {journal} {\bibinfo  {journal} {Journal of Chemical Theory and Computation}\
  }\textbf {\bibinfo {volume} {14}},\ \bibinfo {pages} {486} (\bibinfo {year}
  {2018})}\BibitemShut {NoStop}%
\bibitem [{\citenamefont {Wehner}\ \emph {et~al.}(1996)\citenamefont {Wehner},
  \citenamefont {Michel},\ and\ \citenamefont {Antonsen}}]{wehner_visual_1996}%
  \BibitemOpen
  \bibfield  {author} {\bibinfo {author} {\bibfnamefont {R.}~\bibnamefont
  {Wehner}}, \bibinfo {author} {\bibfnamefont {B.}~\bibnamefont {Michel}}, \
  and\ \bibinfo {author} {\bibfnamefont {P.}~\bibnamefont {Antonsen}},\ }\href
  {http://jeb.biologists.org/content/199/1/129} {\bibfield  {journal} {\bibinfo
   {journal} {Journal of Experimental Biology}\ }\textbf {\bibinfo {volume}
  {199}},\ \bibinfo {pages} {129} (\bibinfo {year} {1996})}\BibitemShut
  {NoStop}%
\bibitem [{\citenamefont {M\"oller}(2000)}]{moller_insect_2000}%
  \BibitemOpen
  \bibfield  {author} {\bibinfo {author} {\bibfnamefont {R.}~\bibnamefont
  {M\"oller}},\ }\href {\doibase 10.1007/PL00007973} {\bibfield  {journal}
  {\bibinfo  {journal} {Biological Cybernetics}\ }\textbf {\bibinfo {volume}
  {83}},\ \bibinfo {pages} {231} (\bibinfo {year} {2000})}\BibitemShut
  {NoStop}%
\bibitem [{\citenamefont {Prodan}\ and\ \citenamefont
  {Kohn}(2005)}]{prodan_nearsightedness_2005}%
  \BibitemOpen
  \bibfield  {author} {\bibinfo {author} {\bibfnamefont {E.}~\bibnamefont
  {Prodan}}\ and\ \bibinfo {author} {\bibfnamefont {W.}~\bibnamefont {Kohn}},\
  }\href {\doibase 10.1073/pnas.0505436102} {\bibfield  {journal} {\bibinfo
  {journal} {Proceedings of the National Academy of Sciences of the United
  States of America}\ }\textbf {\bibinfo {volume} {102}},\ \bibinfo {pages}
  {11635} (\bibinfo {year} {2005})}\BibitemShut {NoStop}%
\bibitem [{\citenamefont {Pauling}(1929)}]{pauling_principles_1929}%
  \BibitemOpen
  \bibfield  {author} {\bibinfo {author} {\bibfnamefont {L.}~\bibnamefont
  {Pauling}},\ }\href {\doibase 10.1021/ja01379a006} {\bibfield  {journal}
  {\bibinfo  {journal} {Journal of the American Chemical Society}\ }\textbf
  {\bibinfo {volume} {51}},\ \bibinfo {pages} {1010} (\bibinfo {year}
  {1929})}\BibitemShut {NoStop}%
\bibitem [{\citenamefont {Molinari}\ \emph {et~al.}(2018)\citenamefont
  {Molinari}, \citenamefont {Mailoa},\ and\ \citenamefont
  {Kozinsky}}]{molinari_effect_2018}%
  \BibitemOpen
  \bibfield  {author} {\bibinfo {author} {\bibfnamefont {N.}~\bibnamefont
  {Molinari}}, \bibinfo {author} {\bibfnamefont {J.}~\bibnamefont {Mailoa}}, \
  and\ \bibinfo {author} {\bibfnamefont {B.}~\bibnamefont {Kozinsky}},\ }\href
  {\doibase 10.1021/acs.chemmater.8b01955} {\bibfield  {journal} {\bibinfo
  {journal} {Chem. Mater.}\ }\textbf {\bibinfo {volume} {30}},\ \bibinfo
  {pages} {6298} (\bibinfo {year} {2018})}\BibitemShut {NoStop}%
\bibitem [{\citenamefont {Wood}\ and\ \citenamefont
  {Marzari}(2007)}]{wood_proton_2007}%
  \BibitemOpen
  \bibfield  {author} {\bibinfo {author} {\bibfnamefont {B.~C.}\ \bibnamefont
  {Wood}}\ and\ \bibinfo {author} {\bibfnamefont {N.}~\bibnamefont {Marzari}},\
  }\href {\doibase 10.1103/PhysRevB.76.134301} {\bibfield  {journal} {\bibinfo
  {journal} {Physical Review B}\ }\textbf {\bibinfo {volume} {76}},\ \bibinfo
  {pages} {134301} (\bibinfo {year} {2007})}\BibitemShut {NoStop}%
\bibitem [{\citenamefont {Morgan}\ and\ \citenamefont
  {Madden}(2014)}]{morgan_relationships_2014}%
  \BibitemOpen
  \bibfield  {author} {\bibinfo {author} {\bibfnamefont {B.~J.}\ \bibnamefont
  {Morgan}}\ and\ \bibinfo {author} {\bibfnamefont {P.~A.}\ \bibnamefont
  {Madden}},\ }\href {\doibase 10.1103/PhysRevLett.112.145901} {\bibfield
  {journal} {\bibinfo  {journal} {Physical Review Letters}\ }\textbf {\bibinfo
  {volume} {112}},\ \bibinfo {pages} {145901} (\bibinfo {year}
  {2014})}\BibitemShut {NoStop}%
\bibitem [{\citenamefont {Okabe}\ \emph {et~al.}(2009)\citenamefont {Okabe},
  \citenamefont {Boots}, \citenamefont {Sugihara},\ and\ \citenamefont
  {Chiu}}]{okabe_spatial_2009}%
  \BibitemOpen
  \bibfield  {author} {\bibinfo {author} {\bibfnamefont {A.}~\bibnamefont
  {Okabe}}, \bibinfo {author} {\bibfnamefont {B.}~\bibnamefont {Boots}},
  \bibinfo {author} {\bibfnamefont {K.}~\bibnamefont {Sugihara}}, \ and\
  \bibinfo {author} {\bibfnamefont {S.~N.}\ \bibnamefont {Chiu}},\ }\href@noop
  {} {\emph {\bibinfo {title} {Spatial {Tessellations}: {Concepts} and
  {Applications} of {Voronoi} {Diagrams}}}}\ (\bibinfo  {publisher} {John Wiley
  \& Sons, Chichester, UK},\ \bibinfo {year} {2009})\BibitemShut {NoStop}%
\bibitem [{Note1()}]{Note1}%
  \BibitemOpen
  \bibinfo {note} {A simplex in $\protect \mathbb {R}^D$ is the convex hull of
  $D+1$ points that do not lie on a hyperplane.}\BibitemShut {Stop}%
\bibitem [{\citenamefont {Van~Dongen}(2008)}]{van_dongen_MCL_2008}%
  \BibitemOpen
  \bibfield  {author} {\bibinfo {author} {\bibfnamefont {S.}~\bibnamefont
  {Van~Dongen}},\ }\href {\doibase 10.1137/040608635} {\bibfield  {journal}
  {\bibinfo  {journal} {SIAM Journal on Matrix Analysis and Applications}\
  }\textbf {\bibinfo {volume} {30}},\ \bibinfo {pages} {121} (\bibinfo {year}
  {2008})}\BibitemShut {NoStop}%
\bibitem [{\citenamefont {Bart\'ok}\ \emph {et~al.}(2013)\citenamefont
  {Bart\'ok}, \citenamefont {Kondor},\ and\ \citenamefont
  {Cs\'anyi}}]{bartok_representing_2013}%
  \BibitemOpen
  \bibfield  {author} {\bibinfo {author} {\bibfnamefont {A.~P.}\ \bibnamefont
  {Bart\'ok}}, \bibinfo {author} {\bibfnamefont {R.}~\bibnamefont {Kondor}}, \
  and\ \bibinfo {author} {\bibfnamefont {G.}~\bibnamefont {Cs\'anyi}},\ }\href
  {\doibase 10.1103/PhysRevB.87.184115} {\bibfield  {journal} {\bibinfo
  {journal} {Physical Review B}\ }\textbf {\bibinfo {volume} {87}},\ \bibinfo
  {pages} {184115} (\bibinfo {year} {2013})}\BibitemShut {NoStop}%
\bibitem [{noa(2018)}]{noauthor_libatoms/quip_2018}%
  \BibitemOpen
  \href {https://github.com/libAtoms/QUIP} {\enquote {\bibinfo {title}
  {{libAtoms}/{QUIP} molecular dynamics framework: http://www.libatoms.org:
  {libAtoms}/{QUIP}},}\ } (\bibinfo {year} {2018}),\ \bibinfo {note}
  {original-date: 2013-07-02T15:21:59Z}\BibitemShut {NoStop}%
\bibitem [{\citenamefont {De}\ \emph {et~al.}(2016)\citenamefont {De},
  \citenamefont {P.~Bart\'ok}, \citenamefont {Cs\'anyi},\ and\ \citenamefont
  {Ceriotti}}]{de_comparing_2016}%
  \BibitemOpen
  \bibfield  {author} {\bibinfo {author} {\bibfnamefont {S.}~\bibnamefont
  {De}}, \bibinfo {author} {\bibfnamefont {A.}~\bibnamefont {P.~Bart\'ok}},
  \bibinfo {author} {\bibfnamefont {G.}~\bibnamefont {Cs\'anyi}}, \ and\
  \bibinfo {author} {\bibfnamefont {M.}~\bibnamefont {Ceriotti}},\ }\href
  {\doibase 10.1039/C6CP00415F} {\bibfield  {journal} {\bibinfo  {journal}
  {Physical Chemistry Chemical Physics}\ }\textbf {\bibinfo {volume} {18}},\
  \bibinfo {pages} {13754} (\bibinfo {year} {2016})}\BibitemShut {NoStop}%
\bibitem [{\citenamefont {Rodriguez}\ and\ \citenamefont
  {Laio}(2014)}]{rodriguez_clustering_2014}%
  \BibitemOpen
  \bibfield  {author} {\bibinfo {author} {\bibfnamefont {A.}~\bibnamefont
  {Rodriguez}}\ and\ \bibinfo {author} {\bibfnamefont {A.}~\bibnamefont
  {Laio}},\ }\href {\doibase 10.1126/science.1242072} {\bibfield  {journal}
  {\bibinfo  {journal} {Science}\ }\textbf {\bibinfo {volume} {344}},\ \bibinfo
  {pages} {1492} (\bibinfo {year} {2014})}\BibitemShut {NoStop}%
\bibitem [{\citenamefont {Allen}\ and\ \citenamefont
  {Tildesley}(1987)}]{allen_computer_1987}%
  \BibitemOpen
  \bibfield  {author} {\bibinfo {author} {\bibfnamefont {M.~P.}\ \bibnamefont
  {Allen}}\ and\ \bibinfo {author} {\bibfnamefont {D.~J.}\ \bibnamefont
  {Tildesley}},\ }\href@noop {} {\emph {\bibinfo {title} {Computer simulation
  of liquids}}}\ (\bibinfo  {publisher} {Clarendon Press, Oxford},\ \bibinfo
  {year} {1987})\BibitemShut {NoStop}%
\bibitem [{\citenamefont {Frenkel}\ and\ \citenamefont
  {Smit}(1996)}]{frenkel_understanding_1996}%
  \BibitemOpen
  \bibfield  {author} {\bibinfo {author} {\bibfnamefont {D.}~\bibnamefont
  {Frenkel}}\ and\ \bibinfo {author} {\bibfnamefont {B.}~\bibnamefont {Smit}},\
  }\href@noop {} {\emph {\bibinfo {title} {Understanding Molecular Simulation:
  From Algorithms to Applications}}}\ (\bibinfo  {publisher} {Academic Press},\
  \bibinfo {address} {San Diego},\ \bibinfo {year} {1996})\BibitemShut
  {NoStop}%
\bibitem [{\citenamefont {Thangadurai}\ \emph {et~al.}(2003)\citenamefont
  {Thangadurai}, \citenamefont {Kaack},\ and\ \citenamefont
  {Weppner}}]{thangadurai_novel_2003}%
  \BibitemOpen
  \bibfield  {author} {\bibinfo {author} {\bibfnamefont {V.}~\bibnamefont
  {Thangadurai}}, \bibinfo {author} {\bibfnamefont {H.}~\bibnamefont {Kaack}},
  \ and\ \bibinfo {author} {\bibfnamefont {W.~J.~F.}\ \bibnamefont {Weppner}},\
  }\href {\doibase 10.1111/j.1151-2916.2003.tb03318.x} {\bibfield  {journal}
  {\bibinfo  {journal} {Journal of the American Ceramic Society}\ }\textbf
  {\bibinfo {volume} {86}},\ \bibinfo {pages} {437} (\bibinfo {year}
  {2003})}\BibitemShut {NoStop}%
\bibitem [{\citenamefont {Knauth}(2009)}]{knauth_inorganic_2009}%
  \BibitemOpen
  \bibfield  {author} {\bibinfo {author} {\bibfnamefont {P.}~\bibnamefont
  {Knauth}},\ }\href {\doibase 10.1016/j.ssi.2009.03.022} {\bibfield  {journal}
  {\bibinfo  {journal} {Solid State Ionics}\ }\textbf {\bibinfo {volume}
  {180}},\ \bibinfo {pages} {911} (\bibinfo {year} {2009})}\BibitemShut
  {NoStop}%
\bibitem [{\citenamefont {Xie}\ \emph {et~al.}(2011)\citenamefont {Xie},
  \citenamefont {Alonso}, \citenamefont {Li}, \citenamefont
  {Fern\'andez-D\'iaz},\ and\ \citenamefont {Goodenough}}]{xie_lithium_2011}%
  \BibitemOpen
  \bibfield  {author} {\bibinfo {author} {\bibfnamefont {H.}~\bibnamefont
  {Xie}}, \bibinfo {author} {\bibfnamefont {J.~A.}\ \bibnamefont {Alonso}},
  \bibinfo {author} {\bibfnamefont {Y.}~\bibnamefont {Li}}, \bibinfo {author}
  {\bibfnamefont {M.~T.}\ \bibnamefont {Fern\'andez-D\'iaz}}, \ and\ \bibinfo
  {author} {\bibfnamefont {J.~B.}\ \bibnamefont {Goodenough}},\ }\href
  {\doibase 10.1021/cm201671k} {\bibfield  {journal} {\bibinfo  {journal}
  {Chemistry of Materials}\ }\textbf {\bibinfo {volume} {23}},\ \bibinfo
  {pages} {3587} (\bibinfo {year} {2011})}\BibitemShut {NoStop}%
\bibitem [{\citenamefont {O'Callaghan}\ \emph {et~al.}(2008)\citenamefont
  {O'Callaghan}, \citenamefont {Powell}, \citenamefont {Titman}, \citenamefont
  {Chen},\ and\ \citenamefont {Cussen}}]{ocallaghan_switching_2008}%
  \BibitemOpen
  \bibfield  {author} {\bibinfo {author} {\bibfnamefont {M.~P.}\ \bibnamefont
  {O'Callaghan}}, \bibinfo {author} {\bibfnamefont {A.~S.}\ \bibnamefont
  {Powell}}, \bibinfo {author} {\bibfnamefont {J.~J.}\ \bibnamefont {Titman}},
  \bibinfo {author} {\bibfnamefont {G.~Z.}\ \bibnamefont {Chen}}, \ and\
  \bibinfo {author} {\bibfnamefont {E.~J.}\ \bibnamefont {Cussen}},\ }\href
  {\doibase 10.1021/cm703677q} {\bibfield  {journal} {\bibinfo  {journal}
  {Chemistry of Materials}\ }\textbf {\bibinfo {volume} {20}},\ \bibinfo
  {pages} {2360} (\bibinfo {year} {2008})}\BibitemShut {NoStop}%
\bibitem [{\citenamefont {O'Callaghan}\ and\ \citenamefont
  {Cussen}(2007)}]{ocallaghan_lithium_2007}%
  \BibitemOpen
  \bibfield  {author} {\bibinfo {author} {\bibfnamefont {M.~P.}\ \bibnamefont
  {O'Callaghan}}\ and\ \bibinfo {author} {\bibfnamefont {E.~J.}\ \bibnamefont
  {Cussen}},\ }\href {\doibase 10.1039/B700369B} {\bibfield  {journal}
  {\bibinfo  {journal} {Chemical Communications}\ }\textbf {\bibinfo {volume}
  {0}},\ \bibinfo {pages} {2048} (\bibinfo {year} {2007})}\BibitemShut
  {NoStop}%
\bibitem [{\citenamefont {Thangadurai}\ \emph {et~al.}(2014)\citenamefont
  {Thangadurai}, \citenamefont {Narayanan},\ and\ \citenamefont
  {Pinzaru}}]{thangadurai_garnet-type_2014}%
  \BibitemOpen
  \bibfield  {author} {\bibinfo {author} {\bibfnamefont {V.}~\bibnamefont
  {Thangadurai}}, \bibinfo {author} {\bibfnamefont {S.}~\bibnamefont
  {Narayanan}}, \ and\ \bibinfo {author} {\bibfnamefont {D.}~\bibnamefont
  {Pinzaru}},\ }\href {\doibase 10.1039/C4CS00020J} {\bibfield  {journal}
  {\bibinfo  {journal} {Chemical Society Reviews}\ }\textbf {\bibinfo {volume}
  {43}},\ \bibinfo {pages} {4714} (\bibinfo {year} {2014})}\BibitemShut
  {NoStop}%
\bibitem [{\citenamefont {Ceriotti}\ \emph {et~al.}(2009)\citenamefont
  {Ceriotti}, \citenamefont {Bussi},\ and\ \citenamefont
  {Parrinello}}]{ceriotti_langevin_2009}%
  \BibitemOpen
  \bibfield  {author} {\bibinfo {author} {\bibfnamefont {M.}~\bibnamefont
  {Ceriotti}}, \bibinfo {author} {\bibfnamefont {G.}~\bibnamefont {Bussi}}, \
  and\ \bibinfo {author} {\bibfnamefont {M.}~\bibnamefont {Parrinello}},\
  }\href {\doibase 10.1103/PhysRevLett.102.020601} {\bibfield  {journal}
  {\bibinfo  {journal} {Physical Review Letters}\ }\textbf {\bibinfo {volume}
  {102}},\ \bibinfo {pages} {020601} (\bibinfo {year} {2009})}\BibitemShut
  {NoStop}%
\bibitem [{\citenamefont {Plimpton}(1995)}]{plimpton_fast_1995}%
  \BibitemOpen
  \bibfield  {author} {\bibinfo {author} {\bibfnamefont {S.}~\bibnamefont
  {Plimpton}},\ }\href {\doibase 10.1006/jcph.1995.1039} {\bibfield  {journal}
  {\bibinfo  {journal} {Journal of Computational Physics}\ }\textbf {\bibinfo
  {volume} {117}},\ \bibinfo {pages} {1} (\bibinfo {year} {1995})}\BibitemShut
  {NoStop}%
\bibitem [{\citenamefont {Mottet}\ \emph {et~al.}(2019)\citenamefont {Mottet},
  \citenamefont {Marcolongo}, \citenamefont {Laino},\ and\ \citenamefont
  {Tavernelli}}]{mottet_doping_2019}%
  \BibitemOpen
  \bibfield  {author} {\bibinfo {author} {\bibfnamefont {M.}~\bibnamefont
  {Mottet}}, \bibinfo {author} {\bibfnamefont {A.}~\bibnamefont {Marcolongo}},
  \bibinfo {author} {\bibfnamefont {T.}~\bibnamefont {Laino}}, \ and\ \bibinfo
  {author} {\bibfnamefont {I.}~\bibnamefont {Tavernelli}},\ }\href {\doibase
  10.1103/PhysRevMaterials.3.035403} {\bibfield  {journal} {\bibinfo  {journal}
  {Physical Review Materials}\ }\textbf {\bibinfo {volume} {3}},\ \bibinfo
  {pages} {035403} (\bibinfo {year} {2019})}\BibitemShut {NoStop}%
\bibitem [{\citenamefont {Jalem}\ \emph {et~al.}(2013)\citenamefont {Jalem},
  \citenamefont {Yamamoto}, \citenamefont {Shiiba}, \citenamefont {Nakayama},
  \citenamefont {Munakata}, \citenamefont {Kasuga},\ and\ \citenamefont
  {Kanamura}}]{jalem_concerted_2013}%
  \BibitemOpen
  \bibfield  {author} {\bibinfo {author} {\bibfnamefont {R.}~\bibnamefont
  {Jalem}}, \bibinfo {author} {\bibfnamefont {Y.}~\bibnamefont {Yamamoto}},
  \bibinfo {author} {\bibfnamefont {H.}~\bibnamefont {Shiiba}}, \bibinfo
  {author} {\bibfnamefont {M.}~\bibnamefont {Nakayama}}, \bibinfo {author}
  {\bibfnamefont {H.}~\bibnamefont {Munakata}}, \bibinfo {author}
  {\bibfnamefont {T.}~\bibnamefont {Kasuga}}, \ and\ \bibinfo {author}
  {\bibfnamefont {K.}~\bibnamefont {Kanamura}},\ }\href {\doibase
  10.1021/cm303542x} {\bibfield  {journal} {\bibinfo  {journal} {Chemistry of
  Materials}\ }\textbf {\bibinfo {volume} {25}},\ \bibinfo {pages} {425}
  (\bibinfo {year} {2013})}\BibitemShut {NoStop}%
\bibitem [{\citenamefont {{Morgan Benjamin
  J.}}(2017)}]{morgan_benjamin_j._lattice-geometry_2017}%
  \BibitemOpen
  \bibfield  {author} {\bibinfo {author} {\bibnamefont {{Morgan Benjamin
  J.}}},\ }\href {\doibase 10.1098/rsos.170824} {\bibfield  {journal} {\bibinfo
   {journal} {Royal Society Open Science}\ }\textbf {\bibinfo {volume} {4}},\
  \bibinfo {pages} {170824} (\bibinfo {year} {2017})}\BibitemShut {NoStop}%
\bibitem [{\citenamefont {Murugan}\ \emph {et~al.}(2007)\citenamefont
  {Murugan}, \citenamefont {Thangadurai},\ and\ \citenamefont
  {Weppner}}]{murugan_fast_2007}%
  \BibitemOpen
  \bibfield  {author} {\bibinfo {author} {\bibfnamefont {R.}~\bibnamefont
  {Murugan}}, \bibinfo {author} {\bibfnamefont {V.}~\bibnamefont
  {Thangadurai}}, \ and\ \bibinfo {author} {\bibfnamefont {W.}~\bibnamefont
  {Weppner}},\ }\href {\doibase 10.1002/anie.200701144} {\bibfield  {journal}
  {\bibinfo  {journal} {Angewandte Chemie International Edition}\ }\textbf
  {\bibinfo {volume} {46}},\ \bibinfo {pages} {7778} (\bibinfo {year}
  {2007})}\BibitemShut {NoStop}%
\bibitem [{\citenamefont {Miara}\ \emph {et~al.}(2013)\citenamefont {Miara},
  \citenamefont {Ong}, \citenamefont {Mo}, \citenamefont {Richards},
  \citenamefont {Park}, \citenamefont {Lee}, \citenamefont {Lee},\ and\
  \citenamefont {Ceder}}]{miara_effect_2013}%
  \BibitemOpen
  \bibfield  {author} {\bibinfo {author} {\bibfnamefont {L.~J.}\ \bibnamefont
  {Miara}}, \bibinfo {author} {\bibfnamefont {S.~P.}\ \bibnamefont {Ong}},
  \bibinfo {author} {\bibfnamefont {Y.}~\bibnamefont {Mo}}, \bibinfo {author}
  {\bibfnamefont {W.~D.}\ \bibnamefont {Richards}}, \bibinfo {author}
  {\bibfnamefont {Y.}~\bibnamefont {Park}}, \bibinfo {author} {\bibfnamefont
  {J.-M.}\ \bibnamefont {Lee}}, \bibinfo {author} {\bibfnamefont {H.~S.}\
  \bibnamefont {Lee}}, \ and\ \bibinfo {author} {\bibfnamefont
  {G.}~\bibnamefont {Ceder}},\ }\href {\doibase 10.1021/cm401232r} {\bibfield
  {journal} {\bibinfo  {journal} {Chemistry of Materials}\ }\textbf {\bibinfo
  {volume} {25}},\ \bibinfo {pages} {3048} (\bibinfo {year}
  {2013})}\BibitemShut {NoStop}%
\bibitem [{\citenamefont {{William W. Pillars}}\ and\ \citenamefont {{Donald R.
  Peacor}}(1973)}]{william_w._pillars_crystal_1973}%
  \BibitemOpen
  \bibfield  {author} {\bibinfo {author} {\bibnamefont {{William W. Pillars}}}\
  and\ \bibinfo {author} {\bibnamefont {{Donald R. Peacor}}},\ }\href
  {https://pubs.geoscienceworld.org/msa/ammin/article-abstract/58/7-8/681/542853}
  {\bibfield  {journal} {\bibinfo  {journal} {American Mineralogist}\ }\textbf
  {\bibinfo {volume} {58}},\ \bibinfo {pages} {681} (\bibinfo {year}
  {1973})}\BibitemShut {NoStop}%
\bibitem [{\citenamefont {Gra\v{z}ulis}\ \emph {et~al.}(2012)\citenamefont
  {Gra\v{z}ulis}, \citenamefont {Da\v{s}kevi\v{c}}, \citenamefont {Merkys},
  \citenamefont {Chateigner}, \citenamefont {Lutterotti}, \citenamefont
  {Quir\'os}, \citenamefont {Serebryanaya}, \citenamefont {Moeck},
  \citenamefont {Downs},\ and\ \citenamefont
  {Le~Bail}}]{grazulis_crystallography_2012}%
  \BibitemOpen
  \bibfield  {author} {\bibinfo {author} {\bibfnamefont {S.}~\bibnamefont
  {Gra\v{z}ulis}}, \bibinfo {author} {\bibfnamefont {A.}~\bibnamefont
  {Da\v{s}kevi\v{c}}}, \bibinfo {author} {\bibfnamefont {A.}~\bibnamefont
  {Merkys}}, \bibinfo {author} {\bibfnamefont {D.}~\bibnamefont {Chateigner}},
  \bibinfo {author} {\bibfnamefont {L.}~\bibnamefont {Lutterotti}}, \bibinfo
  {author} {\bibfnamefont {M.}~\bibnamefont {Quir\'os}}, \bibinfo {author}
  {\bibfnamefont {N.~R.}\ \bibnamefont {Serebryanaya}}, \bibinfo {author}
  {\bibfnamefont {P.}~\bibnamefont {Moeck}}, \bibinfo {author} {\bibfnamefont
  {R.~T.}\ \bibnamefont {Downs}}, \ and\ \bibinfo {author} {\bibfnamefont
  {A.}~\bibnamefont {Le~Bail}},\ }\href {\doibase 10.1093/nar/gkr900}
  {\bibfield  {journal} {\bibinfo  {journal} {Nucleic Acids Research}\ }\textbf
  {\bibinfo {volume} {40}},\ \bibinfo {pages} {D420} (\bibinfo {year}
  {2012})}\BibitemShut {NoStop}%
\bibitem [{\citenamefont {Lichtenstein}\ \emph {et~al.}(1998)\citenamefont
  {Lichtenstein}, \citenamefont {Jones}, \citenamefont {Xu},\ and\
  \citenamefont {Heaney}}]{lichtenstein_anisotropic_1998}%
  \BibitemOpen
  \bibfield  {author} {\bibinfo {author} {\bibfnamefont {A.~I.}\ \bibnamefont
  {Lichtenstein}}, \bibinfo {author} {\bibfnamefont {R.~O.}\ \bibnamefont
  {Jones}}, \bibinfo {author} {\bibfnamefont {H.}~\bibnamefont {Xu}}, \ and\
  \bibinfo {author} {\bibfnamefont {P.~J.}\ \bibnamefont {Heaney}},\ }\href
  {\doibase 10.1103/PhysRevB.58.6219} {\bibfield  {journal} {\bibinfo
  {journal} {Physical Review B}\ }\textbf {\bibinfo {volume} {58}},\ \bibinfo
  {pages} {6219} (\bibinfo {year} {1998})}\BibitemShut {NoStop}%
\bibitem [{\citenamefont {Lichtenstein}\ \emph {et~al.}(2000)\citenamefont
  {Lichtenstein}, \citenamefont {Jones}, \citenamefont {de~Gironcoli},\ and\
  \citenamefont {Baroni}}]{lichtenstein_anisotropic_2000}%
  \BibitemOpen
  \bibfield  {author} {\bibinfo {author} {\bibfnamefont {A.~I.}\ \bibnamefont
  {Lichtenstein}}, \bibinfo {author} {\bibfnamefont {R.~O.}\ \bibnamefont
  {Jones}}, \bibinfo {author} {\bibfnamefont {S.}~\bibnamefont {de~Gironcoli}},
  \ and\ \bibinfo {author} {\bibfnamefont {S.}~\bibnamefont {Baroni}},\ }\href
  {\doibase 10.1103/PhysRevB.62.11487} {\bibfield  {journal} {\bibinfo
  {journal} {Physical Review B}\ }\textbf {\bibinfo {volume} {62}},\ \bibinfo
  {pages} {11487} (\bibinfo {year} {2000})}\BibitemShut {NoStop}%
\bibitem [{\citenamefont {Schink}\ and\ \citenamefont
  {L\"ohneysen}(1983)}]{schink_vacancy-induced_1983}%
  \BibitemOpen
  \bibfield  {author} {\bibinfo {author} {\bibfnamefont {H.~J.}\ \bibnamefont
  {Schink}}\ and\ \bibinfo {author} {\bibfnamefont {H.~v.}\ \bibnamefont
  {L\"ohneysen}},\ }\href {\doibase 10.1016/0038-1098(83)90624-5} {\bibfield
  {journal} {\bibinfo  {journal} {Solid State Communications}\ }\textbf
  {\bibinfo {volume} {47}},\ \bibinfo {pages} {131} (\bibinfo {year}
  {1983})}\BibitemShut {NoStop}%
\bibitem [{\citenamefont {Donduft}\ \emph {et~al.}(1988)\citenamefont
  {Donduft}, \citenamefont {Dimitrijevi\'c},\ and\ \citenamefont
  {Petranovi\'c}}]{donduft_li+_1988}%
  \BibitemOpen
  \bibfield  {author} {\bibinfo {author} {\bibfnamefont {V.}~\bibnamefont
  {Donduft}}, \bibinfo {author} {\bibfnamefont {R.}~\bibnamefont
  {Dimitrijevi\'c}}, \ and\ \bibinfo {author} {\bibfnamefont {N.}~\bibnamefont
  {Petranovi\'c}},\ }\href {\doibase 10.1007/BF01106839} {\bibfield  {journal}
  {\bibinfo  {journal} {Journal of Materials Science}\ }\textbf {\bibinfo
  {volume} {23}},\ \bibinfo {pages} {4081} (\bibinfo {year}
  {1988})}\BibitemShut {NoStop}%
\bibitem [{\citenamefont {Sartbaeva}\ \emph {et~al.}(2005)\citenamefont
  {Sartbaeva}, \citenamefont {Wells}, \citenamefont {Redfern}, \citenamefont
  {Hinton},\ and\ \citenamefont {Reed}}]{sartbaeva_ionic_2005}%
  \BibitemOpen
  \bibfield  {author} {\bibinfo {author} {\bibfnamefont {A.}~\bibnamefont
  {Sartbaeva}}, \bibinfo {author} {\bibfnamefont {S.~A.}\ \bibnamefont
  {Wells}}, \bibinfo {author} {\bibfnamefont {S.~A.~T.}\ \bibnamefont
  {Redfern}}, \bibinfo {author} {\bibfnamefont {R.~W.}\ \bibnamefont {Hinton}},
  \ and\ \bibinfo {author} {\bibfnamefont {S.~J.~B.}\ \bibnamefont {Reed}},\
  }\href {\doibase 10.1088/0953-8984/17/7/004} {\bibfield  {journal} {\bibinfo
  {journal} {Journal of Physics: Condensed Matter}\ }\textbf {\bibinfo {volume}
  {17}},\ \bibinfo {pages} {1099} (\bibinfo {year} {2005})}\BibitemShut
  {NoStop}%
\bibitem [{\citenamefont {Chen}\ \emph {et~al.}(2018)\citenamefont {Chen},
  \citenamefont {Manna}, \citenamefont {Ciobanu},\ and\ \citenamefont
  {Reimanis}}]{chen_thermal_2018}%
  \BibitemOpen
  \bibfield  {author} {\bibinfo {author} {\bibfnamefont {Y.}~\bibnamefont
  {Chen}}, \bibinfo {author} {\bibfnamefont {S.}~\bibnamefont {Manna}},
  \bibinfo {author} {\bibfnamefont {C.~V.}\ \bibnamefont {Ciobanu}}, \ and\
  \bibinfo {author} {\bibfnamefont {I.~E.}\ \bibnamefont {Reimanis}},\ }\href
  {\doibase 10.1111/jace.15173} {\bibfield  {journal} {\bibinfo  {journal}
  {Journal of the American Ceramic Society}\ }\textbf {\bibinfo {volume}
  {101}},\ \bibinfo {pages} {347} (\bibinfo {year} {2018})}\BibitemShut
  {NoStop}%
\bibitem [{\citenamefont {Winkler}(1948)}]{winkler_synthese_1948}%
  \BibitemOpen
  \bibfield  {author} {\bibinfo {author} {\bibfnamefont {H.~G.~F.}\
  \bibnamefont {Winkler}},\ }\href {\doibase 10.1107/S0365110X48000065}
  {\bibfield  {journal} {\bibinfo  {journal} {Acta Crystallographica}\ }\textbf
  {\bibinfo {volume} {1}},\ \bibinfo {pages} {27} (\bibinfo {year}
  {1948})}\BibitemShut {NoStop}%
\bibitem [{kah()}]{kahle_supplementary_2019}%
  \BibitemOpen
  \href@noop {} {}\bibinfo {note} {See Supplemental Material at
  \url{http://link.aps.org/supplemental/ 10.1103/PhysRevMaterials.3.055404} for
  a description of the location of additional data and analysis scripts
  (Sec.~S~I), and additional plots on mean-square displacements, particle
  trajectories and SOAP descriptor clusters (Sec.~S~II)}\BibitemShut {NoStop}%
\bibitem [{\citenamefont {Press}\ \emph {et~al.}(1980)\citenamefont {Press},
  \citenamefont {Renker}, \citenamefont {Schulz},\ and\ \citenamefont
  {B\"ohm}}]{press_neutron_1980}%
  \BibitemOpen
  \bibfield  {author} {\bibinfo {author} {\bibfnamefont {W.}~\bibnamefont
  {Press}}, \bibinfo {author} {\bibfnamefont {B.}~\bibnamefont {Renker}},
  \bibinfo {author} {\bibfnamefont {H.}~\bibnamefont {Schulz}}, \ and\ \bibinfo
  {author} {\bibfnamefont {H.}~\bibnamefont {B\"ohm}},\ }\href {\doibase
  10.1103/PhysRevB.21.1250} {\bibfield  {journal} {\bibinfo  {journal}
  {Physical Review B}\ }\textbf {\bibinfo {volume} {21}},\ \bibinfo {pages}
  {1250} (\bibinfo {year} {1980})}\BibitemShut {NoStop}%
\bibitem [{\citenamefont {Kamaya}\ \emph {et~al.}(2011)\citenamefont {Kamaya},
  \citenamefont {Homma}, \citenamefont {Yamakawa}, \citenamefont {Hirayama},
  \citenamefont {Kanno}, \citenamefont {Yonemura}, \citenamefont {Kamiyama},
  \citenamefont {Kato}, \citenamefont {Hama}, \citenamefont {Kawamoto},\ and\
  \citenamefont {Mitsui}}]{kamaya_lithium_2011}%
  \BibitemOpen
  \bibfield  {author} {\bibinfo {author} {\bibfnamefont {N.}~\bibnamefont
  {Kamaya}}, \bibinfo {author} {\bibfnamefont {K.}~\bibnamefont {Homma}},
  \bibinfo {author} {\bibfnamefont {Y.}~\bibnamefont {Yamakawa}}, \bibinfo
  {author} {\bibfnamefont {M.}~\bibnamefont {Hirayama}}, \bibinfo {author}
  {\bibfnamefont {R.}~\bibnamefont {Kanno}}, \bibinfo {author} {\bibfnamefont
  {M.}~\bibnamefont {Yonemura}}, \bibinfo {author} {\bibfnamefont
  {T.}~\bibnamefont {Kamiyama}}, \bibinfo {author} {\bibfnamefont
  {Y.}~\bibnamefont {Kato}}, \bibinfo {author} {\bibfnamefont {S.}~\bibnamefont
  {Hama}}, \bibinfo {author} {\bibfnamefont {K.}~\bibnamefont {Kawamoto}}, \
  and\ \bibinfo {author} {\bibfnamefont {A.}~\bibnamefont {Mitsui}},\ }\href
  {\doibase 10.1038/nmat3066} {\bibfield  {journal} {\bibinfo  {journal}
  {Nature Materials}\ }\textbf {\bibinfo {volume} {10}},\ \bibinfo {pages}
  {682} (\bibinfo {year} {2011})}\BibitemShut {NoStop}%
\bibitem [{\citenamefont {Kuhn}\ \emph
  {et~al.}(2013{\natexlab{a}})\citenamefont {Kuhn}, \citenamefont {K\"ohler},\
  and\ \citenamefont {Lotsch}}]{kuhn_single-crystal_2013}%
  \BibitemOpen
  \bibfield  {author} {\bibinfo {author} {\bibfnamefont {A.}~\bibnamefont
  {Kuhn}}, \bibinfo {author} {\bibfnamefont {J.}~\bibnamefont {K\"ohler}}, \
  and\ \bibinfo {author} {\bibfnamefont {B.~V.}\ \bibnamefont {Lotsch}},\
  }\href {\doibase 10.1039/c3cp51985f} {\bibfield  {journal} {\bibinfo
  {journal} {Physical Chemistry Chemical Physics}\ }\textbf {\bibinfo {volume}
  {15}},\ \bibinfo {pages} {11620} (\bibinfo {year}
  {2013}{\natexlab{a}})}\BibitemShut {NoStop}%
\bibitem [{\citenamefont {Giannozzi}\ \emph {et~al.}(2009)\citenamefont
  {Giannozzi}, \citenamefont {Baroni}, \citenamefont {Bonini}, \citenamefont
  {Calandra}, \citenamefont {Car}, \citenamefont {Cavazzoni}, \citenamefont
  {Ceresoli}, \citenamefont {Chiarotti}, \citenamefont {Cococcioni},
  \citenamefont {Dabo}, \citenamefont {Corso}, \citenamefont {Gironcoli},
  \citenamefont {Fabris}, \citenamefont {Fratesi}, \citenamefont {Gebauer},
  \citenamefont {Gerstmann}, \citenamefont {Gougoussis}, \citenamefont
  {Kokalj}, \citenamefont {Lazzeri}, \citenamefont {Martin-Samos},
  \citenamefont {Marzari}, \citenamefont {Mauri}, \citenamefont {Mazzarello},
  \citenamefont {Paolini}, \citenamefont {Pasquarello}, \citenamefont
  {Paulatto}, \citenamefont {Sbraccia}, \citenamefont {Scandolo}, \citenamefont
  {Sclauzero}, \citenamefont {Seitsonen}, \citenamefont {Smogunov},
  \citenamefont {Umari},\ and\ \citenamefont
  {Wentzcovitch}}]{giannozzi_quantum_2009}%
  \BibitemOpen
  \bibfield  {author} {\bibinfo {author} {\bibfnamefont {P.}~\bibnamefont
  {Giannozzi}}, \bibinfo {author} {\bibfnamefont {S.}~\bibnamefont {Baroni}},
  \bibinfo {author} {\bibfnamefont {N.}~\bibnamefont {Bonini}}, \bibinfo
  {author} {\bibfnamefont {M.}~\bibnamefont {Calandra}}, \bibinfo {author}
  {\bibfnamefont {R.}~\bibnamefont {Car}}, \bibinfo {author} {\bibfnamefont
  {C.}~\bibnamefont {Cavazzoni}}, \bibinfo {author} {\bibfnamefont
  {D.}~\bibnamefont {Ceresoli}}, \bibinfo {author} {\bibfnamefont {G.~L.}\
  \bibnamefont {Chiarotti}}, \bibinfo {author} {\bibfnamefont {M.}~\bibnamefont
  {Cococcioni}}, \bibinfo {author} {\bibfnamefont {I.}~\bibnamefont {Dabo}},
  \bibinfo {author} {\bibfnamefont {A.~D.}\ \bibnamefont {Corso}}, \bibinfo
  {author} {\bibfnamefont {S.~d.}\ \bibnamefont {Gironcoli}}, \bibinfo {author}
  {\bibfnamefont {S.}~\bibnamefont {Fabris}}, \bibinfo {author} {\bibfnamefont
  {G.}~\bibnamefont {Fratesi}}, \bibinfo {author} {\bibfnamefont
  {R.}~\bibnamefont {Gebauer}}, \bibinfo {author} {\bibfnamefont
  {U.}~\bibnamefont {Gerstmann}}, \bibinfo {author} {\bibfnamefont
  {C.}~\bibnamefont {Gougoussis}}, \bibinfo {author} {\bibfnamefont
  {A.}~\bibnamefont {Kokalj}}, \bibinfo {author} {\bibfnamefont
  {M.}~\bibnamefont {Lazzeri}}, \bibinfo {author} {\bibfnamefont
  {L.}~\bibnamefont {Martin-Samos}}, \bibinfo {author} {\bibfnamefont
  {N.}~\bibnamefont {Marzari}}, \bibinfo {author} {\bibfnamefont
  {F.}~\bibnamefont {Mauri}}, \bibinfo {author} {\bibfnamefont
  {R.}~\bibnamefont {Mazzarello}}, \bibinfo {author} {\bibfnamefont
  {S.}~\bibnamefont {Paolini}}, \bibinfo {author} {\bibfnamefont
  {A.}~\bibnamefont {Pasquarello}}, \bibinfo {author} {\bibfnamefont
  {L.}~\bibnamefont {Paulatto}}, \bibinfo {author} {\bibfnamefont
  {C.}~\bibnamefont {Sbraccia}}, \bibinfo {author} {\bibfnamefont
  {S.}~\bibnamefont {Scandolo}}, \bibinfo {author} {\bibfnamefont
  {G.}~\bibnamefont {Sclauzero}}, \bibinfo {author} {\bibfnamefont {A.~P.}\
  \bibnamefont {Seitsonen}}, \bibinfo {author} {\bibfnamefont {A.}~\bibnamefont
  {Smogunov}}, \bibinfo {author} {\bibfnamefont {P.}~\bibnamefont {Umari}}, \
  and\ \bibinfo {author} {\bibfnamefont {R.~M.}\ \bibnamefont {Wentzcovitch}},\
  }\href {\doibase 10.1088/0953-8984/21/39/395502} {\bibfield  {journal}
  {\bibinfo  {journal} {Journal of Physics: Condensed Matter}\ }\textbf
  {\bibinfo {volume} {21}},\ \bibinfo {pages} {395502} (\bibinfo {year}
  {2009})}\BibitemShut {NoStop}%
\bibitem [{\citenamefont {Perdew}\ \emph {et~al.}(1996)\citenamefont {Perdew},
  \citenamefont {Burke},\ and\ \citenamefont
  {Ernzerhof}}]{perdew_generalized_1996}%
  \BibitemOpen
  \bibfield  {author} {\bibinfo {author} {\bibfnamefont {J.~P.}\ \bibnamefont
  {Perdew}}, \bibinfo {author} {\bibfnamefont {K.}~\bibnamefont {Burke}}, \
  and\ \bibinfo {author} {\bibfnamefont {M.}~\bibnamefont {Ernzerhof}},\ }\href
  {\doibase 10.1103/PhysRevLett.77.3865} {\bibfield  {journal} {\bibinfo
  {journal} {Physical Review Letters}\ }\textbf {\bibinfo {volume} {77}},\
  \bibinfo {pages} {3865} (\bibinfo {year} {1996})}\BibitemShut {NoStop}%
\bibitem [{\citenamefont {Kuhn}\ \emph
  {et~al.}(2013{\natexlab{b}})\citenamefont {Kuhn}, \citenamefont {Duppel},\
  and\ \citenamefont {Lotsch}}]{kuhn_tetragonal_2013}%
  \BibitemOpen
  \bibfield  {author} {\bibinfo {author} {\bibfnamefont {A.}~\bibnamefont
  {Kuhn}}, \bibinfo {author} {\bibfnamefont {V.}~\bibnamefont {Duppel}}, \ and\
  \bibinfo {author} {\bibfnamefont {B.~V.}\ \bibnamefont {Lotsch}},\ }\href
  {\doibase 10.1039/C3EE41728J} {\bibfield  {journal} {\bibinfo  {journal}
  {Energy \& Environmental Science}\ }\textbf {\bibinfo {volume} {6}},\
  \bibinfo {pages} {3548} (\bibinfo {year} {2013}{\natexlab{b}})}\BibitemShut
  {NoStop}%
\bibitem [{\citenamefont {Musaelian}\ and\ \citenamefont
  {Kahle}(2018)}]{sitator}%
  \BibitemOpen
  \bibfield  {author} {\bibinfo {author} {\bibfnamefont {A.}~\bibnamefont
  {Musaelian}}\ and\ \bibinfo {author} {\bibfnamefont {L.}~\bibnamefont
  {Kahle}},\ }\href@noop {} {\enquote {\bibinfo {title} {\texttt{SITATOR}},}\
  }\bibinfo {howpublished} {\url{https://github.com/Linux-cpp-lisp/sitator}}
  (\bibinfo {year} {2018})\BibitemShut {NoStop}%
\bibitem [{\citenamefont {He}\ \emph {et~al.}(2017)\citenamefont {He},
  \citenamefont {Zhu},\ and\ \citenamefont {Mo}}]{he_origin_2017}%
  \BibitemOpen
  \bibfield  {author} {\bibinfo {author} {\bibfnamefont {X.}~\bibnamefont
  {He}}, \bibinfo {author} {\bibfnamefont {Y.}~\bibnamefont {Zhu}}, \ and\
  \bibinfo {author} {\bibfnamefont {Y.}~\bibnamefont {Mo}},\ }\href {\doibase
  10.1038/ncomms15893} {\bibfield  {journal} {\bibinfo  {journal} {Nature
  Communications}\ }\textbf {\bibinfo {volume} {8}},\ \bibinfo {pages} {15893}
  (\bibinfo {year} {2017})}\BibitemShut {NoStop}%
\bibitem [{\citenamefont {Willems}\ \emph {et~al.}(2012)\citenamefont
  {Willems}, \citenamefont {Rycroft}, \citenamefont {Kazi}, \citenamefont
  {Meza},\ and\ \citenamefont {Haranczyk}}]{willems_algorithms_2012}%
  \BibitemOpen
  \bibfield  {author} {\bibinfo {author} {\bibfnamefont {T.~F.}\ \bibnamefont
  {Willems}}, \bibinfo {author} {\bibfnamefont {C.~H.}\ \bibnamefont
  {Rycroft}}, \bibinfo {author} {\bibfnamefont {M.}~\bibnamefont {Kazi}},
  \bibinfo {author} {\bibfnamefont {J.~C.}\ \bibnamefont {Meza}}, \ and\
  \bibinfo {author} {\bibfnamefont {M.}~\bibnamefont {Haranczyk}},\ }\href
  {\doibase 10.1016/j.micromeso.2011.08.020} {\bibfield  {journal} {\bibinfo
  {journal} {Microporous and Mesoporous Materials}\ }\textbf {\bibinfo {volume}
  {149}},\ \bibinfo {pages} {134} (\bibinfo {year} {2012})}\BibitemShut
  {NoStop}%
\bibitem [{\citenamefont {Martin}\ \emph {et~al.}(2012)\citenamefont {Martin},
  \citenamefont {Smit},\ and\ \citenamefont
  {Haranczyk}}]{martin_addressing_2012}%
  \BibitemOpen
  \bibfield  {author} {\bibinfo {author} {\bibfnamefont {R.~L.}\ \bibnamefont
  {Martin}}, \bibinfo {author} {\bibfnamefont {B.}~\bibnamefont {Smit}}, \ and\
  \bibinfo {author} {\bibfnamefont {M.}~\bibnamefont {Haranczyk}},\ }\href
  {\doibase 10.1021/ci200386x} {\bibfield  {journal} {\bibinfo  {journal}
  {Journal of Chemical Information and Modeling}\ }\textbf {\bibinfo {volume}
  {52}},\ \bibinfo {pages} {308} (\bibinfo {year} {2012})}\BibitemShut
  {NoStop}%
\bibitem [{\citenamefont {Satopaa}\ \emph {et~al.}(2011)\citenamefont
  {Satopaa}, \citenamefont {Albrecht}, \citenamefont {Irwin},\ and\
  \citenamefont {Raghavan}}]{satopaa_finding_2011}%
  \BibitemOpen
  \bibfield  {author} {\bibinfo {author} {\bibfnamefont {V.}~\bibnamefont
  {Satopaa}}, \bibinfo {author} {\bibfnamefont {J.}~\bibnamefont {Albrecht}},
  \bibinfo {author} {\bibfnamefont {D.}~\bibnamefont {Irwin}}, \ and\ \bibinfo
  {author} {\bibfnamefont {B.}~\bibnamefont {Raghavan}},\ }in\ \href {\doibase
  10.1109/ICDCSW.2011.20} {\emph {\bibinfo {booktitle} {2011 31st International
  Conference on Distributed Computing Systems Workshops}}}\ (\bibinfo {year}
  {2011})\ pp.\ \bibinfo {pages} {166--171}\BibitemShut {NoStop}%
\bibitem [{\citenamefont {Prandini}\ \emph {et~al.}(2018)\citenamefont
  {Prandini}, \citenamefont {Marrazzo}, \citenamefont {Castelli}, \citenamefont
  {Mounet},\ and\ \citenamefont {Marzari}}]{prandini_precision_2018}%
  \BibitemOpen
  \bibfield  {author} {\bibinfo {author} {\bibfnamefont {G.}~\bibnamefont
  {Prandini}}, \bibinfo {author} {\bibfnamefont {A.}~\bibnamefont {Marrazzo}},
  \bibinfo {author} {\bibfnamefont {I.~E.}\ \bibnamefont {Castelli}}, \bibinfo
  {author} {\bibfnamefont {N.}~\bibnamefont {Mounet}}, \ and\ \bibinfo {author}
  {\bibfnamefont {N.}~\bibnamefont {Marzari}},\ }\href {\doibase
  10.1038/s41524-018-0127-2} {\bibfield  {journal} {\bibinfo  {journal} {npj
  Computational Materials}\ }\textbf {\bibinfo {volume} {4}},\ \bibinfo {pages}
  {72} (\bibinfo {year} {2018})}\BibitemShut {NoStop}%
\bibitem [{\citenamefont {Pizzi}\ \emph {et~al.}(2016)\citenamefont {Pizzi},
  \citenamefont {Cepellotti}, \citenamefont {Sabatini}, \citenamefont
  {Marzari},\ and\ \citenamefont {Kozinsky}}]{pizzi_aiida:_2016}%
  \BibitemOpen
  \bibfield  {author} {\bibinfo {author} {\bibfnamefont {G.}~\bibnamefont
  {Pizzi}}, \bibinfo {author} {\bibfnamefont {A.}~\bibnamefont {Cepellotti}},
  \bibinfo {author} {\bibfnamefont {R.}~\bibnamefont {Sabatini}}, \bibinfo
  {author} {\bibfnamefont {N.}~\bibnamefont {Marzari}}, \ and\ \bibinfo
  {author} {\bibfnamefont {B.}~\bibnamefont {Kozinsky}},\ }\href {\doibase
  10.1016/j.commatsci.2015.09.013} {\bibfield  {journal} {\bibinfo  {journal}
  {Computational Materials Science}\ }\textbf {\bibinfo {volume} {111}},\
  \bibinfo {pages} {218} (\bibinfo {year} {2016})}\BibitemShut {NoStop}%
\bibitem [{\citenamefont {Bussi}\ \emph {et~al.}(2007)\citenamefont {Bussi},
  \citenamefont {Donadio},\ and\ \citenamefont
  {Parrinello}}]{bussi_canonical_2007}%
  \BibitemOpen
  \bibfield  {author} {\bibinfo {author} {\bibfnamefont {G.}~\bibnamefont
  {Bussi}}, \bibinfo {author} {\bibfnamefont {D.}~\bibnamefont {Donadio}}, \
  and\ \bibinfo {author} {\bibfnamefont {M.}~\bibnamefont {Parrinello}},\
  }\href {\doibase 10.1063/1.2408420} {\bibfield  {journal} {\bibinfo
  {journal} {The Journal of Chemical Physics}\ }\textbf {\bibinfo {volume}
  {126}},\ \bibinfo {pages} {014101} (\bibinfo {year} {2007})}\BibitemShut
  {NoStop}%
\bibitem [{\citenamefont {Belsky}\ \emph {et~al.}(2002)\citenamefont {Belsky},
  \citenamefont {Hellenbrandt}, \citenamefont {Karen},\ and\ \citenamefont
  {Luksch}}]{belsky_new_2002}%
  \BibitemOpen
  \bibfield  {author} {\bibinfo {author} {\bibfnamefont {A.}~\bibnamefont
  {Belsky}}, \bibinfo {author} {\bibfnamefont {M.}~\bibnamefont
  {Hellenbrandt}}, \bibinfo {author} {\bibfnamefont {V.~L.}\ \bibnamefont
  {Karen}}, \ and\ \bibinfo {author} {\bibfnamefont {P.}~\bibnamefont
  {Luksch}},\ }\href {\doibase 10.1107/S0108768102006948} {\bibfield  {journal}
  {\bibinfo  {journal} {Acta Crystallographica Section B Structural Science}\
  }\textbf {\bibinfo {volume} {58}},\ \bibinfo {pages} {364} (\bibinfo {year}
  {2002})}\BibitemShut {NoStop}%
\end{thebibliography}%

\end{document}